\documentclass[times,final]{elsarticle}

\usepackage{jcomp}
\usepackage{framed,multirow}

\usepackage{amssymb}
\usepackage{amsmath}
\usepackage{latexsym}
\usepackage{bm}
\usepackage{array}
\usepackage{subfigure}
\usepackage{url}
\usepackage{xcolor}
\usepackage{color}
\usepackage{float}
\usepackage {soul}
\definecolor{zz}{rgb}{.8,.349,.1}
\definecolor{dark-gray}{gray}{0.50}

\usepackage{natbib}
\usepackage{hyperref}
\hypersetup{
    colorlinks = true,	
    linkcolor   = blue,	
    citecolor  = blue,	
    urlcolor   = blue	
}
\usepackage{lineno}

\journal{Journal of Computational Physics}

\begin{document}

\verso{S. Zhao \textit{et al.}}

\begin{frontmatter}

\title{A Sharp and Conservative VOF Method for Multicomponent Liquid--Gas Mass Transfer: Bubble Dissolution and Droplet Evaporation}%

\author[1,3]{Shuo \snm{Zhao}}

\author[2]{Jie \snm{Zhang}\corref{cor1}}
\cortext[cor1]{Corresponding author:}
\ead{j_zhang@xjtu.edu.cn}

\author[1,2]{Ming-Jiu \snm{Ni}\corref{cor1}}
\ead{mjni@ucas.as.cn}

\address[1]{ School of Engineering Science, University of Chinese Academy of Sciences, Beijing 101408, China}
\address[2]{ State Key Laboratory for Strength and Vibration of Mechanical Structures, School of Aerospace, Xi'an Jiaotong University, Xi'an, Shaanxi 710049, China}
\address[3]{Institute of Applied Physics and Computational Mathematics, Beijing 100094, China}


\begin{abstract}
We present a sharp and conservative geometrical VOF--finite-volume method for multicomponent liquid--gas mass transfer across deformable interfaces. The method solves problems in which multiple species are coupled through interfacial mass balances, latent-heat exchange, and vapor--liquid equilibrium. The key novelty is a fully sharp two-field treatment of scalar transport: the species and temperature equations are solved separately in the liquid and gas phases, while the one-sided Robin conditions for species and the two-sided flux jump for temperature are imposed directly on the reconstructed interface through an embedded-boundary discretization. This avoids both volumetric regularization of interfacial source terms and explicit coupling based on previous-time-step interfacial data. A consistent geometrical advection scheme is used for volume, momentum, energy, and species transport, and a sequential coupling strategy is developed to determine the partial interfacial mass fluxes, close the temperature equation, and update the thermodynamic-equilibrium state. The method is validated through single- and multicomponent bubble dissolution, single-component droplet evaporation, non-ideal ethanol--isooctane droplet evaporation, and sessile water--glycerol droplet evaporation. The results demonstrate second-order accuracy, accurate interfacial flux prediction, good mass and energy conservation, and the ability to capture complex multicomponent effects such as gas replacement, azeotropic volatility reversal, and composition-driven Marangoni flow.
\end{abstract}


\end{frontmatter}


\section{Introduction}\label{sec1}

Multiphase flows involving liquid--gas interfaces arise widely in both natural environments and engineering applications, and in many such systems both phases contain multiple chemical components. Representative examples include dissolving gas bubbles composed of $\mathrm{CO_2}$, $\mathrm{N_2}$, and $\mathrm{O_2}$, as well as evaporating fuel droplets consisting of hydrocarbon mixtures and oxygenated additives. In these problems, mass transfer across the liquid--gas interface is essentially multicomponent: different species evaporate or dissolve at different rates according to their own transport properties and thermodynamic affinities, while the associated latent-heat exchange couples species transport to the temperature field. Consequently, multicomponent liquid--gas mass transfer is substantially more complex than its single-component counterpart. Although experiments provide valuable macroscopic observations, they generally cannot resolve the detailed interfacial transport processes and local flow structures with sufficient spatio-temporal resolution. Direct numerical simulation (DNS), by contrast, offers a powerful framework for examining the coupled dynamics of flow, heat transfer, and multicomponent mass transfer in a fully resolved manner.

From the numerical viewpoint, phase-change problems driven by simultaneous heat and mass transfer involve two closely related difficulties. The first is the treatment of the discontinuous flow field induced by interfacial mass transfer, since the singular mass source localized at the moving interface generates a jump in the normal velocity and may lead to numerical instability if it is not handled properly. The second is the accurate imposition of the interfacial thermodynamic conditions. In the single-component case, such as pure-droplet evaporation, the interface acts as a moving internal boundary for the temperature and vapor-concentration fields, and the interfacial closure can often be formulated through a Dirichlet condition for temperature and a Robin condition for vapor concentration \citep{zhao2022boiling}. A widely adopted strategy is then to regularize these discontinuities into volumetric source terms distributed over several grid cells near the interface \citep{irfan2017front,wang2019vaporization,palmore2019volume,kunkelmann2009cfd}. While such diffusive treatments are attractive because of their simplicity and robustness, they inevitably smear the interfacial jumps and thus degrade the local physical fidelity. Sharp-interface methods, in contrast, seek to impose the interfacial conditions directly without introducing an artificial interface thickness. Among them, the ghost-fluid method (GFM) combined with the level-set (LS) approach has become one of the most widely used frameworks \citep{fedkiw1999non,tanguy2014benchmarks,gibou2007level,villegas2016ghost}. More recently, genuinely sharp treatments within a geometrical volume-of-fluid (VOF) framework have also begun to emerge. In particular, as a precursor to the present study, \citet{zhao2022boiling} developed a sharp and conservative finite-volume method based on geometrical VOF and an embedded-boundary method (EBM), in which the interfacial Dirichlet and Robin conditions are incorporated directly into the implicit flux discretization. This formulation combines the conservation advantages of VOF with a sharp treatment of the interfacial transport, and therefore provides a natural starting point for extension to multicomponent problems.

Extending such a sharp VOF formulation from single-component to multicomponent mass transfer is, however, far from straightforward. The essential difficulty is that the interfacial closure changes fundamentally. In the single-component problem, the interfacial thermodynamic state is governed by only one thermal condition and one species relation. In the multicomponent case, by contrast, both parts of the closure become considerably more involved. First, the interfacial temperature is no longer associated with a single latent-heat contribution, but is coupled to the combined effect of all transferring species. Second, each species satisfies its own interfacial mass-balance condition, while the liquid-side and gas-side interfacial compositions are linked by species-dependent vapor--liquid equilibrium relations. As a result, the interfacial temperature, the interfacial species concentrations on both sides of the interface, and the partial mass fluxes must be determined as a coupled, generally nonlinear system. This complexity largely explains why diffusive-interface treatments remain prevalent in VOF-based simulations of multicomponent phase change and mass transfer. For example, \citet{haroun2010volume} proposed a CSF-like treatment in which the interfacial mass flux is introduced into the concentration equation as a distributed source term. Algebraic VOF methods have also been widely adopted because of their relative simplicity and their natural compatibility with smeared-interface formulations; see, for instance, \citet{graveleau2017pore}, \citet{maes2018new,maes2020unified}, and \citet{zanutto2022modeling1,zanutto2022modeling}. Diffusive treatments have likewise been combined with geometrical VOF formulations for interfacial heat and mass transfer \citep{bothe2013volume,fleckenstein2015volume,farsoiya2023direct}. Although these approaches are often effective, their reliance on interfacial smearing becomes a serious limitation in problems featuring very thin concentration boundary layers on the liquid side, where accurate evaluation of the interfacial flux requires extremely fine grids.

Two recent VOF-based multicomponent studies are particularly relevant to the present work. \citet{cipriano2024multicomponent} proposed a second-order procedure in which the partial and total interfacial mass fluxes are evaluated from interfacial information at the previous time step. This treatment avoids solving the fully coupled interfacial problem at the current time level, but effectively relaxes the original implicit coupling into an explicit one. \citet{salimi2024volume} introduced implicit Robin conditions for different species through a ghost-fluid-type strategy; however, the interfacial heat-flux jump in the temperature equation is still treated by a diffusive one-field formulation. Therefore, despite these important advances, a sharp and conservative VOF method that treats \emph{both} the species fields and the temperature field in a fully consistent manner is still lacking.

In this paper, we develop such a method within a geometrical VOF--finite-volume framework. The key idea is to retain the conservative and non-diffusive interface transport of geometrical VOF, while imposing the multicomponent interfacial conditions sharply through an embedded-boundary discretization of the diffusion terms. Specifically, the species equations are solved separately in the two phases, with phase-wise Robin conditions imposed sharply at the interface. The temperature equation is also treated in a fully two-field manner, so that the liquid-side and gas-side diffusion problems are coupled directly through the interfacial heat-flux jump. A coupling strategy is further introduced to determine the partial interfacial mass fluxes from the phase-wise species fields and to use these fluxes consistently in closing the temperature equation and the thermodynamic-equilibrium relations. The resulting framework is sharp, conservative, and provides a unified treatment for dissolving bubbles and evaporating droplets.

The remainder of the paper is organized as follows. Section~\ref{sec2} presents the governing equations together with the interfacial jump and thermodynamic-equilibrium conditions. Section~\ref{sec3} describes the numerical treatment of the flow field and interface transport. Section~\ref{sec4} details the numerical method for the temperature and species equations, with emphasis on the sharp diffusion discretization and the coupling strategy. Section~\ref{sec5} then presents a series of validation and application examples, including static and freely rising bubbles with dissolution, multicomponent dissolution of carbon-dioxide bubbles, evaporation of single-component $n$-heptane droplets, evaporation of non-ideal ethanol--isooctane droplets, and evaporation of sessile water--glycerol droplets. Finally, the main conclusions are summarized in Section~\ref{sec6}.

\section{Problem statement and governing equations}\label{sec2}

\begin{figure}
\centering
\includegraphics[scale=0.5]{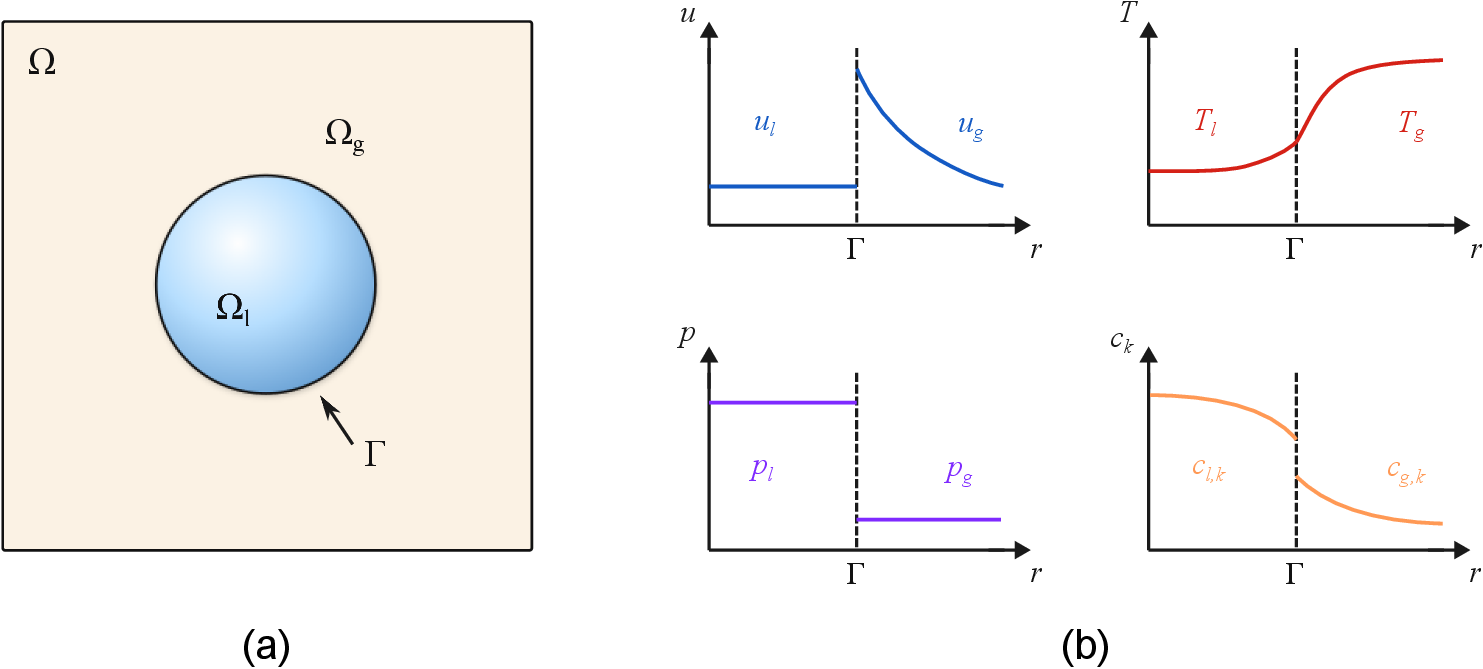}
\caption{Schematic of multicomponent droplet evaporation. (a) Problem configuration. (b) Interfacial jump conditions induced by liquid--gas mass transfer, where $\Gamma$ denotes the interface, $u$ the normal velocity component, $p$ the pressure, $T$ the temperature, and $c_k$ the concentration of the $k$-th species.}
\label{f2.1}
\end{figure}

We consider incompressible two-phase flows involving multicomponent mass transfer across a deformable liquid--gas interface. Depending on the physical configuration, the dispersed phase may be either a multicomponent liquid droplet undergoing evaporation or a multicomponent gas bubble undergoing dissolution. For clarity, Fig.~\ref{f2.1}(a) illustrates the evaporation of a multicomponent droplet, but the formulation developed below applies equally to bubble-dissolution problems. The computational domain, denoted by $\Omega$, consists of a liquid region $\Omega_l$ and a gas region $\Omega_g$, separated by a deformable zero-thickness interface $\Gamma$. The subscripts $l$ and $g$ are used for liquid- and gas-phase quantities, respectively. The liquid phase contains $N_l$ chemical species, while the gas phase is modeled as an ideal mixture composed of $N_v$ vapor species and an inert gas. Owing to interfacial mass transfer, mass, momentum, and energy are continuously exchanged across $\Gamma$, so that the two phases are coupled through both transport processes and thermodynamic equilibrium at the interface.

\subsection{Governing equations}\label{sec2.0}

The present multicomponent phase-change problem follows the same hydrodynamic framework as the single-component formulation developed in our previous work \citep{zhao2022boiling}. The main additional complexity lies in the scalar transport and interfacial closure, because multiple species are exchanged across the liquid--gas interface and are coupled to the temperature field through thermodynamic equilibrium and latent-heat effects. Over the entire fluid domain $\Omega=\Omega_l\cup\Omega_g$, the flow field is governed by the one-fluid incompressible Navier--Stokes equations,
\begin{equation}
\rho \left(\frac{\partial \boldsymbol{u}}{\partial t} + \boldsymbol{u}\cdot\nabla \boldsymbol{u}\right)
=
-\nabla p
+
\nabla\cdot\left[\mu\left(\nabla\boldsymbol{u}+\nabla\boldsymbol{u}^{T}\right)\right]
+
\rho\boldsymbol{g}
+
\sigma\kappa\delta_s\boldsymbol{n}
+
\nabla_t\sigma\,\delta_s,
\label{e1}
\end{equation}
together with the divergence constraint
\begin{equation}
\nabla\cdot\boldsymbol{u}
=
\left(\frac{1}{\rho_g}-\frac{1}{\rho_l}\right)\ddot{m}.
\label{e2}
\end{equation}
Here, $\boldsymbol{u}$ is the velocity ($\mathrm{m/s}$), $p$ the pressure ($\mathrm{Pa}$), $\rho$ the density ($\mathrm{kg/m^3}$), $\mu$ the dynamic viscosity ($\mathrm{Pa\cdot s}$), and $\boldsymbol{g}$ the gravitational acceleration ($\mathrm{m/s^2}$). The last two terms in Eq.~\eqref{e1} represent the normal capillary force and the tangential Marangoni stress, respectively, where $\sigma$ is the surface-tension coefficient ($\mathrm{N/m}$), $\kappa$ the interface curvature ($\mathrm{m^{-1}}$), $\boldsymbol{n}$ the unit normal vector pointing from the gas phase to the liquid phase, $\nabla_t\sigma$ the surface gradient of $\sigma$, and $\delta_s$ the Dirac distribution concentrated on the interface. In Eq.~\eqref{e2}, $\ddot m=\dot m S_{\Gamma}/V$ is the volumetric mass source in an interfacial cell, with $\dot m$ the interfacial mass flux ($\mathrm{kg/(m^2\,s)}$), $S_{\Gamma}$ the interfacial length in two dimensions or area in three dimensions, and $V$ the cell volume. Since $\dot m$ is localized on $\Gamma$, the divergence-free condition is recovered in pure liquid and pure gas cells, consistently with the incompressibility of each bulk phase.

The interface $\Gamma$ is tracked by a geometrical volume-of-fluid (VOF) method. The liquid volume fraction in a control volume is denoted by $f$, and the material properties are defined through the one-fluid interpolation
\[
\rho=f\rho_l+(1-f)\rho_g,
\qquad
\mu=f\mu_l+(1-f)\mu_g,
\]
where the latter is used as a practical mixture law for the viscosity. The volume fraction evolves according to
\begin{equation}
\frac{\partial f}{\partial t}+(\boldsymbol{u}\cdot\nabla)f=-\frac{\ddot m}{\rho},
\label{evof}
\end{equation}
where the right-hand side accounts for the displacement of the interface induced by interfacial mass transfer.

In addition to the flow field, one must also advance the species concentrations and the temperature field. For the $k$-th species, the concentration field satisfies
\begin{equation}
\frac{\partial c_k}{\partial t}+\boldsymbol{u}\cdot\nabla c_k
=
\nabla\cdot(D_k\nabla c_k)+R_k,
\label{e3}
\end{equation}
while the temperature field obeys
\begin{equation}
\rho c_p\left(\frac{\partial T}{\partial t}+\boldsymbol{u}\cdot\nabla T\right)
=
\nabla\cdot(\lambda\nabla T)+Q.
\label{e4}
\end{equation}
Here, $c_k$ denotes the concentration of the $k$-th species ($\mathrm{kg/m^3}$), $D_k$ its diffusion coefficient ($\mathrm{m^2/s}$), $T$ the temperature ($\mathrm{K}$), $c_p$ the specific heat capacity ($\mathrm{J/(kg\cdot K)}$), and $\lambda$ the thermal conductivity ($\mathrm{W/(m\cdot K)}$). The source terms $R_k$ and $Q$ may account for additional effects such as chemical reactions or volumetric heat release, but are set to zero in the present study.

A key point is that Eqs.~\eqref{e3} and \eqref{e4} are not solved in a one-fluid form, unlike the momentum equation. Instead, the species and temperature equations are solved separately in $\Omega_l$ and $\Omega_g$, and the two phase-wise scalar fields are coupled only through boundary and jump conditions imposed on the moving interface $\Gamma$. The formulation is therefore one-fluid for hydrodynamics but two-field for scalar transport. This distinction is essential: the interfacial conditions for temperature and species are enforced directly on the reconstructed interface, rather than being regularized as volumetric source terms. Consequently, the main additional difficulty of the multicomponent problem does not lie in the bulk advection--diffusion equations themselves, but in the interfacial closure that couples temperature, species concentrations, and partial mass fluxes. We now turn to these interfacial conditions.

\subsection{Jump conditions across the interface}\label{sec2.1}

Interfacial mass transfer gives rise to discontinuities in both kinematic and thermodynamic quantities across the moving interface $\Gamma$. These discontinuities, sketched in Fig.~\ref{f2.1}(b), cannot be recovered from the bulk equations alone and must instead be imposed through interfacial jump conditions. In the present multicomponent problem, the relevant conditions involve the velocity and stress fields, the temperature field, and the concentration fields of all transferring species. Unless otherwise stated, the jump of a quantity $\phi$ is defined as $[\phi]_{\Gamma}=\phi_g-\phi_l$ across the interface.

\textit{\textbf{Velocity jump.}}  
The kinematic condition at the interface follows from conservation of mass across $\Gamma$. The interfacial mass flux satisfies
\begin{equation}
\dot m
=
\rho_l(\boldsymbol{u}_l-\boldsymbol{u}_{\Gamma})\cdot\boldsymbol{n}
=
\rho_g(\boldsymbol{u}_g-\boldsymbol{u}_{\Gamma})\cdot\boldsymbol{n},
\label{e5}
\end{equation}
where $\boldsymbol{u}_{\Gamma}$ is the interface velocity. Eq.~\eqref{e5} directly yields the jump in the normal velocity,
\begin{equation}
[\boldsymbol{u}\cdot\boldsymbol{n}]_{\Gamma}
=
\left(\frac{1}{\rho_g}-\frac{1}{\rho_l}\right)\dot m .
\label{e6}
\end{equation}
Thus, even though each bulk phase is incompressible, the one-fluid velocity field is not divergence-free at the interface when $\dot m\neq0$. This velocity jump is the origin of the singular source term in Eq.~\eqref{e2} and must be treated consistently in the VOF transport equation.

\textit{\textbf{Stress jump.}}  
Momentum conservation across the interface gives the normal and tangential stress balances
\begin{equation}
[\boldsymbol{n}\cdot\mathbb{T}\cdot\boldsymbol{n}]_{\Gamma}
=
\sigma\kappa+\dot m[\boldsymbol{u}\cdot\boldsymbol{n}]_{\Gamma},
\label{e7}
\end{equation}
and
\begin{equation}
[\boldsymbol{n}\cdot\mathbb{T}\cdot\boldsymbol{t}]_{\Gamma}
=
\nabla_t\sigma,
\label{e8}
\end{equation}
where $\mathbb{T}=-p\mathbb{I}+\mu(\nabla\boldsymbol{u}+\nabla\boldsymbol{u}^T)$ is the stress tensor, $\mathbb{I}$ is the identity tensor, and $\boldsymbol{t}$ denotes a unit tangent vector along the interface. In Eq.~\eqref{e7}, the first term on the right-hand side is the Laplace contribution, while the second term represents the vapor-recoil effect induced by interfacial mass transfer. Eq.~\eqref{e8} expresses the tangential stress balance, where the Marangoni stress arises from non-uniform temperature or composition along $\Gamma$.

\textit{\textbf{Thermal condition.}}  
The temperature is assumed to be continuous across the interface, whereas the conductive heat flux generally exhibits a jump owing to the latent heat carried by the transferring species. The corresponding interfacial conditions are
\begin{equation}
[T]_{\Gamma}=0,
\label{e9}
\end{equation}
and
\begin{equation}
[\lambda\nabla T\cdot\boldsymbol{n}]_{\Gamma}
=
\sum_{k=1}^{N_t}\dot m_k\mathcal{L}_k,
\label{e10}
\end{equation}
where $N_t$ denotes the number of transferring species, $\dot m_k$ is the partial mass flux of the $k$-th species, and $\mathcal{L}_k$ is the corresponding latent heat ($\mathrm{J/kg}$). Unlike in the single-component case, the thermal jump in a multicomponent system is not determined by a single mass flux, but by the combined latent-heat contribution of all partial interfacial fluxes. This coupling is one of the reasons why the temperature equation cannot be closed independently of the species transport.

\textit{\textbf{Species mass balance.}}  
In contrast to the temperature, the concentration of each species is generally discontinuous across the interface. The liquid-side and gas-side interfacial concentrations are determined by thermodynamic equilibrium and may differ significantly. What must be conserved at $\Gamma$ is therefore not the concentration itself, but the partial mass flux of each transferring species. For the $k$-th species, this interfacial mass balance reads
\begin{equation}
\dot m_k
=
c_{l,k}(\boldsymbol{u}_{l,k}-\boldsymbol{u}_{\Gamma})\cdot\boldsymbol{n}
=
c_{g,k}(\boldsymbol{u}_{g,k}-\boldsymbol{u}_{\Gamma})\cdot\boldsymbol{n},
\label{e11}
\end{equation}
where $c_{l,k}$ and $c_{g,k}$ are the liquid-side and gas-side concentrations, respectively, and $\boldsymbol{u}_{l,k}$ and $\boldsymbol{u}_{g,k}$ are the corresponding species velocities. Since the species velocity generally differs from the bulk velocity, Eq.~\eqref{e11} can be decomposed into convective and diffusive contributions as
\begin{equation}
\begin{aligned}
\dot m_k
&=
c_k(\boldsymbol{u}_k-\boldsymbol{u}_{\Gamma})\cdot\boldsymbol{n}
\\
&=
c_k(\boldsymbol{u}-\boldsymbol{u}_{\Gamma})\cdot\boldsymbol{n}
+
c_k(\boldsymbol{u}_k-\boldsymbol{u})\cdot\boldsymbol{n}
\\
&=
\frac{c_k}{\rho}\dot m+\boldsymbol{J}_k\cdot\boldsymbol{n},
\end{aligned}
\label{e12}
\end{equation}
where the first term on the right-hand side represents convection induced by the total interfacial mass flux $\dot m$, and the second term represents diffusion relative to the bulk motion. In a general multicomponent mixture, $\boldsymbol{J}_k$ should be described by the Maxwell--Stefan formulation \citep{krishna1997maxwell}. In the present study, we restrict attention to cases in which Fickian diffusion provides an acceptable approximation, so that $\boldsymbol{J}_k=-D_k\nabla c_k$. The interfacial condition for the $k$-th species then becomes
\begin{equation}
\dot m_k
=
\frac{c_{l,k}}{\rho_l}\dot m
-
D_{l,k}\left(\frac{\partial c_{l,k}}{\partial n}\right)_{\Gamma}
=
\frac{c_{g,k}}{\rho_g}\dot m
-
D_{g,k}\left(\frac{\partial c_{g,k}}{\partial n}\right)_{\Gamma}.
\label{e14}
\end{equation}
Eq.~\eqref{e14} is a Robin-type condition for the phase-wise concentration fields. It states that the partial interfacial flux of each species results from the balance between convection associated with the total phase-change flux and diffusion driven by the normal concentration gradient. In a multicomponent system, such Robin conditions must be satisfied simultaneously for all transferring species, while the total flux $\dot m=\sum_{k=1}^{N_t}\dot m_k$ depends on the same unknown partial fluxes. This mutual dependence is the main source of the strong coupling between interfacial concentrations and interfacial mass fluxes.

\subsection{Thermodynamic equilibrium conditions}\label{sec2.2}

The interfacial mass-balance conditions introduced above do not determine the interfacial concentrations by themselves. In a multicomponent system, the concentration of each species is generally discontinuous across $\Gamma$, and the liquid-side and gas-side interfacial values must be related through local vapor--liquid equilibrium. This equilibrium depends on the interfacial temperature, pressure, and composition. Therefore, in addition to the flux-balance conditions, thermodynamic closure relations are required to determine the interfacial species concentrations.

Since vapor--liquid equilibrium is more naturally expressed in terms of mole fractions, we denote by $\zeta_{\alpha,k}$ the mole fraction of species $k$ in phase $\alpha$, where $\alpha=l$ or $g$. The interfacial mole fractions are related to the interfacial mass concentrations through
\begin{equation}
\widetilde{\zeta}_{\alpha,k}
=
\frac{\widetilde{n}_{\alpha,k}}
{\sum_i \widetilde{n}_{\alpha,i}}
=
\frac{\widetilde{c}_{\alpha,k}/\mathcal{M}_k}
{\sum_i \widetilde{c}_{\alpha,i}/\mathcal{M}_i},
\label{eq:zeta_def}
\end{equation}
where $\widetilde{(\cdot)}$ denotes an interfacial value $(\cdot)_{\Gamma}$, $\widetilde{n}_{\alpha,k}$ is the molar amount of species $k$ in phase $\alpha$, and $\mathcal{M}_k$ is its molar mass ($\mathrm{kg/kmol}$). The summation is taken over all species present in the corresponding phase, including non-transferring components such as the solvent in the liquid phase or the inert gas in the gas phase.

For the systems considered in this study, the gas phase is assumed to behave as an ideal gas mixture at atmospheric pressure. Under this assumption, the interfacial vapor--liquid equilibrium for an evaporating liquid mixture can be described by the modified Raoult's law \citep{poling2001properties,smith2018introduction},
\begin{equation}
\widetilde{\zeta}_{g,k}P
=
\gamma_k \widetilde{\zeta}_{l,k} P_k^{sat}(\widetilde{T}),
\label{e14-2}
\end{equation}
where $P$ is the total pressure, $P_k^{sat}(\widetilde{T})$ is the saturation vapor pressure of species $k$ at the interfacial temperature $\widetilde{T}$, and $\gamma_k$ is the liquid-phase activity coefficient. The latter accounts for the non-ideal behavior of real liquid mixtures caused by differences in molecular size and intermolecular interactions among the components.

In droplet-evaporation problems, including single-component droplets and non-dilute liquid mixtures, Eq.~\eqref{e14-2} is recast as a gas-side interfacial concentration determined by the liquid-side interfacial composition and temperature. Using the ideal-gas relation for species $k$ in the gas phase, one obtains
\begin{equation}
\widetilde{c}_{g,k}
=
\frac{\mathcal{M}_k P}{\mathcal{R}\widetilde{T}}
\widetilde{\zeta}_{g,k}
=
\frac{\mathcal{M}_k \gamma_k P_k^{sat}(\widetilde{T})}
{\mathcal{R}\widetilde{T}}
\widetilde{\zeta}_{l,k},
\label{e17}
\end{equation}
where $\mathcal{R}$ is the universal gas constant ($\mathrm{J/(mol\cdot K)}$). Eq.~\eqref{e17} provides the gas-side equilibrium concentration of each evaporating species from the liquid-side interfacial composition and the interfacial temperature, and is therefore used as the thermodynamic closure for droplet evaporation.

For bubble-dissolution problems, on the other hand, the transferred gases are usually dilute in the liquid phase, as in the dissolution of $\mathrm{O_2}$, $\mathrm{N_2}$, or $\mathrm{CO_2}$ in water under ambient conditions. In this limit, the equilibrium relation is more conveniently expressed by Henry's law \citep{poling2001properties,smith2018introduction}. Using a pressure-based convention for Henry's constant, the gas-side partial pressure of species $k$ satisfies
\begin{equation}
\widetilde{p}_{g,k}
=
\widetilde{\zeta}_{g,k}P
=
\mathcal{H}_k(\widetilde{T})\widetilde{\zeta}_{l,k},
\label{eq:henry_pressure}
\end{equation}
where $\widetilde{p}_{g,k}$ is the gas-side partial pressure and $\mathcal{H}_k(\widetilde{T})$ is Henry's constant in pressure units (Pa). Combining Eq.~\eqref{eq:henry_pressure} with the ideal-gas relation gives
\begin{equation}
\widetilde{c}_{g,k}
=
\frac{\mathcal{M}_k \mathcal{H}_k(\widetilde{T})}{\mathcal{R}\widetilde{T}}\widetilde{\zeta}_{l,k}
\approx
\frac{\mathcal{M}_s \mathcal{H}_k(\widetilde{T})}{\rho_l\mathcal{R}\widetilde{T}} \widetilde{c}_{l,k},
\label{eq:henry_c_general}
\end{equation}
where $\mathcal{M}_s$ is the molar mass of the solvent, and the liquid-side mole fraction is approximated as $\widetilde{\zeta}_{l,k}
\approx \frac{\widetilde{c}_{l,k}/\mathcal{M}_k} {\rho_l/\mathcal{M}_s} = \frac{\mathcal{M}_s}{\rho_l\mathcal{M}_k} \widetilde{c}_{l,k}$. In the numerical tests below, it is convenient to introduce the dimensionless concentration partition coefficient
\begin{equation}
H_k
=
\frac{\rho_l\mathcal{R}\widetilde{T}}
{\mathcal{M}_s \mathcal{H}_k(\widetilde{T})},
\label{eq:partition_coefficient}
\end{equation}
so that the dilute gas--liquid equilibrium condition can be written simply as
\begin{equation}
\widetilde{c}_{l,k}
=
H_k\widetilde{c}_{g,k}.
\label{e18}
\end{equation}

It should be noted that different conventions for Henry's constant exist in the literature, which may lead to algebraically different but equivalent forms of the gas--liquid equilibrium relation \citep{bothe2013volume,fleckenstein2015volume,deising2016unified,maes2020unified,farsoiya2021bubble}. In the present study, Eq.~\eqref{e17} is used for droplet evaporation, whereas Eq.~\eqref{e18} is adopted for bubble dissolution. Together with the interfacial flux balances in Section~\ref{sec2.1}, these equilibrium relations determine the interfacial temperature, species concentrations, and partial mass fluxes, thereby closing the multicomponent phase-change problem.

\subsection{Difficulties in numerical implementation}\label{sec2.3}

The numerical difficulty of the present problem is twofold. At the hydrodynamic level, interfacial mass transfer introduces a singular volumetric source term in the projection step and a jump in the normal velocity across $\Gamma$, which makes the construction of a continuous VOF advection velocity nontrivial. These issues are already present in single-component phase-change problems and can be treated within the sharp VOF framework developed in our previous work \citep{zhao2022boiling}. They will therefore only be briefly recalled in Section~\ref{sec3}.

The genuinely new difficulty lies in the scalar interfacial closure. In a multicomponent system, each transferring species satisfies a Robin-type interfacial mass balance, Eq.~\eqref{e14}, while the temperature field satisfies the heat-flux jump condition \eqref{e10}. These conditions are not independent: the thermal jump depends on the sum of all partial mass fluxes $\dot m_k$, whereas each $\dot m_k$ depends on the phase-wise concentration fields and on the interfacial equilibrium state. In addition, the liquid-side and gas-side interfacial concentrations are generally discontinuous and must be linked through thermodynamic-equilibrium relations such as Eqs.~\eqref{e17} and \eqref{e18}. Thus, the interfacial temperature, the phase-wise interfacial concentrations, and the partial mass fluxes form a coupled local closure problem.

This coupling prevents a direct extension of single-component phase-change algorithms. In particular, the heat-flux jump cannot determine the individual partial mass fluxes, and the species equations cannot be closed without the thermodynamic interfacial state. The objective of the present work is to construct a sharp two-field formulation in which the species Robin conditions, the thermal flux jump, and the thermodynamic-equilibrium relations are imposed directly on the reconstructed interface.


\section{Numerical methods for fluid flow and interface transport}\label{sec3}

The numerical method is implemented in the open-source framework \textit{Basilisk}, originally developed by Popinet \citep{popinet2009accurate,popinet2014basilisk,popinet2015quadtree}. \textit{Basilisk} discretizes the governing equations by a finite-volume method on Cartesian meshes. In the present implementation, the primary variables, including the velocity $\boldsymbol{u}$, pressure $p$, temperature $T$, and species concentrations $c_k$, are stored at cell centers. The liquid--gas interface is represented by a geometrical volume-of-fluid (VOF) method based on piecewise-linear interface construction (PLIC), which preserves a sharp interface and ensures conservation of the phase volume. The quad-/octree adaptive mesh refinement (AMR) capability of \textit{Basilisk} is also used to refine the grid locally near the interface and in regions of strong vortical or scalar-gradient activity, thereby improving computational efficiency without sacrificing local resolution.

The present multicomponent formulation builds on our previous sharp-interface method for single-component phase change \citep{zhao2022boiling}. At the hydrodynamic level, the same one-fluid projection framework is retained. Nevertheless, interfacial mass transfer introduces two issues that must be treated consistently before the phase-wise scalar equations can be solved sharply. First, the interfacial mass flux appears as a singular volumetric source term and generates a jump in the normal velocity across $\Gamma$, as expressed by Eq.~\eqref{e6}. This modifies both the divergence constraint and the pressure projection. Second, the same velocity jump makes the VOF transport nontrivial, because geometrical interface advection requires a continuous velocity field. Any inconsistency in the evaluation or incorporation of the interfacial mass flux would immediately affect the interface position, and would subsequently contaminate the transport of temperature and species. Therefore, although the main methodological contribution of the present work lies in the sharp treatment of the scalar equations, the flow solver and the interface-transport algorithm form the necessary conservative foundation of the method.

For this reason, the present section briefly recalls the numerical treatment of the flow field and of the VOF equation in the presence of phase change. Particular emphasis is placed on the construction of a consistent geometrical advection strategy, because the same phase fluxes and reconstructed interface geometry will later be used to advect momentum, thermal energy, and species mass in Section~\ref{sec4}. This consistency is essential for maintaining conservation across the coupled interface-motion, flow, heat-transfer, and multicomponent mass-transfer problems.

\subsection{Flow field}\label{sec3.1}

As shown by Eqs.~\eqref{e5}--\eqref{e8}, interfacial mass transfer induces jumps in both the normal velocity and the interfacial stress. At the hydrodynamic level, these effects are treated within the same one-fluid projection framework as in our previous single-component formulation \citep{zhao2022boiling}. The bulk properties of each phase are assumed to remain constant during phase change, while the surface-tension coefficient may vary along the interface owing to local temperature and composition variations. Under these assumptions, the Navier--Stokes equations \eqref{e1} are advanced by a second-order time-staggered approximate projection method. Since the overall procedure follows \citet{zhao2022boiling}, only the main update steps are recalled here:
\begin{equation}
\rho_c^{n+\frac{1}{2}}
\left(
\frac{\boldsymbol{u}^{*}-\boldsymbol{u}^{n}}{\Delta t}
+
\boldsymbol{u}^{n+\frac{1}{2}} \cdot \nabla \boldsymbol{u}^{n+\frac{1}{2}}
\right)_c
=
\nabla_c \cdot
\left[
\mu_f^{n+\frac{1}{2}}
(\nabla \boldsymbol{u} + \nabla \boldsymbol{u}^T)^{*}
\right]
+
\left[
\left(
\sigma \kappa \delta_{s} \boldsymbol{n}
+
\nabla_t \sigma \delta_s
\right)^{n-\frac{1}{2}}
-
\nabla p^{n}
\right]_{f\rightarrow c},
\label{e22}
\end{equation}
\begin{equation}
\boldsymbol{u}_{c}^{**}
=
\boldsymbol{u}_c^*
-
\frac{\Delta t}{\rho_c^{n+\frac{1}{2}}}
\left[
\left(
\sigma \kappa \delta_{s} \boldsymbol{n}
+
\nabla_t \sigma \delta_s
\right)^{n-\frac{1}{2}}
-
\nabla p^{n}
\right]_{f\rightarrow c},
\label{e23}
\end{equation}
\begin{equation}
\boldsymbol{u}_{f}^{n+1}
=
\boldsymbol{u}_{c\rightarrow f}^{**}
+
\frac{\Delta t}{\rho_f^{n+\frac{1}{2}}}
\left[
\left(
\sigma \kappa \delta_{s} \boldsymbol{n}
+
\nabla_t \sigma \delta_s
\right)^{n+\frac{1}{2}}
-
\nabla p^{n+1}
\right]_f,
\label{e24}
\end{equation}
\begin{equation}
\boldsymbol{u}_{c}^{n+1}
=
\boldsymbol{u}_{c}^{**}
+
\frac{\Delta t}{\rho_c^{n+\frac{1}{2}}}
\left[
\left(
\sigma \kappa \delta_{s} \boldsymbol{n}
+
\nabla_t \sigma \delta_s
\right)^{n+\frac{1}{2}}
-
\nabla p^{n+1}
\right]_{f\rightarrow c}.
\label{e25}
\end{equation}

Here, the velocity and pressure are advanced from time level $n$ to $n+1$. The subscripts $c$ and $f$ denote cell-centered and face-centered quantities, respectively, while the notation $f\rightarrow c$ represents second-order arithmetic interpolation from face centers to cell centers. The viscous term is treated implicitly using a standard finite-volume discretization and is not repeated here. When Marangoni effects are considered, the tangential surface-tension force is discretized consistently with the height-function-based formulation proposed by \citet{seric2018direct} and further developed by \citet{tripathi2018motion}; the corresponding implementation details are summarized in~\ref{app1}.

A point that deserves emphasis is the treatment of advection. In the present method, the convective transport in the momentum equation is discretized by the same geometrical advection strategy as that used for the VOF equation. This idea is later extended to the temperature and species equations. Such consistency is important because, in the presence of phase change, the interface motion, the flow field, and the scalar transport are all driven by the same interfacial mass flux. Using compatible conservative advection schemes for all transported quantities prevents spurious inconsistencies among the volume, momentum, energy, and species balances, and improves the overall conservation properties of the method \citep{zhao2022boiling,cipriano2024multicomponent}. The geometrical advection scheme is briefly introduced in Section~\ref{sec3.2} for the VOF field and then revisited in Section~\ref{sec4.1} for the conservative transport of momentum, thermal energy, and species mass.

Because a collocated arrangement is employed, both velocity and pressure are stored at cell centers, whereas the pressure gradient and surface-tension force are evaluated at cell faces. Following the balanced-force formulation of \citet{francois2006balanced}, the capillary force is discretized at the same location and time level as the pressure gradient, thereby minimizing parasitic currents near the interface. The interface curvature is computed by the height-function technique. Taking the discrete divergence of Eq.~\eqref{e24} yields the Poisson equation for the updated pressure,
\begin{equation}
\nabla_c \cdot
\left[
\frac{\Delta t}{\rho^{n+\frac{1}{2}}}
\left(
\left(
\sigma \kappa \delta_{s} \boldsymbol{n}
+
\nabla_t \sigma \delta_s
\right)^{n+\frac{1}{2}}
-
\nabla p^{n+1}
\right)
\right]_f
=
\nabla_c \cdot \boldsymbol{u}_{f}^{n+1}
-
\nabla_c \cdot \boldsymbol{u}_{c\rightarrow f}^{**},
\label{e26}
\end{equation}
where the capillary and Marangoni terms, as well as the intermediate velocity $\boldsymbol{u}_{c\rightarrow f}^{**}$, are already known. The remaining term $\nabla_c \cdot \boldsymbol{u}_{f}^{n+1}$ is prescribed by the continuity constraint \eqref{e2}, in which the non-zero divergence arises from the interfacial mass-transfer source. The resulting Poisson equation is solved by the multigrid algorithm available in \textit{Basilisk}. It is worth recalling that, in the presence of phase change, the approximate projection commonly used on collocated grids may generate spurious oscillations in the cell-centered velocity field. This occurs because the divergence constraint is no longer zero but contains a singular interfacial source. The correction proposed in \citet{zhao2022boiling} is adopted here without modification, and the reader is referred to that work for implementation details.

\subsection{Interface tracking}\label{sec3.2}

The interface position is updated by solving the VOF equation \eqref{evof}. In the presence of phase change, however, this update is more delicate than in ordinary incompressible two-phase flows. As discussed in Section~\ref{sec2.1}, the interfacial mass flux generates a jump in the normal velocity across $\Gamma$, so that the one-fluid velocity $\boldsymbol{u}$ is not directly suitable for advecting the VOF field. In addition, the source term associated with interfacial mass transfer may drive the volume fraction outside the admissible range $0\leq f\leq1$ if it is not incorporated carefully. Both issues may compromise the robustness of interface transport and the conservation of the phase volume.

To overcome these difficulties, we adopt the velocity-decomposition strategy used by \citet{scapin2020volume}, \citet{malan2021geometric}, and \citet{zhao2022boiling}. The basic idea is to separate from the one-fluid velocity the irrotational contribution associated with the Stefan flow, so as to construct a continuous advection velocity near the interface. Specifically, the velocity field is decomposed as $\boldsymbol{u}=\boldsymbol{u}_D+\boldsymbol{u}_S$, where $\boldsymbol{u}_S$ denotes the Stefan velocity induced by the interfacial mass flux and $\boldsymbol{u}_D$ is the remaining phase velocity. The Stefan velocity is obtained from a velocity potential $\psi$ satisfying
\begin{equation}
\nabla^2 \psi
=
\left(
\frac{1}{\rho_g}-\frac{1}{\rho_l}
\right)\ddot m,
\label{e28}
\end{equation}
\begin{equation}
\boldsymbol{u}_S=\nabla\psi.
\label{e29}
\end{equation}
Once $\boldsymbol{u}_S$ has been computed throughout the domain, the phase velocity $\boldsymbol{u}_D=\boldsymbol{u}-\boldsymbol{u}_S$ becomes continuous across the interface and may be used to define a geometrically consistent interface transport velocity.

A further complication is that the source term in Eq.~\eqref{evof} directly modifies the phase volume inside interfacial cells. To avoid treating this source term explicitly in the transport step, we define the interface velocity as
\[
\boldsymbol{u}_{\Gamma}
=
\boldsymbol{u}_D+\frac{\dot m}{\rho}\boldsymbol{n},
\]
so that the VOF equation can be recast into the pure advection form
\begin{equation}
\frac{\partial f}{\partial t}
+
\boldsymbol{u}_{\Gamma}\cdot\nabla f
=
0.
\label{e30}
\end{equation}
The interface can then be advanced by the standard geometrical VOF procedure using the continuous velocity $\boldsymbol{u}_{\Gamma}$. Equivalently, one may advect $f$ with $\boldsymbol{u}_D$ and account for the interface displacement induced by the right-hand side of Eq.~\eqref{evof} through a geometrical correction during reconstruction \citep{malan2021geometric}. In the present implementation, these two viewpoints are consistent and lead to the same interface update.

\begin{figure}
\centering
\includegraphics[scale=0.7]{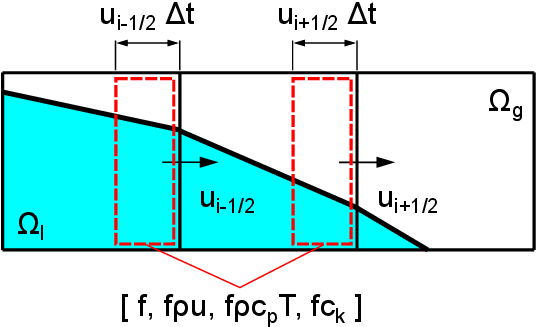}
\caption{Schematic of the geometrical conservative transport used in the direction-split advection step, illustrated here for the $x$ direction. The red dashed strips denote the swept liquid volumes associated with the face displacements $u_{i-\frac{1}{2}}\Delta t$ and $u_{i+\frac{1}{2}}\Delta t$. The bracketed quantities indicate that the same geometrically reconstructed phase fluxes are used consistently to transport volume, momentum, thermal energy, and species mass.}
\label{f3.2}
\end{figure}

Equation~\eqref{e30} is discretized by the conservative directional-splitting scheme of \citet{weymouth2010conservative}. For a two-dimensional cell $(i,j)$, the update of the volume fraction from time level $n-1/2$ to $n+1/2$ reads
\begin{subequations}
\begin{align}
\frac{f_{i,j}^{*}-f_{i,j}^{n-\frac{1}{2}}}{\Delta t}
+
\frac{F_{i+\frac{1}{2},j}^{n}-F_{i-\frac{1}{2},j}^{n}}{\Delta x}
&=
f_c
\frac{u_{i+\frac{1}{2},j}^{n}-u_{i-\frac{1}{2},j}^{n}}{\Delta x},
\\
\frac{f_{i,j}^{n+\frac{1}{2}}-f_{i,j}^{*}}{\Delta t}
+
\frac{G_{i,j+\frac{1}{2}}^{*}-G_{i,j-\frac{1}{2}}^{*}}{\Delta y}
&=
f_c
\frac{v_{i,j+\frac{1}{2}}^{n}-v_{i,j-\frac{1}{2}}^{n}}{\Delta y}.
\end{align}
\label{e31}
\end{subequations}
Here, $F$ and $G$ are the geometrical volume fluxes through the cell faces in the $x$- and $y$-directions, respectively, computed from the reconstructed interface and the normal face velocities $(u_f,v_f)$. The terms on the right-hand side account for the local volume variation associated with the non-zero discrete divergence that appears during directional splitting. Following \citet{weymouth2010conservative}, the quantity $f_c$ is chosen according to the upwind phase indicator: $f_c=1$ when $f_{i,j}^{n-1/2}>0.5$, and $f_c=0$ otherwise. This choice guarantees exact conservation in the incompressible limit and improves the robustness of the VOF update in the presence of phase change. Figure~\ref{f3.2} illustrates the geometrical transport procedure for liquid-phase advection in the $x-$ direction. The red dashed strips denote the liquid volumes swept through the left and right cell faces during one direction-split advection substep. The consistent geometrical construction is subsequently used to transport other conservative quantities, as indicated in the figure. Before each substep, the interface is reconstructed from the most recent volume-fraction field by the PLIC method, and the fluxes $F$ and $G$ are then evaluated geometrically from the reconstructed interface \citep{popinet2009accurate}. This geometrical procedure has been extended in \textit{Basilisk} to adaptive Cartesian meshes. 

Specifically, the same geometrical transport philosophy is also used, in a consistent manner, for the advection terms appearing in the momentum, temperature, and species equations. This consistency is particularly important in the present phase-change problem, because the interface motion and the transport of all conserved quantities are controlled by the same phase fluxes and the same reconstructed interfacial geometry. As a result, numerical diffusion of the interface is avoided, and the conservation properties of the coupled volume, momentum, energy, and species transport are improved.

Finally, in order to reduce the splitting error, the sequence of advection directions is alternated from one time step to the next. Having introduced the numerical treatment of the flow field and interface transport, we now turn to the temperature and species equations, where the principal difficulties identified in Section~\ref{sec2.3} arise and where the main methodological contribution of the present work lies.

\section{Numerical method for the temperature and species equations}\label{sec4}

We now turn to the central part of the present methodology: the sharp and conservative treatment of the temperature and species equations in multicomponent phase change. As discussed in Sections~\ref{sec1} and \ref{sec2}, many VOF-based methods avoid imposing the interfacial scalar conditions directly by regularizing the jump conditions into volumetric source terms distributed over several grid cells near the interface. Such diffusive treatments have been used for species transport \citep{haroun2010volume,marschall2012numerical,deising2016unified} and for the interfacial heat-flux jump \citep{sato2013sharp,scapin2020volume}. Although these approaches are often robust and relatively simple to implement, they inevitably smear the interfacial transport and may weaken both local accuracy and scalar conservation near $\Gamma$. This limitation becomes particularly restrictive in multicomponent problems, where the interfacial mass flux is not a single quantity but the sum of several partial species fluxes coupled to the temperature field through latent-heat exchange and thermodynamic equilibrium.

The objective of this section is therefore to develop a genuinely sharp two-field formulation for scalar transport. In contrast to the one-fluid treatment used for the hydrodynamics, the temperature and species equations are solved separately in the liquid and gas phases. The two phase-wise scalar fields are then coupled only through boundary and jump conditions imposed directly on the reconstructed moving interface. In this way, the interface remains sharp not only in the VOF representation, but also in the transport of temperature and species concentrations. The main numerical difficulty arises in the diffusion step, where the heat-flux jump condition for temperature, the Robin conditions for species concentrations, and the thermodynamic-equilibrium relations must be incorporated consistently into the finite-volume discretization.
\begin{figure}
\centering
\includegraphics[scale=0.48]{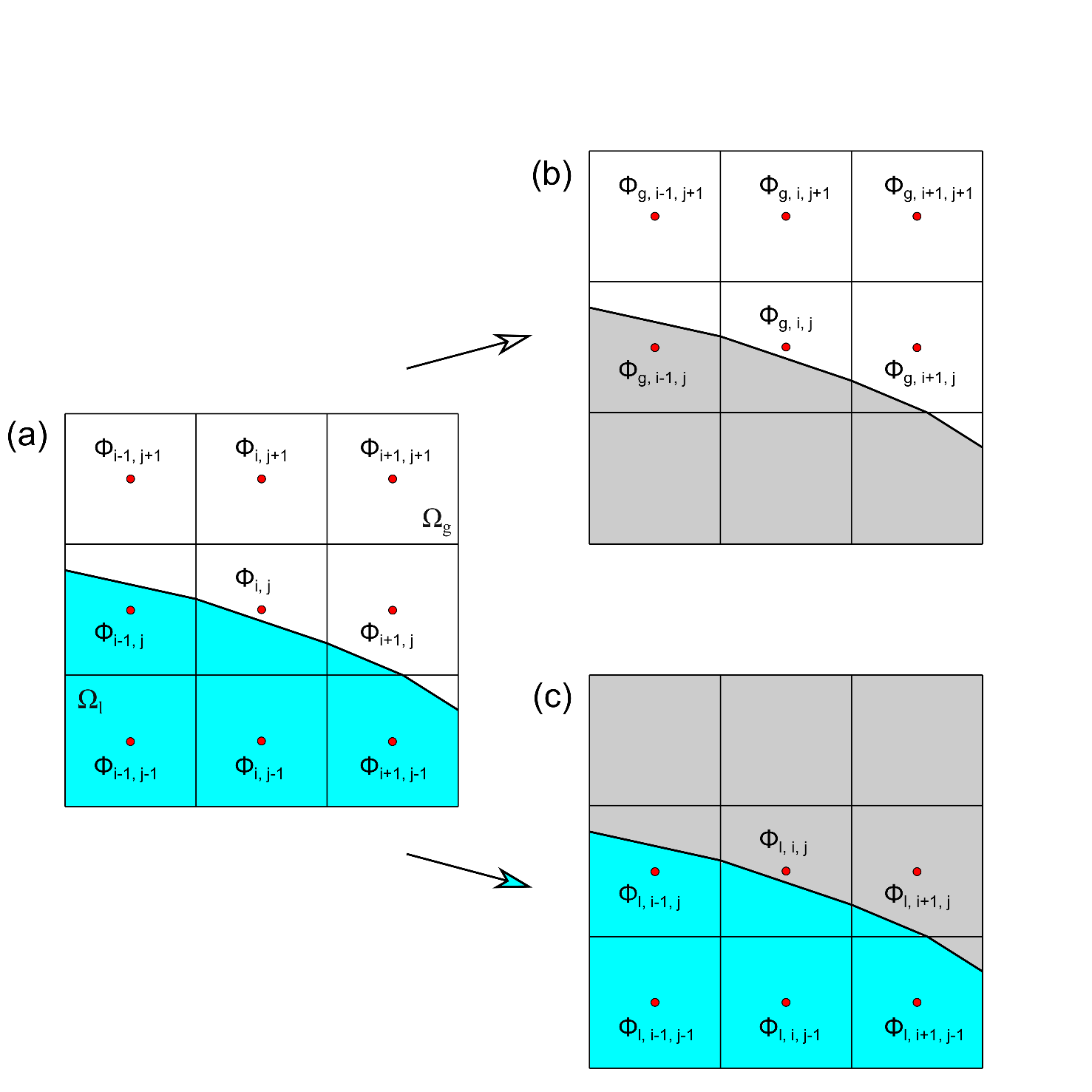}
\caption{Schematic of the sharp two-field scalar formulation. The temperature and species, denoted by $\Phi$ here, are solved separately in the liquid and gas phases, and the two phase-wise fields are coupled only through interfacial jump conditions and thermodynamic-equilibrium relations imposed directly on the reconstructed interface $\Gamma$.}
\label{f4.1}
\end{figure}

Within the present framework, the temperature and species equations are solved separately in $\Omega_l$ and $\Omega_g$, as sketched in Fig.~\ref{f4.1}. After rewriting the convective terms in conservative form and accounting for the non-zero divergence induced by interfacial mass transfer, the phase-wise temperature equation can be written as
\begin{equation}
\frac{T_c^{n+\frac{1}{2}}-T_c^{n-\frac{1}{2}}}{\Delta t}
+
\nabla_c \cdot (T\boldsymbol{u})^{n}
=
\nabla_c \cdot
\left[
\left(
\frac{\lambda}{\rho c_p}
\right)_f
\nabla_f T
\right]^{n+\frac{1}{2}}
+
\left(
\frac{\ddot m}{\rho}T_{\Gamma}
\right)^{n-\frac{1}{2}},
\qquad
\mathrm{in}\quad \Omega_l~\mathrm{and}~\Omega_g,
\label{e32}
\end{equation}
and, for the $k$-th species concentration,
\begin{equation}
\frac{c_{k,c}^{n+\frac{1}{2}}-c_{k,c}^{n-\frac{1}{2}}}{\Delta t}
+
\nabla_c \cdot (c_k\boldsymbol{u})^{n}
=
\nabla_c \cdot
\left[
(D_k)_f\nabla_f c_k
\right]^{n+\frac{1}{2}}
+
\left(
\frac{\ddot m}{\rho}c_{k,\Gamma}
\right)^{n-\frac{1}{2}},
\qquad
\mathrm{in}\quad \Omega_l~\mathrm{and}~\Omega_g.
\label{e33}
\end{equation}

In both equations, the last term originates from rewriting the convective derivative in conservative form when $\nabla\cdot\boldsymbol{u}\neq0$. Physically, it accounts for the scalar transport associated with the local volume variation generated by interfacial mass transfer. The interfacial values $T_{\Gamma}$ and $c_{k,\Gamma}$ are used because this volume variation is localized at the moving interface.

Although Eqs.~\eqref{e32} and \eqref{e33} have similar conservative forms, their numerical treatment is not a standard advection--diffusion problem. The advection terms must be discretized consistently with the geometrical VOF transport introduced in Section~\ref{sec3}, so that the transport of volume, momentum, thermal energy, and species mass relies on the same reconstructed interface and phase fluxes. More importantly, the diffusion terms must be discretized together with the interfacial conditions, because the thermal flux jump and the species Robin conditions enter the finite-volume formulation through the diffusive fluxes at $\Gamma$. This is the point at which the present method differs most clearly from diffusive-interface formulations and from partially sharp VOF approaches.

Accordingly, the remainder of this section is organized around three ingredients. Section~\ref{sec4.1} describes the geometrical conservative discretization of the advection terms. Section~\ref{sec4.2} presents the embedded-boundary discretization of the diffusion operator, with emphasis on the sharp treatment of both Robin and flux-jump conditions. Section~\ref{sec4.3} then introduces the coupling strategy used to determine the partial interfacial mass fluxes and to close the thermodynamic-equilibrium relations. Together, these ingredients form a sharp and conservative scalar-transport framework for multicomponent liquid--gas phase change.

\subsection{Geometrical conservative discretization of the advection term}\label{sec4.1}

As discussed in Section~\ref{sec3.2}, the geometrical VOF algorithm provides a conservative and non-diffusive update of the phase volume. In the present method, the same geometrical transport principle is extended to the advection of momentum, thermal energy, and species mass. This consistency is essential once the temperature and species equations are solved separately in the liquid and gas phases: the scalar fields must be transported with the same phase fluxes and the same reconstructed interface geometry as the VOF field. Otherwise, inconsistencies may arise near $\Gamma$, thereby degrading both interface sharpness and conservation.

To this end, we introduce the phase-wise conservative variables
\[
\vartheta =
\left[
f \rho_l \boldsymbol{u}_l,\,
(1-f)\rho_g \boldsymbol{u}_g,\,
f \rho_l c_{p,l} T_l,\,
(1-f)\rho_g c_{p,g} T_g,\,
f c_{l,k},\,
(1-f)c_{g,k}
\right],
\]
which respectively represent the momentum, thermal energy, and species mass carried by the two phases. For all these quantities, the advection operator is discretized in the same directional-splitting form as that used for the VOF equation. In two dimensions, the update from time level $n-1/2$ to the intermediate advected state is written as
\begin{subequations}
\begin{align}
\frac{\vartheta_{i,j}^{*}-\vartheta_{i,j}^{n-\frac{1}{2}}}{\Delta t}
+
\frac{\mathcal{F}_{i+\frac{1}{2},j}^{n}-\mathcal{F}_{i-\frac{1}{2},j}^{n}}{\Delta x}
&=
\vartheta_c
\frac{u_{i+\frac{1}{2},j}^{n}-u_{i-\frac{1}{2},j}^{n}}{\Delta x},
\\
\frac{\vartheta_{i,j}^{**}-\vartheta_{i,j}^{*}}{\Delta t}
+
\frac{\mathcal{G}_{i,j+\frac{1}{2}}^{*}-\mathcal{G}_{i,j-\frac{1}{2}}^{*}}{\Delta y}
&=
\vartheta_c
\frac{v_{i,j+\frac{1}{2}}^{n}-v_{i,j-\frac{1}{2}}^{n}}{\Delta y},
\end{align}
\label{e34}
\end{subequations}
where $\mathcal{F}=u_f\vartheta$ and $\mathcal{G}=v_f\vartheta$ denote the conservative fluxes through the cell faces in the $x$- and $y$-directions, respectively. The face velocities $(u_f,v_f)$ are the same as those used in the VOF update, and the volume fraction $f$ has already been advanced by the geometrical scheme described in Section~\ref{sec3.2}. Thus, the transport of all conservative quantities is based on exactly the same phase fluxes as the transport of the interface itself.

As illustrated in Fig.~\ref{f3.2}, the same geometrically swept phase volumes are used to advect the phase-wise conservative variables $\vartheta$. Thus, the swept liquid volume carries not only liquid volume, but also the corresponding liquid momentum, thermal energy, and species mass. The primitive variables required at the cell faces are reconstructed by the second-order Bell--Colella--Glaz (BCG) upwind scheme \citep{bell1989second}, whereas the transported conservative quantities are evaluated from the geometrically reconstructed phase fluxes. In this way, the advection of all variables remains fully compatible with the PLIC reconstruction of the interface. The quantity $\vartheta_c$ on the right-hand side of Eq.~\eqref{e34} plays the same role as $f_c$ in the Weymouth--Yue VOF scheme: it accounts for the local volume variation associated with a non-zero discrete divergence during directional splitting. Its value is chosen according to the phase to which the conservative variable belongs. For liquid-phase quantities, $\vartheta_c=\vartheta^{n-1/2}$ when $f_{i,j}^{\,n-1/2}>0.5$, and $\vartheta_c=0$ otherwise. For gas-phase quantities, $\vartheta_c=\vartheta^{n-1/2}$ when $1-f_{i,j}^{\,n-1/2}>0.5$, and $\vartheta_c=0$ otherwise. This choice applies the compressibility correction consistently with the locally dominant phase and preserves the conservative character of the split update.

After advection has been completed in all directions, the intermediate conservative variables $\vartheta^{**}$ are converted back to primitive variables by dividing by the updated phase fraction, namely by $f^{\,n+1/2}$ for liquid-phase quantities and by $1-f^{\,n+1/2}$ for gas-phase quantities. This gives the intermediate fields $[\boldsymbol{u}_l,\boldsymbol{u}_g,T_l,T_g,c_{l,k},c_{g,k}]^{**}$. For the momentum equation, however, the hydrodynamic update is ultimately performed in one-fluid form. The mass-averaged velocity entering the projection step is therefore reconstructed as
\begin{equation}
\boldsymbol{u}
=
\frac{
f\rho_l\boldsymbol{u}_l+(1-f)\rho_g\boldsymbol{u}_g
}
{
f\rho_l+(1-f)\rho_g
},
\label{e35}
\end{equation}
after which the projection method described in Section~\ref{sec3.1} is used to obtain $\boldsymbol{u}^{n+1}$ and $p^{n+1}$.

The above geometrical advection scheme ensures that the transport of momentum, thermal energy, and species mass is fully consistent with the transport of the interface. This consistency is a key ingredient of the present method, because it preserves conservation and avoids introducing additional numerical diffusion near $\Gamma$ before the interfacial conditions are imposed through the diffusion operator. Once the advection step has been completed, the remaining task is to discretize the diffusion terms in Eqs.~\eqref{e32} and \eqref{e33}, thereby correcting the intermediate scalar fields to their values at time level $n+1/2$, \textit{i.e.} $T^{**}\rightarrow T^{n+1/2}$ and $c_k^{**}\rightarrow c_k^{n+1/2}$. It is precisely in this diffusion step that the thermal and species interfacial conditions are imposed sharply.

\subsection{Sharp and conservative discretization of the diffusion term}\label{sec4.2}

We now arrive at the core of the present numerical method. After the conservative advection step described above, the remaining task is to solve the diffusion part of the temperature and species equations while enforcing the interfacial conditions sharply on the moving VOF interface. This step is the essential difficulty of the multicomponent phase-change problem, because the interfacial conditions are not regularized as volumetric source terms, but are incorporated directly into the finite-volume discretization on the geometrically reconstructed interface.

The key point is that the scalar transport equations are solved separately in the liquid and gas phases. Consequently, the reconstructed VOF interface $\Gamma$ acts as an embedded internal boundary for each phase, and the diffusion operator must be discretized in the presence of this boundary. In the present multicomponent problem, two types of interfacial conditions must be handled. For each species equation, the interface imposes a Robin condition on the phase-wise concentration field. For the temperature equation, the interface imposes the continuity of temperature together with a two-sided heat-flux jump. The latter case is more demanding, because the liquid-side and gas-side diffusion problems are coupled through the interfacial heat fluxes. This is precisely the obstacle that is avoided in existing VOF-based multicomponent methods.
As illustrated in Fig.~\ref{f4.1}, the scalar field is split into two phase-wise fields. This phase-wise representation provides a natural way to distinguish the two types of interfacial conditions treated in this section. A Robin condition closes a \emph{one-sided} boundary-value problem: depending on which phase is solved, only the field in Fig.~\ref{f4.1}(b) or that in Fig.~\ref{f4.1}(c) is involved. By contrast, a jump condition closes a \emph{two-sided} coupled problem: the gas-side and liquid-side fields in Figs.~\ref{f4.1}(b) and \ref{f4.1}(c) must be solved together, because the interfacial values and the one-sided normal fluxes are linked by the jump relations. For both the temperature and species equations, the diffusion correction step can be written in the generic form
\begin{equation}
\frac{\phi^{n+\frac{1}{2}}-\phi^{**}}{\Delta t}
=
\nabla\cdot(\mathcal{D}\nabla\phi)^{n+\frac{1}{2}},
\label{e36}
\end{equation}
where $\phi^{**}$ denotes the primitive variable after the advection step, and $\mathcal{D}$ is the diffusion coefficient appearing in the bulk scalar equation. For a species equation, $\phi=c_k$ and $\mathcal{D}=D_k$, whereas for the temperature equation, $\phi=T$ and $\mathcal{D}=\lambda/(\rho c_p)$.

For a species equation, the interfacial mass balance can be cast into the one-sided Robin form
\begin{equation}
\left.
\mathcal{K}\frac{\partial \phi}{\partial n}
\right|_{\Gamma}
+
\alpha\phi_{\Gamma}
=
\beta,
\qquad
\mathrm{on}\quad \Gamma,
\label{e361}
\end{equation}
where $\alpha$ and $\beta$ are known coefficients once the interfacial coupling quantities have been specified, and $\mathcal{K}$ is the coefficient entering the interfacial flux condition. The normal derivative is taken along the outward normal of the corresponding phase-wise control volume. This condition closes a one-sided boundary-value problem: once $\alpha$ and $\beta$ are specified, the diffusion solve in a given phase involves only unknowns from that phase. For the present species equations, $\mathcal{K}=D_k$, $\alpha=-\dot{m}/\rho$, $\beta=-\dot{m}_k$, and Eq.~\eqref{e361} represents the generic Robin form of the interfacial mass-balance condition derived from Eq.~\eqref{e14}.

For the temperature equation, the interface instead imposes two-sided jump conditions,
\begin{equation}
\left\{
\begin{array}{ll}
\phi_{l,\Gamma}-\phi_{g,\Gamma}=\alpha, \vspace{0.8em}\\
\left.
\mathcal{K}_l\dfrac{\partial \phi_l}{\partial n}
\right|_{\Gamma}
-
\left.
\mathcal{K}_g\dfrac{\partial \phi_g}{\partial n}
\right|_{\Gamma}
=
\beta,
\end{array}
\right.
\qquad
\mathrm{on}\quad \Gamma.
\label{e37}
\end{equation}
These conditions define a two-sided coupled boundary-value problem, because the liquid- and gas-side interfacial values and one-sided normal fluxes must be determined simultaneously. In the present work, $\mathcal{K}_l=\lambda_l$, $\mathcal{K}_g=\lambda_g$, $\alpha=0$, and $\beta=\sum_{k=1}^{N_t}\dot m_k\mathcal{L}_k$. Therefore, Eq.~\eqref{e37} enforces temperature continuity and the latent-heat-induced jump of the physical conductive heat flux.


\subsubsection{Discretization of the species equations}\label{sec4.2.1}

We first consider the species equations. For each transferring species, the concentration field is solved in one phase at a time, and the reconstructed interface $\Gamma$ provides a Robin boundary condition of the form \eqref{e361}. The task is therefore to construct a phase-wise finite-volume diffusion operator that incorporates the Robin condition sharply at the reconstructed interface, rather than replacing it by a smeared volumetric source term.

Consider one phase, denoted by $\Omega_{\alpha}$ with $\alpha=l$ or $g$, in an interfacial cell cut by the reconstructed interface $\Gamma$. Integrating Eq.~\eqref{e36} over the phase-wise control volume $d\Omega_{\alpha}$ and applying Gauss's theorem gives
\begin{equation}
\int_{d\Omega_{\alpha}}
\nabla\cdot(\mathcal{D}\nabla\phi)\,dV
=
\int_{d\partial\Omega_{\alpha}}
\mathcal{D}\nabla\phi\cdot\boldsymbol{n}\,dS
=
\left(
\underbrace{
\sum_{f=1}^{nf}
(\mathcal{D}\nabla\phi)_f\cdot\boldsymbol{n}_f S_f
}_{\text{cell faces}}
+
\underbrace{
\vphantom{\sum_{f=1}^{nf}}
(\mathcal{D}\nabla\phi)_{\Gamma}\cdot\boldsymbol{n}_{\Gamma}S_{\Gamma}
}_{\text{embedded boundary}}
\right),
\label{e38}
\end{equation}
where $\boldsymbol{n}_f$ and $\boldsymbol{n}_{\Gamma}$ are the outward unit normals of the Cartesian face and the embedded boundary, respectively, $S_f$ and $S_{\Gamma}$ are the corresponding surface measures, and $nf$ denotes the number of Cartesian faces intersecting the phase-wise control volume. The embedded boundary is the piecewise-linear interface reconstructed by the PLIC algorithm; hence its centroid, normal direction, and surface measure are all known geometrically.

\begin{figure}
\centering
\includegraphics[scale=0.6]{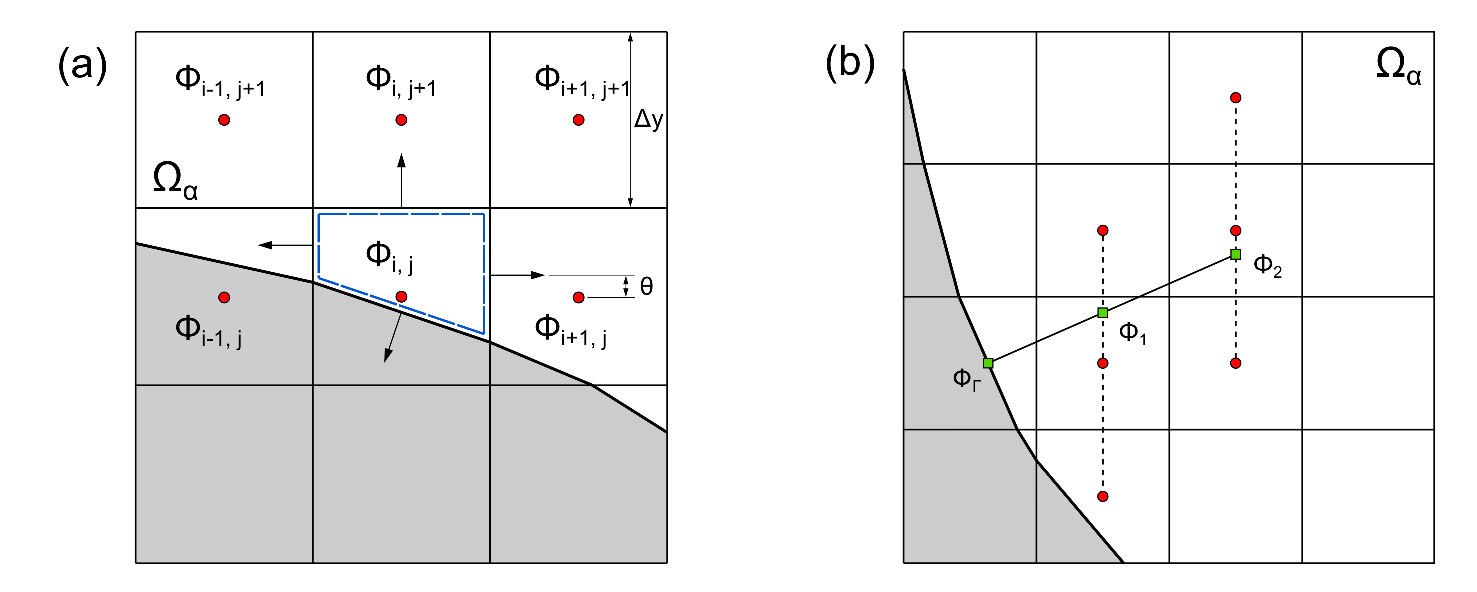}
\caption{Schematic of the embedded-boundary treatment for a Robin condition. (a) Phase-wise finite-volume discretization in an interfacial cell $(i, j)$ cut by $\Gamma$. (b) One-sided second-order reconstruction of the normal derivative at the embedded boundary.}
\label{f3.2.1}
\end{figure}

For cells away from the interface, Eq.~\eqref{e38} reduces to the standard second-order central-difference discretization. The nontrivial case is an interfacial cell, where the phase-wise control volume is bounded by full Cartesian faces, partial Cartesian faces, and the embedded boundary, as illustrated in Fig.~\ref{f3.2.1}(a). The corresponding fluxes are evaluated as follows.

\begin{itemize}
\item \textbf{Full cell faces.}
For a complete Cartesian face, such as $(i,j+1/2)$, the gradient is evaluated by the standard second-order central difference.

\item \textbf{Partial cell faces.}
For a cut Cartesian face, such as $(i\pm1/2,j)$, the flux must be evaluated at the centroid of the \emph{partial} face rather than at the center of the original full face. Denoting by $\theta$ the offset between these two points in the transverse direction, a second-order interpolation gives
\begin{equation}
\left( \frac{\partial \phi}{\partial n} \right)_{i+\frac{1}{2},j}
=
\left(1-\frac{\theta}{\Delta y}\right)
\frac{\phi_{i+1,j}-\phi_{i,j}}{\Delta x}
+
\frac{\theta}{\Delta y}
\frac{\phi_{i+1,j+1}-\phi_{i,j+1}}{\Delta x}
+O(\Delta^2),
\label{e39}
\end{equation}
where the normal direction corresponds to the local face-normal direction.

\item \textbf{Embedded boundary.}
The flux through the reconstructed interface cannot be obtained from a standard finite-difference stencil, because the interfacial value $\phi_{\Gamma}$ is not known \emph{a priori}. Instead, a one-sided second-order interpolation is constructed along the local normal direction. As sketched in Fig.~\ref{f3.2.1}(b), a line normal to the interface is extended into the interior of the same phase, and two auxiliary values $\phi_1$ and $\phi_2$ are reconstructed from neighboring cell-centered values by second-order interpolation. Denoting by $d_1$ and $d_2$ the distances from the interface centroid to these two auxiliary points, the normal derivative along the outward normal at the embedded boundary is approximated by
\begin{equation}
\left.
\frac{\partial \phi}{\partial n}
\right|_{\Gamma}
=
\frac{1}{d_2-d_1}
\left[
\frac{d_2}{d_1}(\phi_{\Gamma}-\phi_1)
-
\frac{d_1}{d_2}(\phi_{\Gamma}-\phi_2)
\right]
+O(\Delta^2).
\label{e40}
\end{equation}
\end{itemize}

The Robin boundary condition is imposed by combining Eq.~\eqref{e40} with Eq.~\eqref{e361}. For a species equation, the interfacial flux coefficient in Eq.~\eqref{e361} is identical to the diffusion coefficient, namely $\mathcal{K}=\mathcal{D}=D_k$. The local system is therefore
\begin{equation}
\left\{
\begin{aligned}
&
\left.
\mathcal{D}\frac{\partial \phi}{\partial n}
\right|_{\Gamma}
+
\alpha\phi_{\Gamma}
=
\beta,
\\[0.8em]
&
\left.
\frac{\partial \phi}{\partial n}
\right|_{\Gamma}
=
\frac{1}{d_2-d_1}
\left[
\frac{d_2}{d_1}(\phi_{\Gamma}-\phi_1)
-
\frac{d_1}{d_2}(\phi_{\Gamma}-\phi_2)
\right],
\end{aligned}
\right.
\label{e41}
\end{equation}
where $\phi_1$ and $\phi_2$ are reconstructed from cell-centered values in the same phase. Eq.~\eqref{e41} provides a local algebraic relation between the unknown interfacial value $\phi_{\Gamma}$ and the one-sided normal derivative. Eliminating $\phi_{\Gamma}$ from Eq.~\eqref{e41} gives the embedded-boundary flux directly as
\begin{equation}
\left.
\mathcal{D}\frac{\partial \phi}{\partial n}
\right|_{\Gamma}
=
\mathcal{D}
\frac{A\beta-\alpha B}
{\alpha+\mathcal{D}A},
\label{e41a}
\end{equation}
where the reconstruction coefficients are defined by
\begin{equation}
A
=
\frac{1}{d_2-d_1}
\left(
\frac{d_2}{d_1}
-
\frac{d_1}{d_2}
\right),
\qquad
B
=
\frac{1}{d_2-d_1}
\left(
\frac{d_2}{d_1}\phi_1
-
\frac{d_1}{d_2}\phi_2
\right).
\label{e41b}
\end{equation}
Therefore, the embedded-boundary contribution in Eq.~\eqref{e38} is expressed entirely in terms of phase-wise unknowns and can be incorporated directly into the implicit finite-volume diffusion operator.

Two remarks are worth emphasizing. First, the scheme remains sharp because the reconstruction of $\phi_1$ and $\phi_2$ uses only information from the same phase, and no interpolation across the interface is involved. Second, the scheme remains conservative because the interfacial contribution enters Eq.~\eqref{e38} as an actual flux through the reconstructed embedded boundary, rather than as a regularized source distributed over neighboring cells.

\subsubsection{Discretization of the temperature equation}\label{sec4.2.2}

We next consider the temperature equation, for which the interfacial conditions in Eq.~\eqref{e37} define a two-sided coupled problem. Unlike the Robin case, the liquid- and gas-side diffusion operators cannot be closed independently, because the interfacial temperature and the one-sided conductive fluxes must be determined simultaneously. We therefore construct the embedded-boundary fluxes on the two sides of $\Gamma$ in a coupled manner.

\begin{figure}
\centering
\includegraphics[scale=0.6]{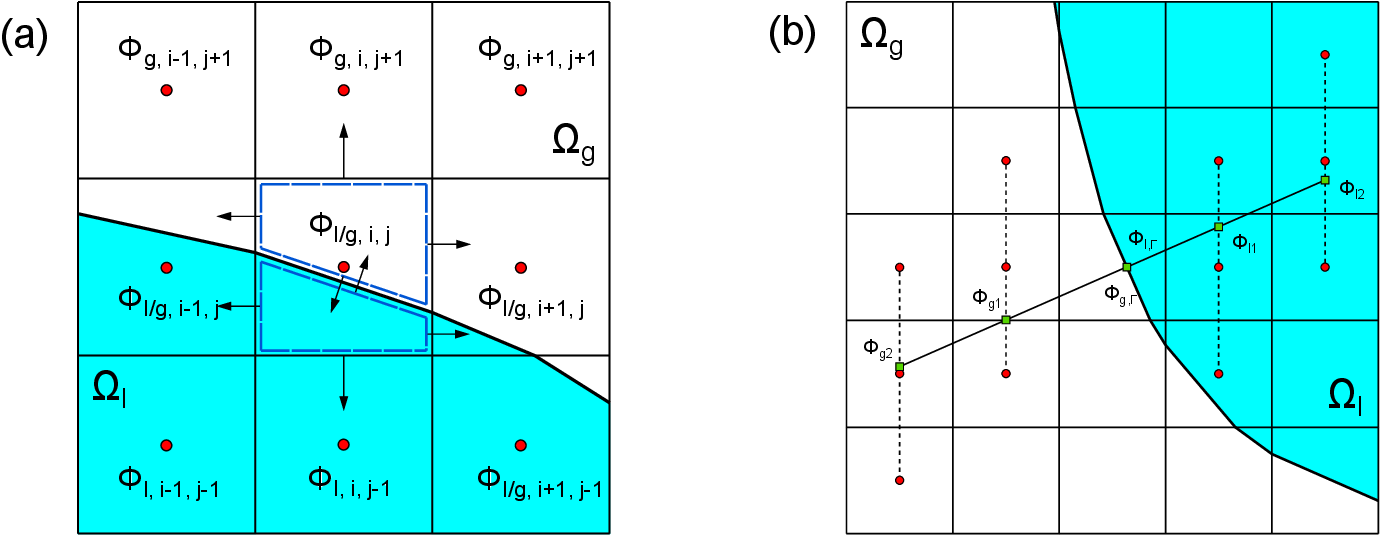}
\caption{Schematic of the embedded-boundary treatment for a flux-jump condition. (a) Phase-wise finite-volume discretization in an interfacial cell $(i, j)$ cut by $\Gamma$. (b) Two-sided second-order reconstruction of the normal derivatives at the embedded boundary, where the liquid- and gas-side reconstructions are coupled simultaneously through the interfacial jump conditions.}
\label{f3.2.2}
\end{figure}

For a cell cut by $\Gamma$, the diffusion operator is integrated separately over the liquid and gas portions, as illustrated in Fig.~\ref{f3.2.2}(a). Adding the liquid- and gas-side finite-volume balances over the cell gives
\begin{equation}
\begin{aligned}
\int_{d\Omega} \nabla \cdot (\mathcal{D}\nabla \phi)\, dV
&=
\int_{d\Omega_l} \nabla \cdot (\mathcal{D}_l\nabla \phi_l)\, dV
+
\int_{d\Omega_g} \nabla \cdot (\mathcal{D}_g\nabla \phi_g)\, dV
\\
&=
\int_{d\partial\Omega_l} \mathcal{D}_l \nabla \phi_l \cdot \boldsymbol{n}\, dS
+
\int_{d\partial\Omega_g} \mathcal{D}_g \nabla \phi_g \cdot \boldsymbol{n}\, dS
\\
&=
\left(
\underbrace{
\sum_{f\in\partial d\Omega_l}
(\mathcal{D}_l \nabla \phi_l )_f \cdot \boldsymbol{n}_f S_f
+
\sum_{f\in\partial d\Omega_g}
(\mathcal{D}_g \nabla\phi_g)_f \cdot \boldsymbol{n}_f S_f
}_{\text{cell faces}}
+
\underbrace{
\vphantom{\sum_{f\in\partial d\Omega_l}^{nf}}
(\mathcal{D}_g\nabla \phi_g)_{\Gamma} \cdot \boldsymbol{n}_{\Gamma} S_{\Gamma}
-
(\mathcal{D}_l \nabla \phi_l)_{\Gamma} \cdot \boldsymbol{n}_{\Gamma} S_{\Gamma}
}_{\text{embedded boundary}}
\right).
\end{aligned}
\label{e42}
\end{equation}
At first sight, it may be tempting to merge the liquid- and gas-side diffusion contributions within the same interfacial cell and to combine the two embedded-boundary contributions, shown by the second lower bracket in Eq.~\eqref{e42}, into a single net interfacial term $[\mathcal{D}\nabla\phi\cdot\boldsymbol{n}]_{\Gamma}S_{\Gamma}$, which is exactly prescribed by the jump condition. Such a treatment would greatly simplify the discretization by collapsing the two phase-wise diffusion problems into a single control-volume balance. The interfacial jump would then be represented only through a net cell-wise contribution, which is essentially the idea behind diffusive-interface or smeared-source treatments. Although the jump condition can be enforced in an integral sense, the sharp separation between the two phase-wise scalar fields is lost on the remaining cell faces. Therefore, this treatment destroys the two-field character of the formulation and fails to preserve sharp scalar discontinuities across the interface.

For a genuinely sharp scheme, the liquid and gas diffusion operators must be discretized independently. The full and partial Cartesian faces are treated in the same way as in the Robin case, using only phase-consistent stencils. The essential difficulty lies in the embedded-boundary fluxes on the two sides of $\Gamma$: neither the interfacial values $\phi_{l,\Gamma}$ and $\phi_{g,\Gamma}$ nor the one-sided normal derivatives $(\partial\phi_l/\partial n)_\Gamma$ and $(\partial\phi_g/\partial n)_\Gamma$ are known \emph{a priori}. To close the problem, we combine the interfacial jump conditions with the one-sided second-order reconstructions from both phases, as sketched in Fig.~\ref{f3.2.2}(b). All normal derivatives in the following local system are taken along the same normal direction as in Eq.~\eqref{e37}, namely from the gas phase to the liquid phase:
\begin{equation}
\left\{
\begin{array}{l}
\phi_{l,\Gamma}-\phi_{g,\Gamma}=\alpha, \vspace{0.8em}\\
\mathcal{K}_l \left.\dfrac{\partial\phi_l}{\partial n}\right|_{\Gamma}
-
\mathcal{K}_g \left.\dfrac{\partial\phi_g}{\partial n}\right|_{\Gamma}
=\beta, \vspace{0.8em}\\
\left.\dfrac{\partial\phi_l}{\partial n}\right|_{\Gamma}
=
\dfrac{-1}{d_{l,2}-d_{l,1}}
\left[
\dfrac{d_{l,2}}{d_{l,1}}(\phi_{l,\Gamma}-\phi_{l,1})
-
\dfrac{d_{l,1}}{d_{l,2}}(\phi_{l,\Gamma}-\phi_{l,2})
\right], \vspace{0.8em}\\
\left.\dfrac{\partial\phi_g}{\partial n}\right|_{\Gamma}
=
\dfrac{1}{d_{g,2}-d_{g,1}}
\left[
\dfrac{d_{g,2}}{d_{g,1}}(\phi_{g,\Gamma}-\phi_{g,1})
-
\dfrac{d_{g,1}}{d_{g,2}}(\phi_{g,\Gamma}-\phi_{g,2})
\right].
\end{array}
\right.
\label{e43}
\end{equation}
Here, $\phi_{\alpha,1}$ and $\phi_{\alpha,2}$ are reconstructed from cell-centered values in phase $\alpha$, and $d_{\alpha,1}$ and $d_{\alpha,2}$ are the corresponding distances from the interface centroid along the one-sided normal reconstruction line. Eq.~\eqref{e43} forms a local algebraic system for the interfacial values $\phi_{l,\Gamma}$ and $\phi_{g,\Gamma}$ together with the two one-sided normal derivatives. For compactness, we introduce the coefficients $A_{\alpha}$ and $B_{\alpha}$ associated with the one-sided reconstructions,
\begin{equation}
A_{\alpha}
=
\frac{1}{d_{\alpha,2}-d_{\alpha,1}}
\left(
\frac{d_{\alpha,2}}{d_{\alpha,1}}
-
\frac{d_{\alpha,1}}{d_{\alpha,2}}
\right),
\qquad
B_{\alpha}
=
\frac{1}{d_{\alpha,2}-d_{\alpha,1}}
\left(
\frac{d_{\alpha,2}}{d_{\alpha,1}}\phi_{\alpha,1}
-
\frac{d_{\alpha,1}}{d_{\alpha,2}}\phi_{\alpha,2}
\right).
\label{e43a}
\end{equation}
Eliminating the interfacial values from Eq.~\eqref{e43} gives the normal diffusive fluxes directly as
\begin{equation}
\left\{
\begin{aligned}
&
\left.
\mathcal{D}_l \frac{\partial \phi_l}{\partial n}
\right|_{\Gamma}
=
\mathcal{D}_l
\frac{
A_l \beta
+
\mathcal{K}_g \left(A_g B_l - A_l B_g - A_l A_g \alpha \right)
}
{\mathcal{K}_l A_l + \mathcal{K}_g A_g},
\\[0.8em]
&
\left.
\mathcal{D}_g \frac{\partial \phi_g}{\partial n}
\right|_{\Gamma}
=
\mathcal{D}_g
\frac{
\mathcal{K}_l \left(A_g B_l - A_l B_g - A_l A_g \alpha \right)
-
A_g \beta
}
{\mathcal{K}_l A_l + \mathcal{K}_g A_g}.
\end{aligned}
\right.
\label{e43c}
\end{equation}

The two embedded-boundary contributions obtained from Eq.~\eqref{e43c} are then inserted separately into the liquid- and gas-side diffusion operators, with the signs determined by the corresponding phase-wise outward normals. This construction solves the jump-condition problem without collapsing the two phases into a single scalar field. The coupling between the phases enters only through the interfacial jump relations, while each phase-wise diffusion operator remains sharply separated. Since the one-sided reconstructions use only same-phase cell-centered values, no cross-interface averaging or volumetric regularization is introduced. This is fundamentally different from a diffusive one-field treatment and also more rigorous than an explicit coupling based on previous-time-step interfacial data.

Once the fluxes through full cell faces, partial cell faces, and the embedded boundary have been evaluated, the diffusion correction step for both the temperature and species equations is completed, and the intermediate variables are updated from $\phi^{**}$ to $\phi^{n+1/2}$. The extension of the above discretization to three dimensions is straightforward in principle: the same embedded-boundary formulation applies, while the second-order interpolations on partial faces and along the interface normal are generalized in the corresponding geometrical manner.

\subsection{Coupling strategy for the species and temperature equations}\label{sec4.3}

We now describe how the phase-wise species equations and the temperature equation are coupled through the interfacial mass fluxes and thermodynamic-equilibrium relations. This coupling step closes the sharp two-field formulation developed in Sections~\ref{sec4.1} and \ref{sec4.2}, and provides the remaining ingredient required to advance the multicomponent phase-change problem over one time step.

For single-component phase change, the interfacial mass flux may often be determined either from the temperature gradients near the interface or from the vapor-concentration field. In the present multicomponent problem, however, the temperature equation is not a convenient primary closure for the individual species fluxes. The heat-flux jump condition constrains only the combined latent-heat contribution of all transferring species, and therefore cannot uniquely determine the partial mass fluxes $\dot m_k$. Moreover, in many bubble-dissolution problems, the temperature variation induced by interfacial mass transfer is negligible and the process may be treated as nearly isothermal. For these reasons, the present method first determines the partial interfacial mass fluxes from the species fields, and then uses the resulting $\dot m_k$ to close the temperature equation and update the thermodynamic-equilibrium state.

To formulate this procedure in a unified way for both bubble dissolution and droplet evaporation, we distinguish between the \emph{continuous phase} and the \emph{dispersed phase}. The continuous phase is the phase that contains at least one non-transferring component, so that the interfacial concentrations of the transferring species can be prescribed there through thermodynamic equilibrium. In bubble dissolution, the continuous phase is the liquid, which contains the solvent. In droplet evaporation, the continuous phase is the gas, which contains the inert gas. The dispersed phase is the complementary phase. With this convention, the coupling procedure within one time step is organized as follows.

\paragraph{\textbf{Step 1}: solve the species equations in the continuous phase.}
At the beginning of the time step, the interfacial concentrations of the transferring species on the continuous-phase side are known from the previous thermodynamic-equilibrium update. They are imposed as Dirichlet conditions on $\Gamma$, namely $c_{c,k,\Gamma}$, where the subscript $c$ denotes the continuous phase in this subsection. The species equations are then solved in the continuous phase using the sharp embedded-boundary discretization described in Section~\ref{sec4.2}. Once the continuous-phase concentration fields have been updated, the one-sided normal gradients at the interface are reconstructed and used to evaluate the corresponding diffusive fluxes.

\paragraph{\textbf{Step 2}: compute the partial mass fluxes $\dot m_k$.}
For each transferring species, the continuous-phase interfacial mass balance follows from Eq.~\eqref{e14}. Using $\dot m=\sum_{i=1}^{N_t}\dot m_i$, where $N_t$ denotes the number of transferring species, the balance can be written as the linear system
\begin{equation}
\dot m_k
-
\frac{c_{c,k,\Gamma}}{\rho_c}
\sum_{i=1}^{N_t}\dot m_i
=
-
D_{c,k}
\left.
\frac{\partial c_{c,k}}{\partial n}
\right|_{\Gamma},
\qquad
k=1,\ldots,N_t.
\label{e44}
\end{equation}
All partial mass fluxes $\dot m_k$ are therefore obtained simultaneously from the continuous-phase concentration gradients. The system is well posed because the continuous phase contains at least one non-transferring component; hence the sum of the interfacial concentrations of the transferring species does not exhaust the total density of that phase.

\paragraph{\textbf{Step 3}: solve the species equations in the dispersed phase.}
Once the partial fluxes $\dot m_k$ and the total flux $\dot m$ are known, the interfacial condition on the dispersed-phase side becomes a Robin condition,
\begin{equation}
\dot m_k
=
\frac{c_{d,k,\Gamma}}{\rho_d}\dot m
-
D_{d,k}
\left.
\frac{\partial c_{d,k}}{\partial n}
\right|_{\Gamma},
\label{e45}
\end{equation}
where the subscript $d$ denotes the dispersed phase. This condition is imposed sharply through the embedded-boundary formulation of Section~\ref{sec4.2.1}, and the dispersed-phase concentration field is updated accordingly. The corresponding interfacial concentrations $c_{d,k,\Gamma}$ are reconstructed as part of the Robin boundary treatment.

\paragraph{\textbf{Step 4}: solve the temperature equation.}
The partial mass fluxes obtained in Step~2 determine the interfacial heat-flux jump through
\[
[\lambda\nabla T\cdot\boldsymbol{n}]_{\Gamma}
=
\sum_{k=1}^{N_t}\dot m_k\mathcal{L}_k.
\]
The temperature equation is then solved separately in the liquid and gas phases using the sharp two-sided flux-jump treatment developed in Section~\ref{sec4.2.2}. In this way, the temperature field is updated consistently with the same set of interfacial mass fluxes that closes the species equations.

\paragraph{\textbf{Step 5}: update the interfacial equilibrium state.}
After the dispersed-phase concentrations and, when needed, the temperature field have been updated, the interfacial values on the dispersed-phase side are available from the sharp boundary treatment. The interfacial temperature $T_{\Gamma}$ and the dispersed-phase interfacial compositions are then substituted into the thermodynamic-equilibrium relations introduced in Section~\ref{sec2.2}, namely Eq.~\eqref{e17} for droplet evaporation or Eq.~\eqref{e18} for bubble dissolution. This gives the updated interfacial concentrations on the continuous-phase side, which will be used as Dirichlet conditions in Step~1 of the next time step. Thus, the species equations, the temperature equation, the interfacial mass balances, and the thermodynamic-equilibrium relations are closed consistently within a sequential sharp-interface coupling procedure.

For bubble-dissolution problems at normal temperature and pressure, the temperature variation induced by interfacial mass transfer is often negligible. In the absence of external heat sources or imposed thermal boundary fluxes, the system may then be treated as isothermal, and Step~4 can be omitted without modifying the species-coupling procedure. Finally, because the dispersed phase is treated as incompressible in the present study, its density must remain constant at the discrete level. When applying the Robin condition \eqref{e45}, the density appearing in the dispersed-phase boundary condition is therefore evaluated from the sum of the interfacial species concentrations obtained from the previous time step. This treatment keeps the total concentration in the dispersed phase consistent with the imposed constant density.

\subsection{Overall solution procedure}\label{sec4.4}

The complete algorithm within one time step is summarized below. The ordering is chosen according to the coupling structure of the problem: the interface geometry is first updated, the species equations are then used to determine the interfacial mass fluxes, and these fluxes are subsequently used to close the temperature equation and the flow-field update.

\begin{enumerate}
\item \textbf{Interface update.}  
Using the velocity field at time level $n$, solve the VOF transport equation as described in Section~\ref{sec3.2} to obtain the updated volume fraction $f^{n+1/2}$ and reconstruct the interface $\Gamma^{n+1/2}$.

\item \textbf{Material-property update.}  
Update the phase indicator, density, viscosity, and other piecewise-constant material properties from the new volume fraction field $f^{n+1/2}$.

\item \textbf{Species transport in the continuous phase.}  
For each transferring species, solve Eq.~\eqref{e33} in the continuous phase using the interfacial concentrations provided by the thermodynamic-equilibrium relation from the previous coupling update. These values are imposed as Dirichlet conditions at $\Gamma^{n+1/2}$, and the continuous-phase concentration field is updated to $c_{c,k}^{n+1/2}$.

\item \textbf{Evaluation of interfacial mass fluxes.}  
Reconstruct the one-sided concentration gradients on the continuous-phase side of the interface and solve the coupled interfacial system \eqref{e44} to obtain the partial mass fluxes $\dot m_k^{\,n+1/2}$ and the total mass flux $\dot m^{\,n+1/2} = \sum_{k=1}^{N_t}\dot m_k^{\,n+1/2}$, where $N_t$ is the number of transferring species.

\item \textbf{Species transport in the dispersed phase.}  
With the interfacial mass fluxes now known, solve Eq.~\eqref{e33} in the dispersed phase by imposing the corresponding Robin conditions \eqref{e45}. The dispersed-phase concentration field is updated to $c_{d,k}^{n+1/2}$, and the dispersed-phase interfacial concentrations are reconstructed from the sharp embedded-boundary treatment.

\item \textbf{Temperature update.}  
If thermal effects are considered, solve Eq.~\eqref{e32} in the liquid and gas phases using the interfacial temperature-continuity condition and the heat-flux jump condition \eqref{e9}--\eqref{e10}, whose right-hand side is determined by the partial mass fluxes obtained in Step~4. This gives the temperature field $T^{n+1/2}$ and the interfacial temperature $T_{\Gamma}^{n+1/2}$. If the process is treated as isothermal, this step is omitted.

\item \textbf{Update of the interfacial equilibrium state.}  
Use the reconstructed interfacial temperature and dispersed-phase interfacial concentrations together with the thermodynamic-equilibrium relation, for example Eq.~\eqref{e17} for droplet evaporation or Eq.~\eqref{e18} for bubble dissolution, to update the interfacial concentrations on the continuous-phase side. These values serve as Dirichlet conditions for the next time step.

\item \textbf{Flow-field update.}  
With the interfacial mass fluxes known, solve the one-fluid momentum equations and apply the projection step described in Section~\ref{sec3.1} to update the velocity and pressure fields from $(\boldsymbol{u}^{n},p^{n})$ to $(\boldsymbol{u}^{n+1},p^{n+1})$.

\item \textbf{Advance to the next time step.}  
Proceed to the next step in time and repeat the above procedure.
\end{enumerate}

The above sequence highlights the coupling logic of the method. The continuous-phase species equations first provide the one-sided concentration gradients needed to determine the partial interfacial mass fluxes. These fluxes then close the dispersed-phase species equations, the temperature equation, and finally the hydrodynamic update. This ordering keeps interface motion, multicomponent mass transfer, heat transfer, and flow evolution consistent within the same geometrical VOF--finite-volume framework.

\section{Numerical tests}\label{sec5}

In this section, a series of numerical tests is presented to validate the proposed sharp-interface method and to demonstrate its applicability to representative liquid--gas mass-transfer problems. The tests are organized progressively. We first isolate the sharp treatment of the species equation using a static dissolving bubble with an analytical solution. We then consider moving and deformable interfaces, multicomponent bubble dissolution, and droplet evaporation, in order to assess the coupled treatment of interface transport, species transfer, heat transfer, and thermodynamic equilibrium.

\subsection{Dissolution of a static bubble with a single transferring component}\label{sec5.1}

We first consider the dissolution of a static spherical bubble involving a single transferring component. This test is designed to validate the embedded-boundary treatment developed in Sections~\ref{sec4.2} and \ref{sec4.3} for the species equation, with particular emphasis on two aspects: the second-order discretization of the diffusion operator in the presence of a sharp interface, and the accurate evaluation of the interfacial mass flux from one-sided concentration gradients. Since the bubble remains fixed and no thermal coupling is involved, this benchmark isolates the interfacial mass-transfer process from the additional complications associated with interface advection, hydrodynamic motion, and temperature evolution. It therefore provides a clean test of the phase-wise concentration solver before more complex multicomponent and thermally coupled configurations are considered.

Consider a spherical bubble of radius $R_0$, initially at rest in an unbounded liquid. The concentration inside the bubble is initially higher than that in the surrounding liquid, so that the transferring component diffuses outward across the interface. Under spherical symmetry, the gas-side and liquid-side concentration fields satisfy
\begin{subequations}
\begin{align}
\frac{\partial c_g}{\partial t}
&=
D_g \frac{1}{r^2}
\frac{\partial}{\partial r}
\left(
r^2 \frac{\partial c_g}{\partial r}
\right),
\qquad
\mathrm{in}\quad \Omega_g,
\\[0.5em]
\frac{\partial c_l}{\partial t}
&=
D_l \frac{1}{r^2}
\frac{\partial}{\partial r}
\left(
r^2 \frac{\partial c_l}{\partial r}
\right),
\qquad
\mathrm{in}\quad \Omega_l,
\end{align}
\label{e467}
\end{subequations}
subject to the interfacial, initial, and far-field conditions
\begin{equation}
\left\{
\begin{array}{l}
c_l=Hc_g,
\qquad
\mathrm{on}\quad \Gamma, \vspace{0.5em}\\
D_g\dfrac{\partial c_g}{\partial r}
=
D_l\dfrac{\partial c_l}{\partial r},
\qquad
\mathrm{on}\quad \Gamma, \vspace{0.5em}\\
c_g(r,t=0)=c_{g,0},
\qquad
r<R_0, \vspace{0.5em}\\
c_l(r,t=0)=c_{l,0},
\qquad
r>R_0, \vspace{0.5em}\\
c_l(r\rightarrow\infty,t)=0,
\qquad
t>0.
\end{array}
\right.
\label{e48}
\end{equation}
Here, $H$ is the concentration partition coefficient linking the liquid-side and gas-side interfacial concentrations.

For this problem, \citet{farsoiya2021bubble} derived an analytical solution for the transient concentration fields inside and outside the bubble using the residue theorem. The solution reads
\begin{subequations}
\begin{align}
c_g(r,t)
&=
-\frac{2c_{g,0}}{\pi r}
\int_0^{\infty}
\frac{1}{x}
\mathrm{Im}
\left[
\frac{\sinh(\lambda_g(-x)r)}
{\zeta(-x)}
\right]
e^{-xt}\,dx,
\qquad
\mathrm{in}\quad \Omega_g,
\\[0.5em]
c_l(r,t)
&=
-\frac{c_{g,0}}{\pi r}
\int_0^{\infty}
\frac{1}{x}
\mathrm{Im}
\left[
\frac{\xi(-x)}
{\zeta(-x)}
e^{-\lambda_l(-x)r}
\right]
e^{-xt}\,dx,
\qquad
\mathrm{in}\quad \Omega_l,
\end{align}
\label{e490}
\end{subequations}
where
\begin{equation}
\begin{aligned}
& \lambda_{g/l}(s)=\sqrt{\frac{s}{D_{g/l}}},\\[0.5em]
& \zeta(s)
=
\frac{\xi(s)\exp[-\lambda_l(s)R_0]}{HR_0}
+
\frac{2}{R_0}\sinh[\lambda_g(s)R_0],\\[0.5em]
& \xi(s)
=
\frac{
2D_g
\left[
\lambda_g(s)R_0\cosh[\lambda_g(s)R_0]
-
\sinh[\lambda_g(s)R_0]
\right]
}
{
D_l\exp[-\lambda_l(s)R_0]
[1+\lambda_l(s)R_0]
}.
\end{aligned}
\label{e50}
\end{equation}
Here, $\mathrm{Im}(\cdot)$ denotes the imaginary part of a complex number.

\begin{figure}[!h]
\centering
\includegraphics[scale=0.35]{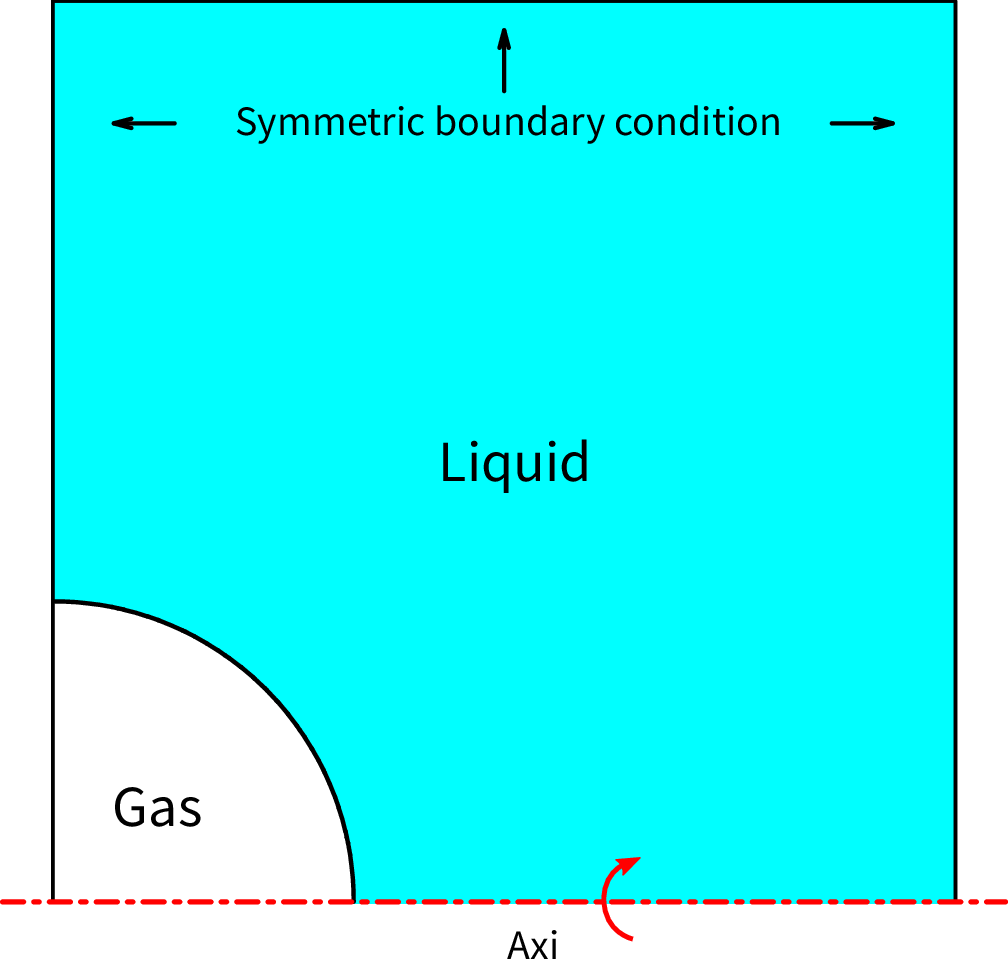}
\caption{Axisymmetric configuration for the dissolution of a static spherical bubble with a single transferring component. The initial bubble radius is $R_0$, and the computational domain size is $10R_0\times10R_0$.}
\label{f4.1.1}
\end{figure}

\begin{figure}[!h]
\centering
\includegraphics[scale=0.345]{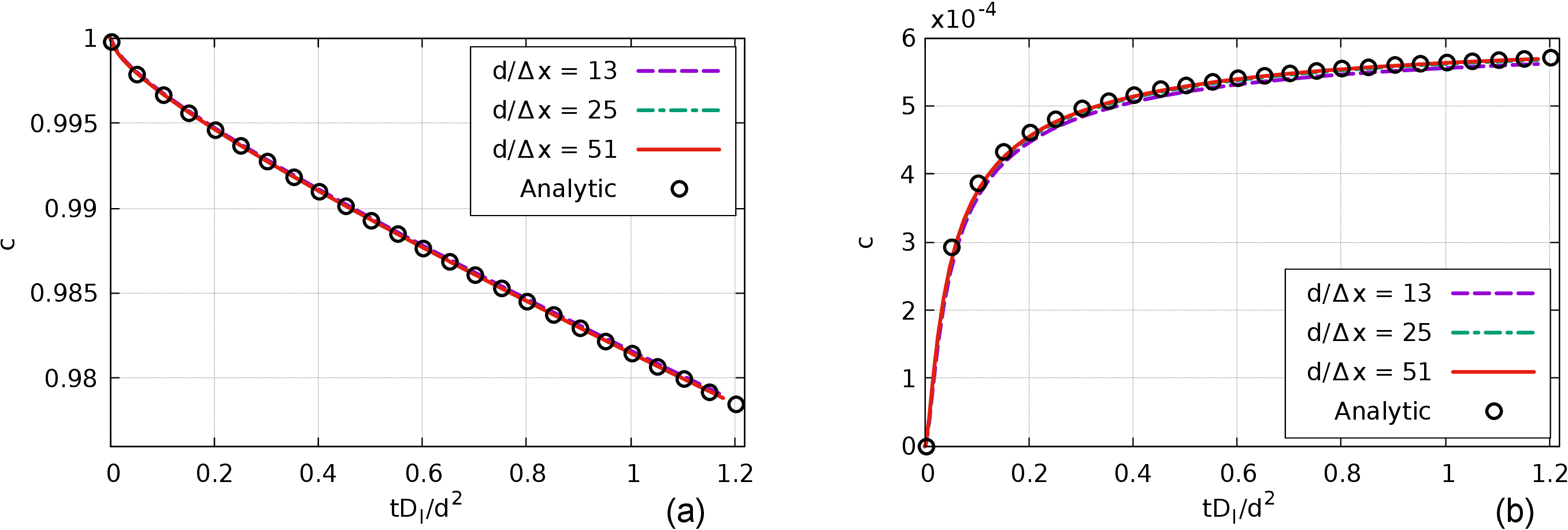}
\caption{Temporal evolution of the concentration at two representative locations for different spatial resolutions: (a) $r=0.5R_0$ inside the bubble and (b) $r=1.5R_0$ in the surrounding liquid. Symbols denote the numerical results, and solid lines denote the analytical solution.}
\label{f4.1.2}
\end{figure}

To validate the present method, simulations are performed in a two-dimensional axisymmetric domain, as shown in Fig.~\ref{f4.1.1}. The bubble radius is $R_0$, and the computational domain size is $10R_0\times10R_0$. The parameters are chosen as $c_{g,0}=1$, $c_{l,0}=0$, $D_l=0.1$, $D_g=1$, and $H=0.001$. This configuration is demanding because the equilibrium relation imposes a sharp concentration discontinuity at the interface, while the diffusivities on the two sides of the interface differ by one order of magnitude. Figure~\ref{f4.1.2} presents the temporal evolution of the concentration at two representative locations, $r=0.5R_0$ inside the bubble and $r=1.5R_0$ in the surrounding liquid. Three spatial resolutions are considered, corresponding to $d/\Delta=13$, 25, and 51, where $d=2R_0$ is the bubble diameter and $\Delta$ is the minimum cell size near the interface. Even on the coarsest mesh, the numerical solution captures the transient evolution reasonably well. As the mesh is refined, the numerical results converge systematically toward the analytical solution on both sides of the interface.

\begin{figure}[!h]
\centering
\includegraphics[scale=0.33]{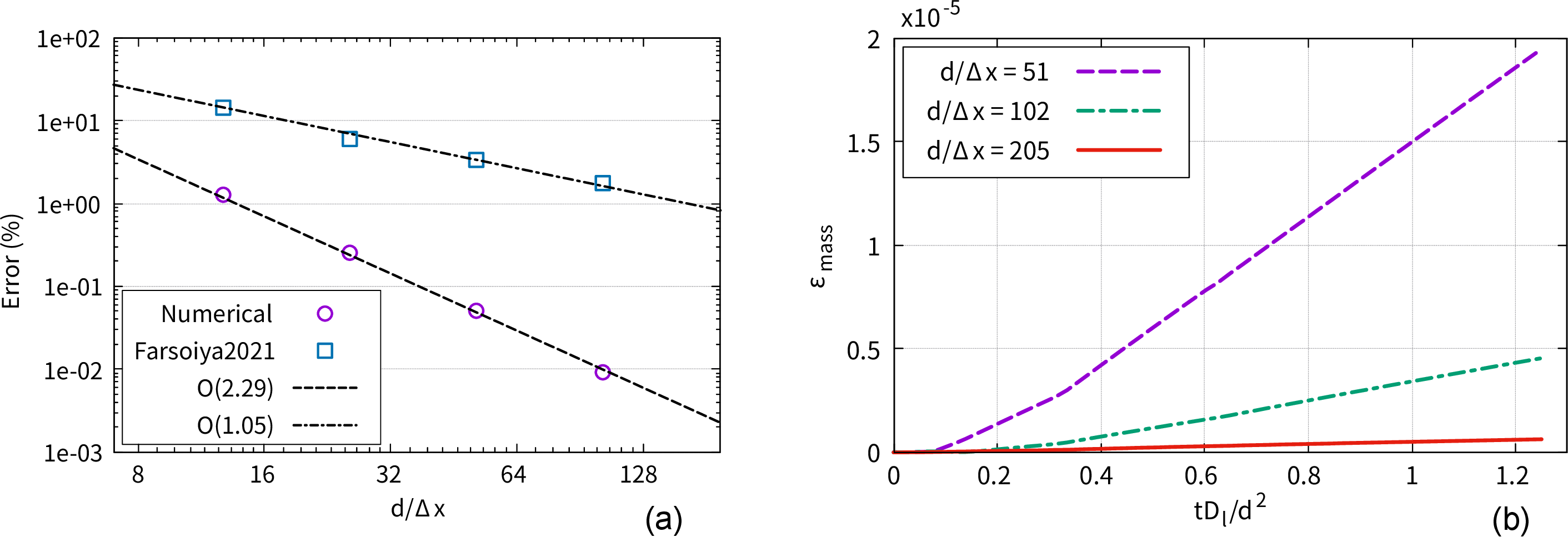}
\caption{Validation of the static-bubble dissolution problem. (a) Relative concentration error at the final time as a function of grid resolution, compared with the one-fluid formulation of \citet{farsoiya2021bubble}. (b) Global mass-conservation error, defined by Eq.~\eqref{e54}, for different grid resolutions.}
\label{f4.1.3}
\end{figure}

To quantify the spatial accuracy, Fig.~\ref{f4.1.3}(a) shows the relative error at the final time, defined as $|(c_{sim}-c_{theo})/c_{theo}|$, where $c_{sim}$ and $c_{theo}$ denote the numerical and analytical concentrations, respectively. For comparison, the results obtained by the one-fluid formulation of \citet{farsoiya2021bubble} are also shown. The present sharp two-field method exhibits a clear second-order convergence trend, whereas the one-fluid formulation is only first-order accurate. This improvement confirms that the embedded-boundary discretization provides a substantially more accurate treatment of the interfacial concentration jump and the associated diffusive flux.

Finally, we assess global mass conservation. The relative mass error is defined as
\begin{equation}
\varepsilon_{mass}
=
\left|
\frac{
M_l+M_g+M_{out}-M_0
}
{M_0}
\right|,
\label{e54}
\end{equation}
where
\begin{equation}
M_g=\int_{\Omega_g}c_g\,dV,
\qquad
M_l=\int_{\Omega_l}c_l\,dV,
\qquad
M_{out}
=
\int_0^t
\int_{\partial\Omega}
\left(
c_l\boldsymbol{u}\cdot\boldsymbol{n}
-
D_l\nabla c_l\cdot\boldsymbol{n}
\right)
dS\,dt',
\label{e55}
\end{equation}
and $M_0$ is the initial total mass of the transferring component in the computational domain. The outward normal on the outer boundary is denoted by $\boldsymbol{n}$. The results shown in Fig.~\ref{f4.1.3}(b) indicate excellent mass conservation: the mass error remains below $0.002\%$ and decreases monotonically with mesh refinement.

\subsection{Rising bubble with dissolution}\label{sec5.2}

We next consider the dissolution of a freely rising bubble with its volume kept constant. This test is intended to validate the coupled flow--mass-transfer solver developed in Sections~\ref{sec3} and \ref{sec4} in a configuration where interfacial transport is governed simultaneously by convection and diffusion. Compared with the static-bubble benchmark in Section~\ref{sec5.1}, the present problem involves a moving interface, a wake-induced concentration field, and a concentration boundary layer whose thickness depends strongly on the Schmidt number. It therefore provides a more stringent assessment of the sharp treatment of the species equation and of the evaluation of the interfacial mass flux in a hydrodynamically evolving flow.

\begin{figure}[!h]
\centering
\includegraphics[scale=0.35]{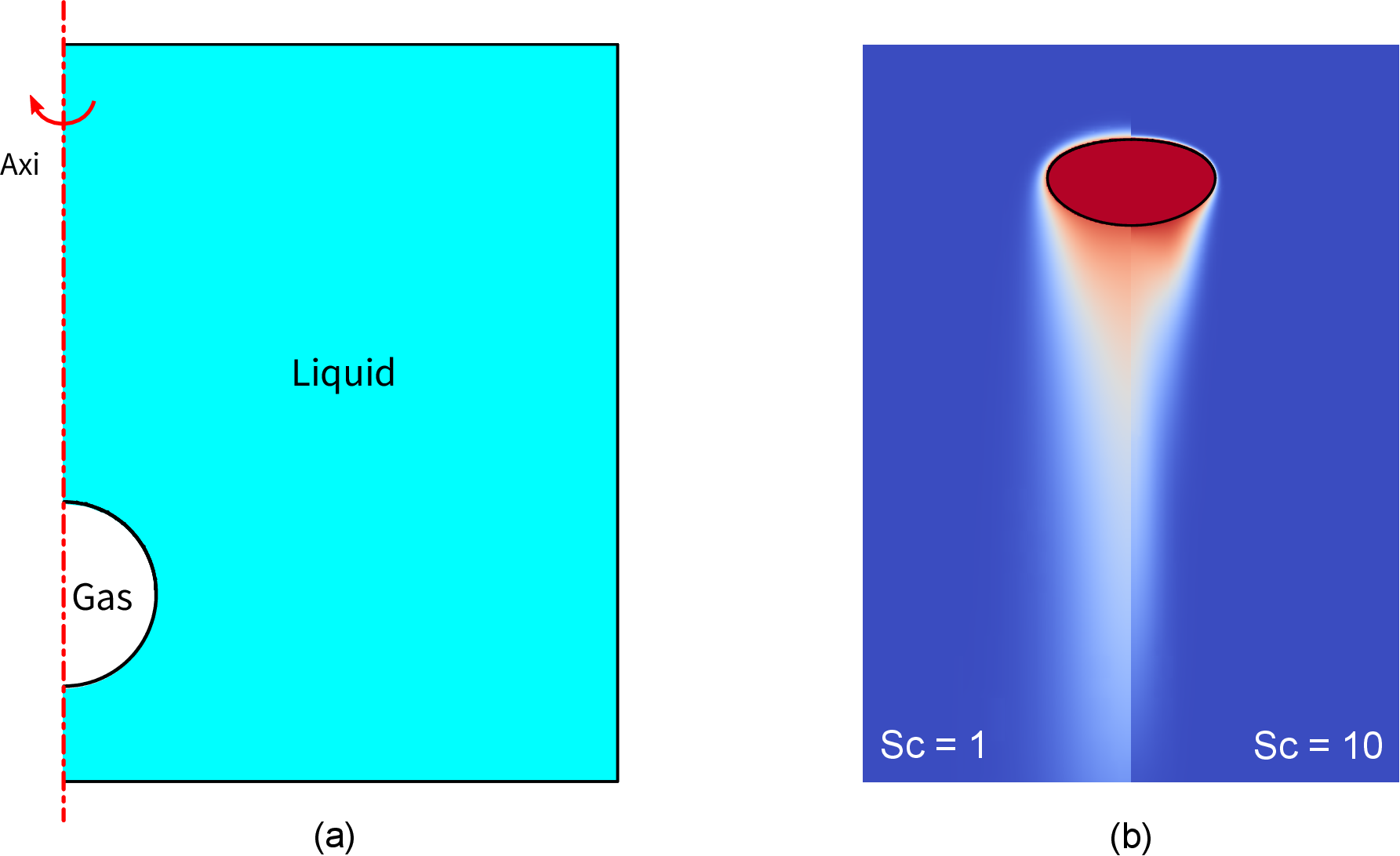}
\caption{Dissolution of a freely rising bubble. (a) Axisymmetric computational configuration. (b) Instantaneous concentration contours for case 3 with $Sc=1$ (left) and case 4 with $Sc=10$ (right), showing the thinning of the concentration boundary layer with increasing $Sc$.}
\label{f4.2.1}
\end{figure}

The problem is solved in a two-dimensional axisymmetric domain, as sketched in Fig.~\ref{f4.2.1}(a). A spherical bubble of initial radius $R_0$ rises freely in an initially quiescent liquid, while one transferring component dissolves from the bubble into the surrounding liquid. The computational domain size is chosen as $20R_0\times20R_0$, which is sufficiently large to avoid any significant influence of the outer boundaries on the bubble motion and on the concentration field during the time interval of interest.

 The bubble dynamics are characterized by the Morton number $Mo=g\mu_l^4/(\rho_l\sigma^3)$ and the Bond number $Bo=\rho_l g d_0^2/\sigma$, where $d_0=2R_0$ is the initial bubble diameter. The mass-transfer process is characterized by the Schmidt number $Sc=\nu_l/D_l$ and the Peclet number $Pe=Ud/D_l$, where $U$ and $d$ denote the instantaneous bubble velocity and equivalent diameter, respectively. To enable direct comparison with previous numerical studies \citep{deising2016unified,farsoiya2021bubble}, we consider the set of cases listed in Table~\ref{t4.1}. In all cases, the diffusivity ratio and the concentration partition coefficient are fixed as $D_g/D_l=100$ and $H=1/30$, respectively.

\begin{table*}[!t]
\caption{\label{t4.1}Test cases for the dissolution of a freely rising bubble considered in Section~\ref{sec5.2}.}
\centering
\footnotesize
\begin{tabular}{lp{1.6cm}p{1.6cm}p{1.6cm}}
\hline
\rule{0pt}{10pt}
Case & Mo & Bo & Sc \\ [3pt]
\hline
\rule{0pt}{10pt}
1 & $10^{-4}$ & 1 & 1 \\
\rule{0pt}{10pt}
2 & $5\times 10^{-4}$ & 3.125 & 1 \\
\rule{0pt}{10pt}
3 & $5\times 10^{-7}$ & 3.125 & 1 \\
\rule{0pt}{10pt}
4 & $5\times 10^{-7}$ & 3.125 & 10 \\
\rule{0pt}{10pt}
5 & $5\times 10^{-7}$ & 3.125 & 100 \\ [3pt]
\hline
\end{tabular}
\end{table*}

The dissolution rate is quantified by the Sherwood number, $Sh=k_L d / D_l$, where $k_L$ is the liquid-side mass-transfer coefficient. In the present simulations, $k_L$ is evaluated from the total interfacial mass flux as $k_L = \int_{\Gamma}\dot m\,dA/A(Hc_{g,in}-c_{l,bulk})$, where $A$ is the bubble surface area, and $c_{g,in}$ and $c_{l,bulk}$ denote the average concentration inside the bubble and the bulk concentration in the surrounding liquid, respectively.

Figure~\ref{f4.2.1}(b) first illustrates the influence of the Schmidt number on the concentration boundary layer. At the same instant, the case with $Sc=10$ exhibits a much thinner concentration boundary layer than that with $Sc=1$, as expected. This trend is numerically important, because a larger Schmidt number produces a sharper interfacial concentration gradient and therefore requires a finer mesh for accurate flux evaluation. In the present simulations, a resolution of $d/\Delta=102$ is employed for $Sc=1$, whereas a finer resolution of $d/\Delta=205$ is used for the higher-Schmidt-number cases with $Sc=10$ and $100$.

\begin{figure}[!h]
\centering
\includegraphics[scale=0.34]{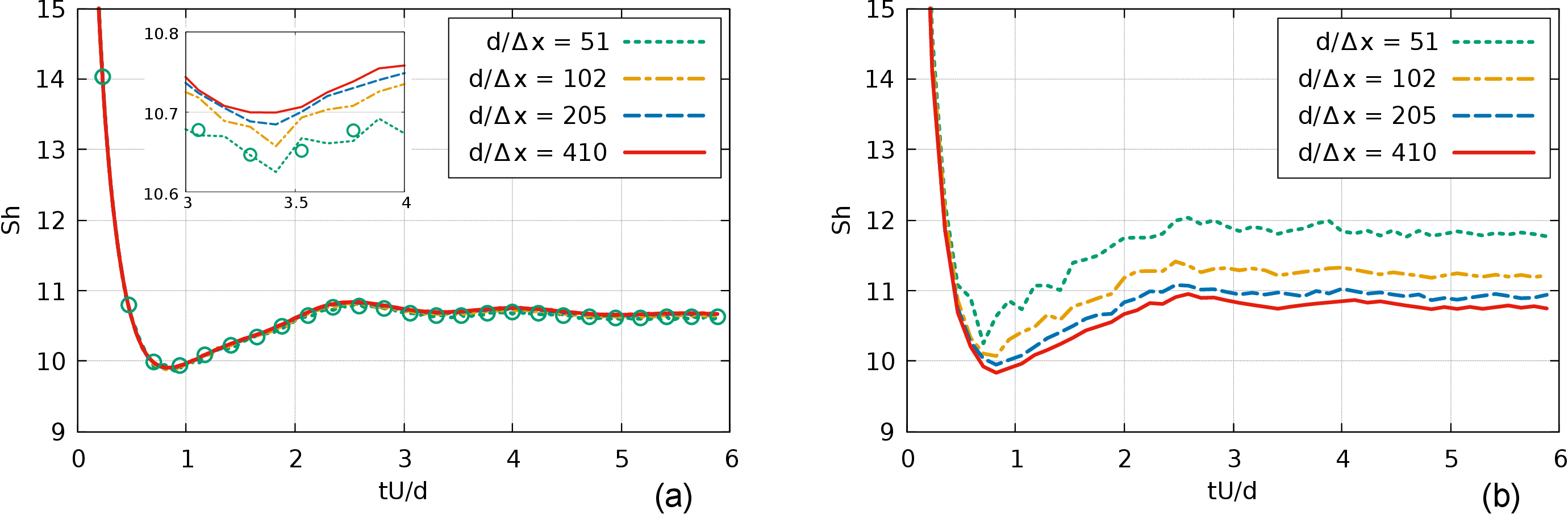}
\caption{Temporal evolution of the Sherwood number for case 3 at different grid resolutions: (a) present sharp two-field method and (b) one-fluid formulation of \citet{farsoiya2021bubble}. In panel (a), circles denote the corresponding three-dimensional results at the coarsest resolution, and the inset provides an enlarged view for $3\le tU/d\le4$. Time is normalized by the bubble equivalent diameter and terminal rising velocity.}
\label{f4.2.2}
\end{figure}

Figure~\ref{f4.2.2} shows the temporal evolution of $Sh$ for case 3. After a short transient, the Sherwood number gradually approaches a nearly constant value as the bubble reaches its terminal rising velocity. The figure also compares the present sharp two-field results, shown in Fig.~\ref{f4.2.2}(a), with the one-fluid results of \citet{farsoiya2021bubble}, shown in Fig.~\ref{f4.2.2}(b). As in the static-bubble benchmark, the present method exhibits a noticeably faster convergence under mesh refinement, indicating that the interfacial mass flux is resolved more accurately by the sharp embedded-boundary treatment. For additional verification, a three-dimensional simulation at the coarsest resolution is also included in Fig.~\ref{f4.2.2}(a) using circles. The resulting values agree closely with the corresponding axisymmetric results, further supporting the consistency of the present formulation.

\begin{figure}[!h]
\centering
\includegraphics[scale=0.34]{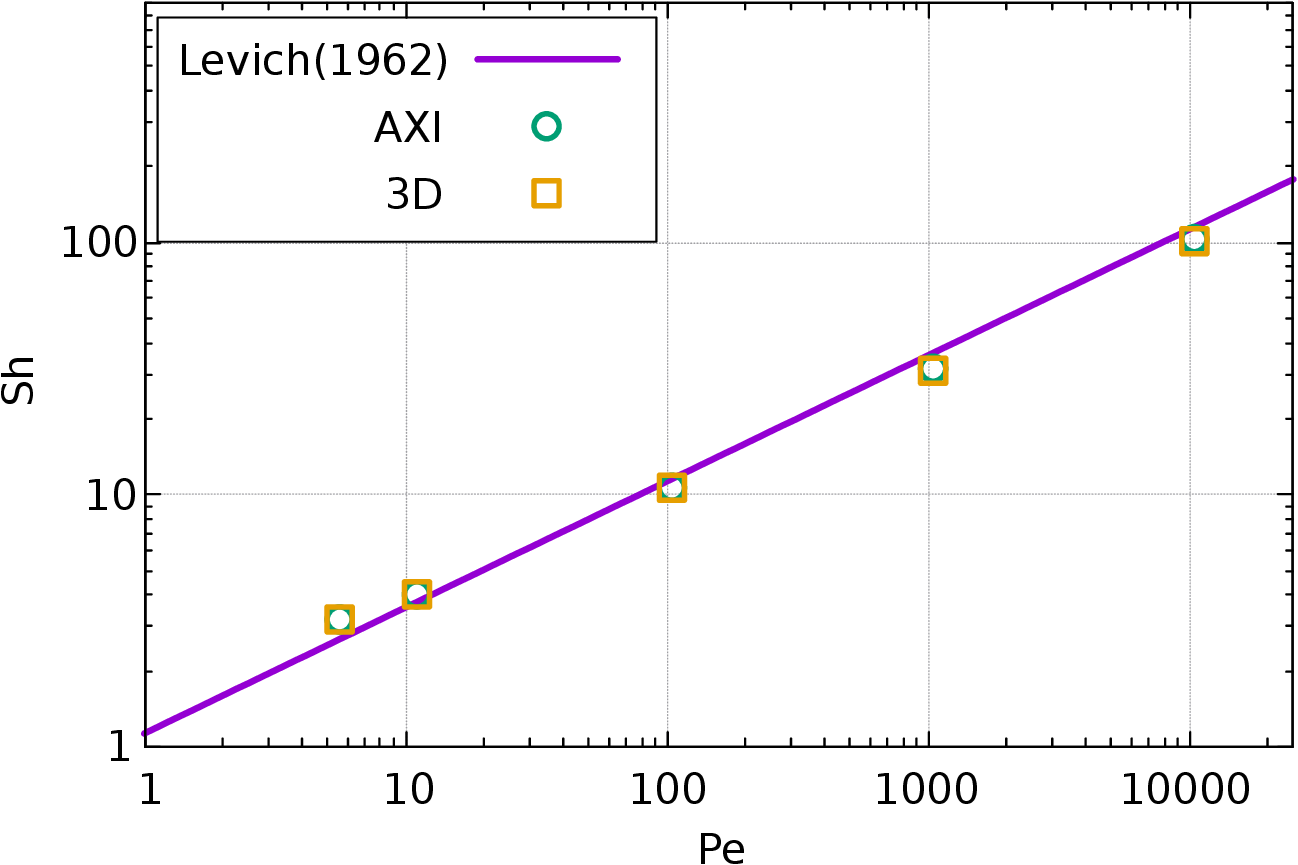}
\caption{Steady Sherwood number as a function of the Peclet number for all simulated cases, compared with the theoretical prediction $Sh=2\sqrt{Pe/\pi}$.}
\label{f4.2.3}
\end{figure}

For comparison with theory, we recall that \citet{boussinesq1905calcul,levivc1962physicochemical} derived, in the potential-flow limit, the classical prediction $Sh=2\sqrt{Pe/\pi}$. Figure~\ref{f4.2.3} compares the steady Sherwood numbers obtained from the simulations with this theoretical prediction, where $Pe$ is evaluated using the terminal bubble velocity. The numerical results from both the axisymmetric and three-dimensional simulations collapse well in the $Sh$--$Pe$ plane and agree closely with the theoretical scaling. This demonstrates that the present method accurately captures the coupled convection--diffusion mass-transfer process over a broad range of flow and transport conditions.

\begin{figure}[!h]
\centering
\includegraphics[scale=0.34]{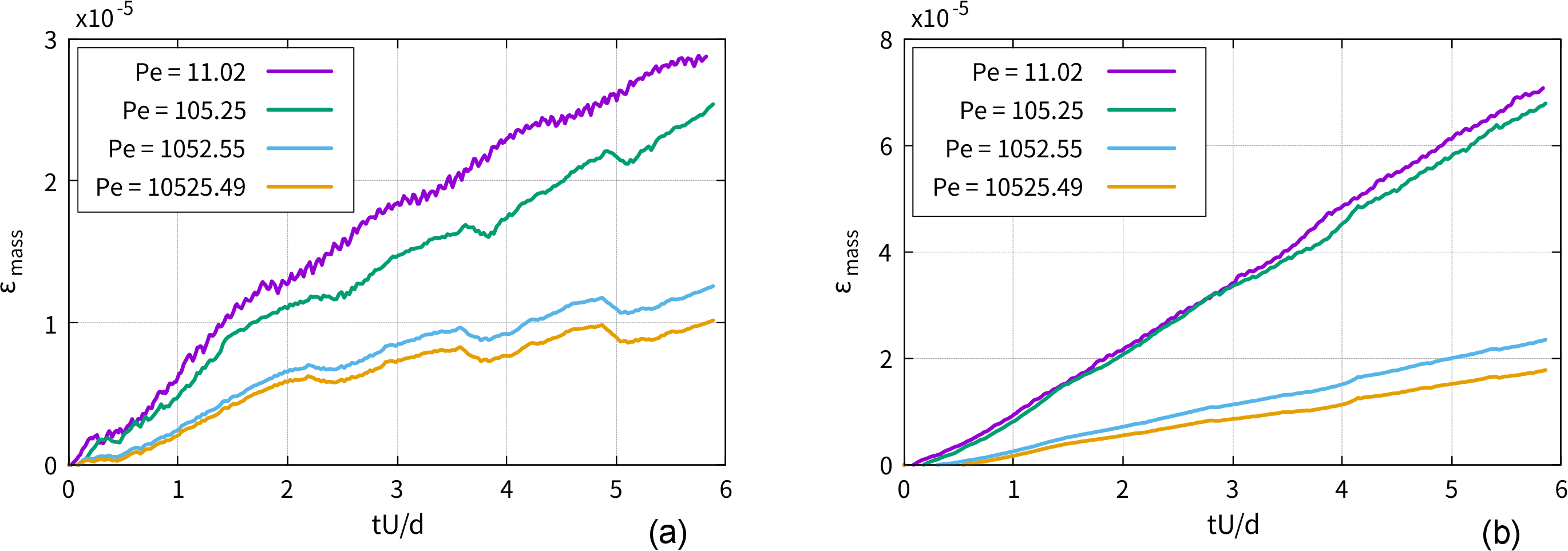}
\caption{Global mass-conservation error for the rising-bubble dissolution problem: (a) axisymmetric simulations and (b) three-dimensional simulations.}\label{f4.2.4}
\end{figure}

Finally, global mass conservation is examined in Fig.~\ref{f4.2.4}, with panel (a) showing the axisymmetric results and panel (b) the three-dimensional results. In all cases, the overall mass error remains below $0.008\%$ and decreases monotonically as the mesh is refined. This confirms that the present sharp interfacial treatment preserves not only the local accuracy of the mass-transfer rate, but also the global conservation property of the concentration field.

\subsection{Dissolution of a freely rising carbon-dioxide bubble}\label{sec5.3}

We now consider a more practical multicomponent mass-transfer problem, namely the dissolution of a carbon-dioxide bubble rising in water. This example is designed to assess the full coupling strategy developed in Sections~\ref{sec4.2}--\ref{sec4.4}. Unlike the preceding single-component tests, the present case involves several species that may be transferred in opposite directions across the same interface: $\mathrm{CO_2}$ dissolves from the bubble into the surrounding water, whereas dissolved $\mathrm{N_2}$ and $\mathrm{O_2}$ in the liquid may diffuse into the bubble. The bubble volume also evolves continuously during the dissolution process. This test therefore involves multicomponent interfacial equilibrium, phase-wise concentration fields, multiple partial mass fluxes, and their feedback on the bubble dynamics. It provides a demanding and practically relevant validation case, since the dissolution of carbon-dioxide bubbles is important in many environmental and industrial processes.

Carbon dioxide is highly soluble in water. However, experiments by \citet{Hosoda2014Mass,hosoda2015dissolution} showed that a pure carbon-dioxide bubble does not simply disappear monotonically. Instead, after a long dissolution process, the bubble may approach a quasi-equilibrium state in which its volume decreases only very slowly. The physical reason is that, while $\mathrm{CO_2}$ leaves the bubble and dissolves into the surrounding liquid, trace amounts of dissolved $\mathrm{N_2}$ and $\mathrm{O_2}$ in water enter the bubble through the same interface. As a result, $\mathrm{N_2}$ and $\mathrm{O_2}$ gradually replace $\mathrm{CO_2}$ and eventually become the dominant components inside the bubble. This behavior is particularly suitable for testing the present method, because it requires the accurate prediction of three species, opposite transfer directions, and long-time evolution of the interfacial composition. Earlier numerical attempts, such as that of \citet{fleckenstein2015volume}, did not yet provide fully satisfactory results for this problem. We therefore re-examine it here using the present sharp and conservative multicomponent framework.

As an independent reference for the present simulations, we use the semi-empirical model of \citet{Hosoda2014Mass,hosoda2015dissolution}, which has been extensively compared with experimental measurements. Assuming that the process is isothermal, the molar amount $n_i$ of component $i$ inside the bubble may be estimated from
\begin{equation}
\frac{dn_i}{dt}
=
-\frac{\pi d^2 k_{L,i}}{\mathcal{M}_i}
\left(
c_{l,i,\Gamma}-c_{l,i,bulk}
\right),
\label{e58}
\end{equation}
and the bubble diameter is evaluated as
\begin{equation}
d
=
\left[
\frac{6}{\pi}
\left(
\sum_{i=1}^{N_t}
\frac{n_i\mathcal{M}_i}{\rho_g}
\right)
\right]^{1/3},
\label{e59}
\end{equation}
where $\mathcal{M}_i$ is the molar mass of component $i$, and $N_t=3$ is the number of transferring species in the present problem. For small bubbles with limited deformation, the liquid-side mass-transfer coefficient is estimated by $k_{L,i} = 2\sqrt{UD_{l,i}/\pi d}$, which corresponds to the classical convection--diffusion correlation introduced in Section~\ref{sec5.2}. The liquid-side interfacial concentration is approximated by the equilibrium relation $c_{l,i,\Gamma} \approx c_L\mathcal{M}_i \zeta_{g,i}P/\mathcal{H}_i$, where $c_L=55.4~\mathrm{kmol/m^3}$ is the molar concentration of pure water, $\zeta_{g,i}$ is the mole fraction of component $i$ inside the bubble, $P$ is the pressure, and $\mathcal{H}_i$ is Henry's constant. In the reduced model, the bubble rise velocity is estimated from
\begin{equation}
U
=
\sqrt{
\frac{4(\rho_l-\rho_g)gd}{3C_D\rho_l}
},
\qquad
C_D
=
\max
\left[
\min
\left(
\frac{16}{Re}\left(1+0.15Re^{0.687}\right),
\frac{48}{Re}
\right),
\frac{8}{3}\frac{Eo}{Eo+4}
\right],
\label{e61}
\end{equation}
where $C_D$ is the drag coefficient. It should be emphasized that the reduced model used in the experiments accounts for gas compressibility through an equation of state, whereas the present simulations assume a constant gas density and do not solve a gas equation of state. Therefore, a direct quantitative comparison with the experimental measurements themselves is not attempted here. Instead, the reduced model is used as a reference to assess whether the present numerical method captures the dominant multicomponent dissolution mechanism.

\begin{table*}[!t]
\caption{\label{t4.2}Physical properties of the liquid--gas system used in the carbon-dioxide bubble dissolution problem of Section~\ref{sec5.3}.}
\centering
\footnotesize
\begin{tabular}{lp{1.8cm}p{1.8cm}p{1.8cm}p{1.8cm}p{3cm}}
\hline
\rule{0pt}{10pt}
Phase & $\rho$ $(\mathrm{kg/m^3})$ & $\mu$ $(\mathrm{Pa\cdot s})$ & $D$ $(\mathrm{m^2/s})$ & $\sigma$ $(\mathrm{N/m})$ & $H_{\mathrm{CO_2/N_2/O_2}}$ \\ [3pt]
\hline
\rule{0pt}{10pt}
Gas & 1.8 & $7\times 10^{-5}$ & $1.7\times 10^{-5}$ & 0.07 & 0.827 /\ 0.016 /\ 0.031 \\
\rule{0pt}{10pt}
Liquid & 1000 & $10^{-3}$ & $2\times 10^{-8}$ &  &  \\ [3pt]
\hline
\end{tabular}
\end{table*}

The numerical simulations are performed in a two-dimensional axisymmetric domain similar to that used in Section~\ref{sec5.2}, except that the domain is enlarged to $80d_0\times80d_0$ in order to capture the long-time dissolution dynamics. The initial bubble diameter is $d_0=1~\mathrm{mm}$. Before considering the full three-component problem, we first examine the dissolution of a pure carbon-dioxide bubble using the physical parameters listed in Table~\ref{t4.2}. This preliminary step serves two purposes: it validates the present method in a simpler configuration with variable bubble volume, and it helps identify the grid resolution required for the high-Schmidt-number transport in the liquid phase.

\begin{figure}[!h]
\centering
\includegraphics[scale=0.35]{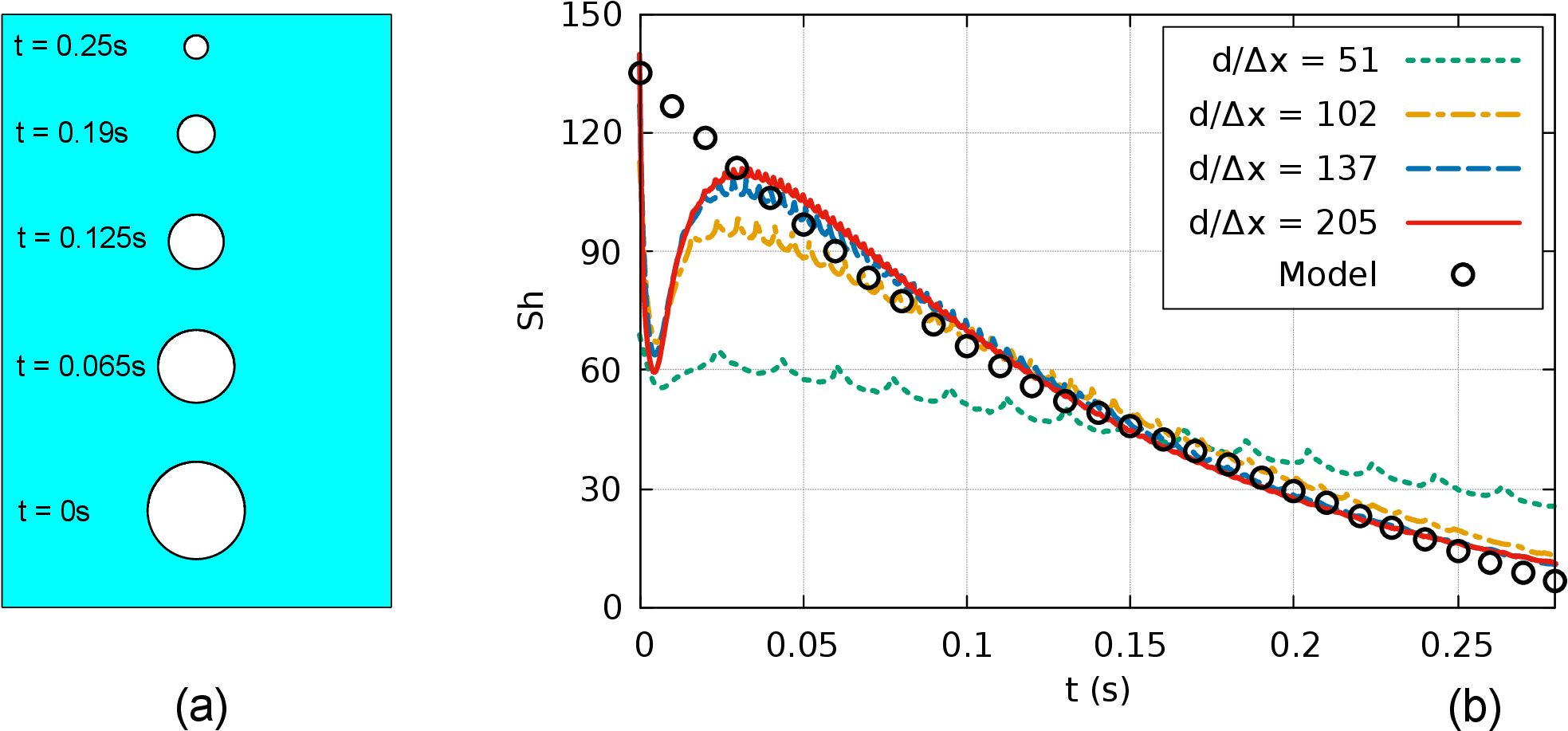}
\caption{Dissolution of a pure carbon-dioxide bubble. (a) Bubble shapes at selected instants during the free-rise dissolution process. (b) Temporal evolution of the Sherwood number for different grid resolutions; circles denote the reduced-model prediction obtained from Eqs.~\eqref{e58}--\eqref{e61}.}
\label{f4.3.1}
\end{figure}

Figure~\ref{f4.3.1}(a) shows the bubble shape at several instants during the rise of an initially stationary carbon-dioxide bubble in water. The bubble shrinks continuously as $\mathrm{CO_2}$ dissolves into the surrounding liquid. Figure~\ref{f4.3.1}(b) presents the corresponding time evolution of the Sherwood number for several grid resolutions. Because of the high Schmidt number in the liquid phase, accurate convergence requires a fine mesh; the present results indicate that approximately 200 grid points per initial bubble diameter are sufficient. During the initial acceleration stage, the Sherwood number rises rapidly and differs noticeably from the semi-empirical prediction. This discrepancy is expected for two reasons. First, the bubble is initially at rest and undergoes a finite acceleration process, whereas the reduced model uses a quasi-steady estimate of the bubble rise velocity. Second, the liquid-side concentration boundary layer has not yet fully developed, so the quasi-steady mass-transfer correlation used for $k_{L,i}$ is not immediately applicable. After this initial transient, the numerical results agree well with the reduced model obtained by solving Eqs.~\eqref{e58}--\eqref{e61}, confirming that the present method captures the coupling between bubble shrinkage and interfacial mass transfer. The small oscillations visible in the numerical curves are associated with weak surface disturbances and with variations in the discrete interfacial gradient as the bubble crosses successive grid layers. Their period decreases under mesh refinement, which is consistent with a discretization-induced effect.

Having established the resolution requirement for the pure carbon-dioxide case, we next consider the full three-component problem with a fixed resolution of $d/\Delta=205$. Initially, the carbon-dioxide concentration inside the bubble is set to $c_{g,\mathrm{CO_2}}=1.8~\mathrm{kg/m^3}$ and to zero in the surrounding liquid. Conversely, nitrogen and oxygen are absent from the bubble initially, while their liquid-phase concentrations are set to $c_{l,\mathrm{N_2}}=0.00714~\mathrm{kg/m^3}$ and $c_{l,\mathrm{O_2}}=0.00864~\mathrm{kg/m^3}$, corresponding to equilibrium with air at standard atmospheric pressure. All other physical parameters are the same as those listed in Table~\ref{t4.2}. Since the pure-carbon-dioxide results show that the early transient is affected by both the initial acceleration of the bubble and the development of the liquid-side concentration boundary layer, a preprocessing stage is first performed. During this stage, the gas composition and bubble volume are kept fixed, while the flow and concentration fields are allowed to develop. The full dissolution process is activated only after the bubble reaches a nearly steady rise velocity.

\begin{figure}[!h]
\centering
\includegraphics[scale=0.34]{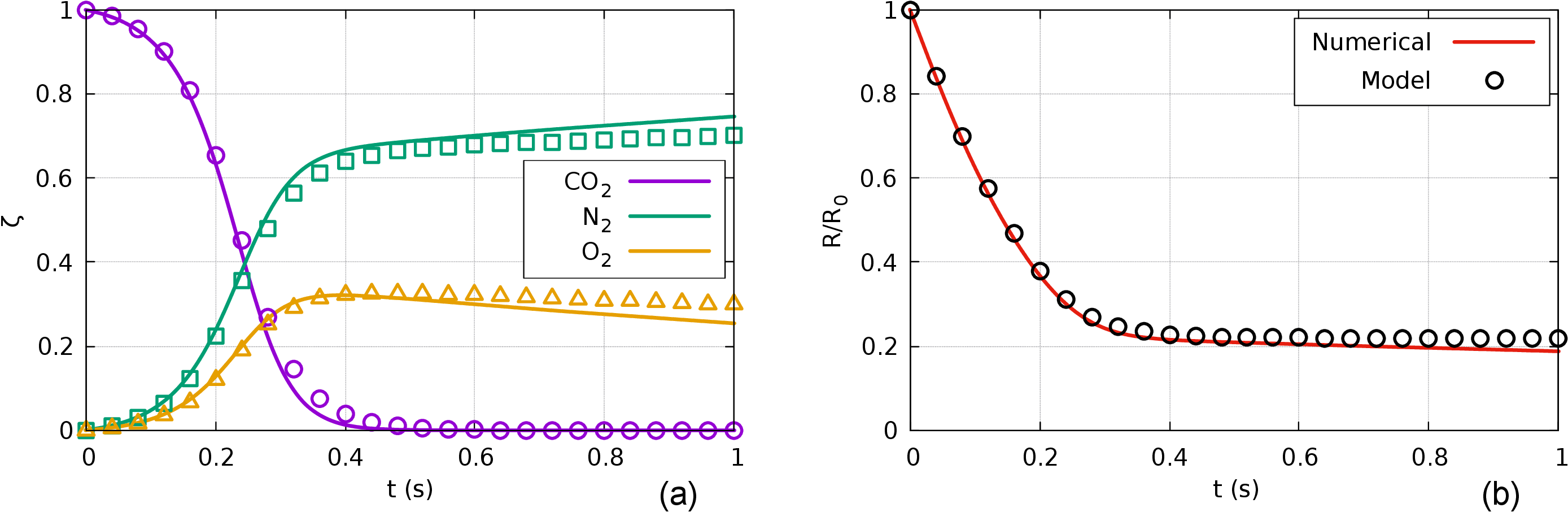}
\caption{Dissolution of a three-component carbon-dioxide bubble. Temporal evolution of (a) the mole fractions of $\mathrm{CO_2}$, $\mathrm{N_2}$, and $\mathrm{O_2}$ inside the bubble and (b) the bubble radius. Symbols denote the reduced-model prediction obtained from Eqs.~\eqref{e58}--\eqref{e61}.}
\label{f4.3.2}
\end{figure}

Figure~\ref{f4.3.2} presents the temporal evolution of the mole fractions of the three gas components inside the bubble and of the bubble radius. As $\mathrm{CO_2}$ dissolves into the surrounding liquid, Fig.~\ref{f4.3.2}(a) shows that its mole fraction decreases rapidly. At the same time, $\mathrm{N_2}$ and $\mathrm{O_2}$ enter the bubble from the liquid phase, and their mole fractions increase continuously. The corresponding bubble shrinkage is shown in Fig.~\ref{f4.3.2}(b). Once most of the $\mathrm{CO_2}$ has been depleted, $\mathrm{N_2}$ and $\mathrm{O_2}$ become the dominant components inside the bubble, and the rate of volume reduction decreases markedly. Eventually, the $\mathrm{CO_2}$ content becomes very small, while the mole fractions of $\mathrm{N_2}$ and $\mathrm{O_2}$ approach approximately $78\%$ and $22\%$, respectively. Correspondingly, the bubble radius tends toward a nearly constant value. This behavior is consistent with the experimentally observed long-time replacement of carbon dioxide by nitrogen and oxygen \citep{Hosoda2014Mass,hosoda2015dissolution}. The numerical predictions of both the gas composition and the bubble radius agree well with the reduced model over most of the dissolution process. The discrepancy becomes more visible only in the final stage, when the bubble has lost most of its initial volume. This deviation is expected, because the number of grid cells inside the bubble then becomes limited, while the simplifying assumptions underlying the reduced model also become less accurate in this regime.

Finally, we evaluate the global mass-conservation error of carbon dioxide over the whole computational domain. During the entire dissolution process, the relative error remains below $0.02\%$, which further demonstrates that the present multicomponent sharp-interface method preserves the conservation of the transferred species even in a long-time simulation involving evolving bubble composition and changing bubble volume.

\subsection{Evaporation of an n-heptane droplet}\label{sec5.4}

We next turn to droplet evaporation. Unlike the bubble-dissolution cases considered above, droplet evaporation involves a strong coupling between vapor transport and heat transfer: the interfacial mass flux is controlled by vapor--liquid equilibrium, while the latent heat of evaporation modifies the temperature field and feeds back on the evaporation rate. This test follows the finite-domain droplet-evaporation benchmark of \citet{pathak2018steady}, whose transient solutions are used as reference data for the droplet radius, temperature field, and vapor-concentration field. The benchmark is therefore used here to validate the thermally coupled evaporation formulation before considering multicomponent liquid mixtures.

\begin{table*}[!t]
\caption{\label{t5.4}Physical properties of the n-heptane droplet and surrounding gas used in Section~\ref{sec5.4}.}
\centering
\footnotesize
\begin{tabular}{lp{1.2cm}p{1.6cm}p{1.5cm}p{1.5cm}p{1.5cm}p{1.4cm}p{1.2cm}p{1.3cm}}
\hline
\rule{0pt}{10pt}
& $\rho$ \newline $(\mathrm{kg/m^3})$
& $\mu$ \newline $(\mathrm{Pa\cdot s})$
& $c_p$ \newline $(\mathrm{J/(kg\,K)})$
& $\lambda$ \newline $(\mathrm{W/(m\,K)})$
& $\mathcal{M}$ \newline $(\mathrm{kg/kmol})$
& $\mathcal{L}$ \newline $(\mathrm{J/kg})$
& $\sigma$ \newline $(\mathrm{N/m})$
& $D$ \newline $(\mathrm{m^2/s})$ \\ [3pt]
\hline
\rule{0pt}{10pt}
Gas & 17.51 & $1\times10^{-5}$ & 1053 & 0.04428 & 29 &  &  & $6.77\times10^{-7}$ \\
\rule{0pt}{10pt}
Liquid & 626.7 & $1\times10^{-3}$ & 2505 & 0.1121 & 100 & $3.23\times10^{5}$ & 0.02 &  \\ [3pt]
\hline
\end{tabular}
\end{table*}

\begin{figure}[!h]
\centering
\includegraphics[scale=0.34]{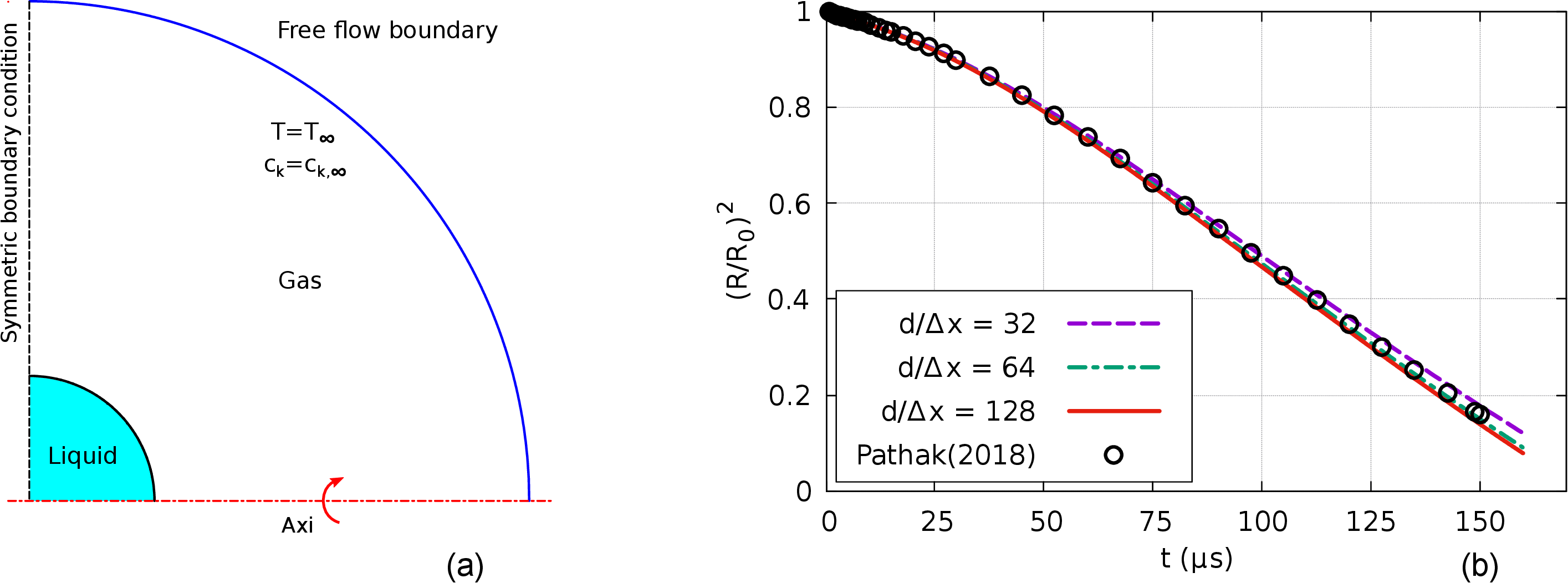}
\caption{Evaporation of a single-component n-heptane droplet. (a) Axisymmetric computational configuration. (b) Temporal evolution of the droplet radius at different spatial resolutions, compared with the reference solution of \citet{pathak2018steady}.}
\label{f5.4.1}
\end{figure}

We consider the evaporation of a stationary spherical droplet in the absence of gravity. The problem is solved in a two-dimensional axisymmetric domain, as shown in Fig.~\ref{f5.4.1}(a). The initial droplet radius is $R_0=5~\mathrm{\mu m}$, and the computational domain size is $4R_0\times4R_0$. To reduce boundary-induced anisotropy, a circular outer boundary of radius $4R_0$ is used to impose the far-field conditions. The initial droplet temperature is $363~\mathrm{K}$, whereas the ambient gas temperature is $536~\mathrm{K}$. The far-field temperature and vapor concentration are fixed at $T_{\infty}=536~\mathrm{K}$ and $c_{v,\infty}=0$, respectively. The physical properties are listed in Table~\ref{t5.4}. The saturation vapor pressure of n-heptane is evaluated from the Antoine correlation listed in~\ref{app2}.

\begin{figure}[!h]
\centering
\includegraphics[scale=0.34]{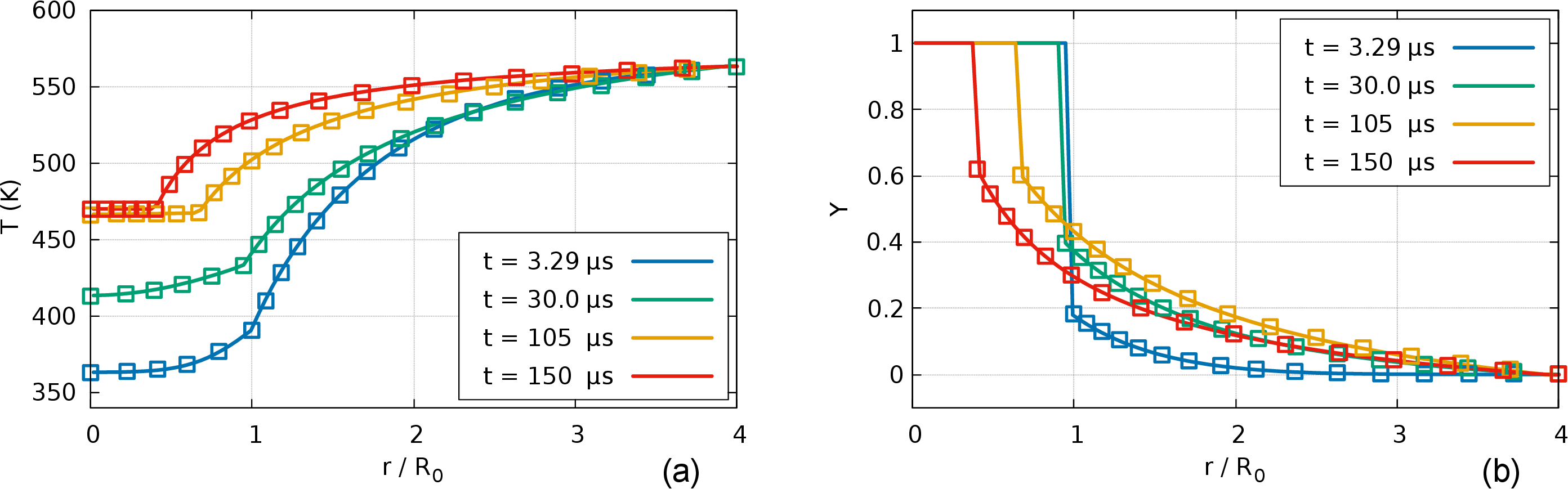}
\caption{Cross-sectional distributions of (a) temperature and (b) vapor concentration during n-heptane droplet evaporation at selected times. Solid lines denote the numerical results, and symbols denote the reference solutions of \citet{pathak2018steady}.}
\label{f5.4.2}
\end{figure}

Figure~\ref{f5.4.1}(b) compares the temporal evolution of the droplet radius obtained at different spatial resolutions with the reference solution of \citet{pathak2018steady}. The numerical results agree well with the reference prediction, and the agreement improves systematically as the mesh is refined. This confirms that the present method accurately captures the evaporation rate of a thermally coupled single-component droplet. Figure~\ref{f5.4.2} further compares the cross-sectional distributions of temperature and vapor concentration with the reference solutions at several representative times. The numerical profiles reproduce the reference results very well for both the thermal field and the vapor-concentration field. These results demonstrate that the coupled solution of heat conduction, vapor diffusion, and interfacial evaporation is accurately captured by the present sharp-interface formulation.

Finally, we examine the conservation of mass and energy during the evaporation process. The relative total-energy error is defined as
\begin{equation}
\varepsilon_{energy}
=
\left|
\frac{
E_l+E_g+E_{pc}+E_{out}-E_{in}-E_0
}
{E_0}
\right|,
\label{e5.52}
\end{equation}
where
\begin{equation}
\left\{
\begin{array}{ll}
E_l
=
\displaystyle
\int_{\Omega_l}
\rho_l c_{p,l}T\,dV, \vspace{1em}\\
E_g
=
\displaystyle
\int_{\Omega_g}
\rho_g c_{p,g}T\,dV, \vspace{1em}\\
E_{pc}
=
\displaystyle
\int_0^t
\int_{\Gamma}
\left[
\sum_{k=1}^{N_t}\dot m_k\mathcal{L}_k
+
\dot m(c_{p,l}-c_{p,g})T_{\Gamma}
\right]
\,dS\,dt', \vspace{1em}\\
E_{out}
=
\displaystyle
\int_0^t
\int_{\partial\Omega}
\rho_g c_{p,g}T\boldsymbol{u}\cdot\boldsymbol{n}
\,dS\,dt', \vspace{1em}\\
E_{in}
=
\displaystyle
\int_0^t
\int_{\partial\Omega}
\lambda_g
\frac{\partial T}{\partial n}
\,dS\,dt'.
\end{array}
\right.
\label{e5.53}
\end{equation}
Here, $E_l$ and $E_g$ are the total internal energies of the liquid and gas phases, respectively, $E_{pc}$ is the energy consumed by phase change, $E_{out}$ is the energy leaving the computational domain through convection, $E_{in}$ is the energy entering the domain through thermal conduction, and $E_0$ is the initial total internal energy. For the present single-component case, $N_t=1$.

\begin{figure}[!h]
\centering
\includegraphics[scale=0.34]{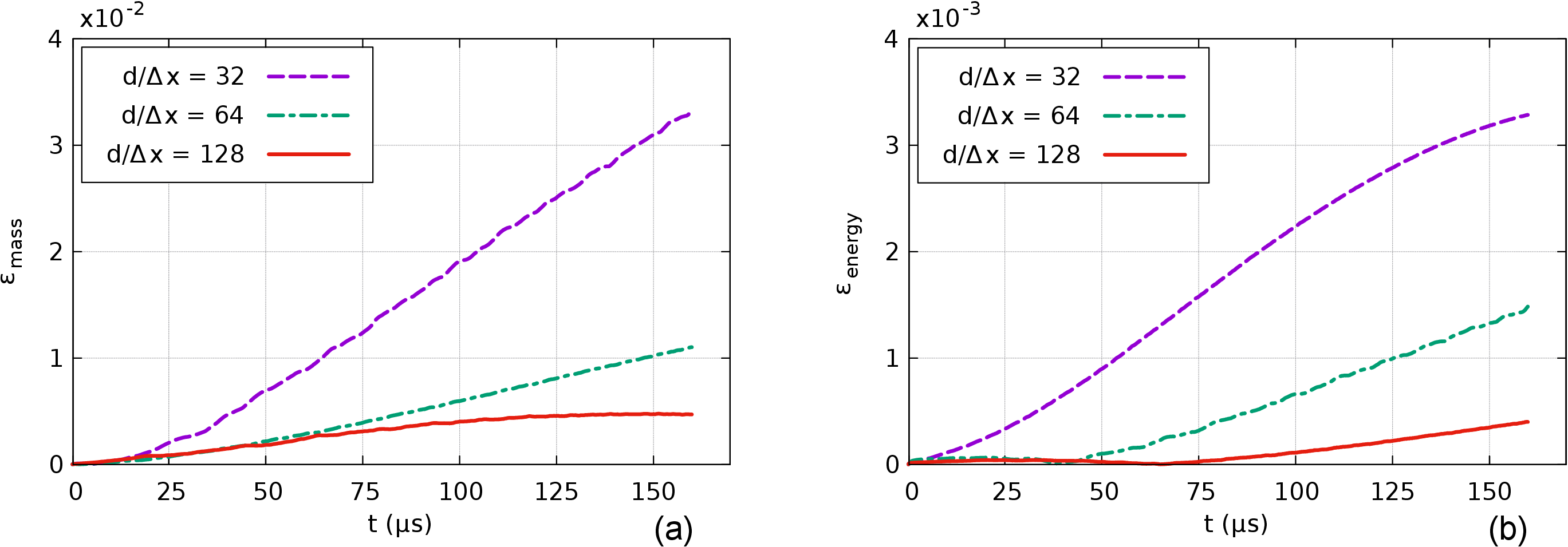}
\caption{Temporal evolution of the (a) relative mass error and (b) relative energy error for different spatial resolutions during n-heptane droplet evaporation.}
\label{f5.4.3}
\end{figure}

Figure~\ref{f5.4.3} shows the temporal evolution of the relative mass and energy errors for different spatial resolutions. Both errors remain small throughout the simulation, confirming that the present method preserves the conservative properties of the thermally coupled single-component evaporation problem.

\subsection{Evaporation of ethanol--isooctane mixture droplets}\label{sec5.5}

We now turn to the evaporation of multicomponent droplets. This example is intended to validate the thermodynamic coupling strategy developed in Sections~\ref{sec4.2}--\ref{sec4.4} for non-ideal liquid mixtures. Compared with the single-component n-heptane case considered in Section~\ref{sec5.4}, the present problem involves two volatile liquid components whose relative evaporation rates are determined not only by their saturation vapor pressures, but also by the activity coefficients associated with liquid-mixture non-ideality. The ethanol--isooctane system provides a representative test case, since isooctane is a major component of gasoline, ethanol is a widely used fuel additive, and their mixture exhibits pronounced non-ideal behavior with an azeotropic point.

\begin{table*}[!t]
\caption{\label{t4.3}Physical properties of the ethanol--isooctane droplets and the surrounding gas used in Section~\ref{sec5.5}.}
\centering
\footnotesize
\begin{tabular}{lp{1.6cm}p{1.6cm}p{2cm}p{2cm}p{1.65cm}}
\hline
\rule{0pt}{10pt}
Phase & $\rho\ (\mathrm{kg/m^3})$ & $\mu\ (\mathrm{Pa\cdot s})$ & $c_p\ (\mathrm{J/(kg\cdot K)})$ & $\lambda\ (\mathrm{W/(m\cdot K)})$ & $D\ (\mathrm{m^2/s})$ \\ [3pt]
\hline
\rule{0pt}{10pt}
Y90 & 730 & $1.0\times10^{-3}$ & 2800 & 0.155 & $2.5\times10^{-9}$ \\
\rule{0pt}{10pt}
Y35 & 700 & $1.0\times10^{-3}$ & 2500 & 0.110 & $2.0\times10^{-9}$ \\
\rule{0pt}{10pt}
Y10 & 670 & $1.0\times10^{-3}$ & 2250 & 0.095 & $2.9\times10^{-9}$ \\
\rule{0pt}{10pt}
Gas & 1.0 & $1.8\times10^{-5}$ & 1010 & 0.030 & $1.0\times10^{-5}$ \\ [3pt]
\hline
\end{tabular}
\end{table*}

To isolate the interfacial mass-transfer characteristics of the non-ideal mixture, we consider stationary droplets and neglect gravity. The problem is solved in a two-dimensional axisymmetric domain, following the same configuration as in the preceding droplet-evaporation case. The initial droplet radius is $R_0=150~\mathrm{\mu m}$, and the computational domain size is $10R_0 \times 10R_0$. The spatial resolution corresponds to approximately 102 grid cells across the initial droplet diameter, namely $d/ \Delta= 102$, where $d=2R_0$. The initial temperatures of the liquid and gas phases are set to $331~\mathrm{K}$ and $372~\mathrm{K}$, respectively, while the far-field conditions are prescribed as $T_{\infty}=372~\mathrm{K}$ and $c_{k,\infty}=0$. Three initial ethanol mass fractions are considered, namely $Y_e=0.9$, $0.35$, and $0.1$, which are denoted hereafter by Y90, Y35, and Y10, respectively. The corresponding material properties are listed in Table~\ref{t4.3}. In all cases, the latent heats of vaporization are taken as $\mathcal{L}_e=8.5\times10^{5}~\mathrm{J/kg}$ for ethanol and $\mathcal{L}_{iso}=3\times10^{5}~\mathrm{J/kg}$ for isooctane, and the surface-tension coefficient is fixed at $\sigma=0.02~\mathrm{N/m}$. The saturation vapor pressures of ethanol and isooctane are evaluated from the Antoine correlations listed in~\ref{app2}. The liquid-phase activity coefficients are evaluated using the UNIFAC group-contribution method \citep{poling2001properties,smith2018introduction}. 

\begin{figure}[!h]
\centering
\includegraphics[scale=0.36]{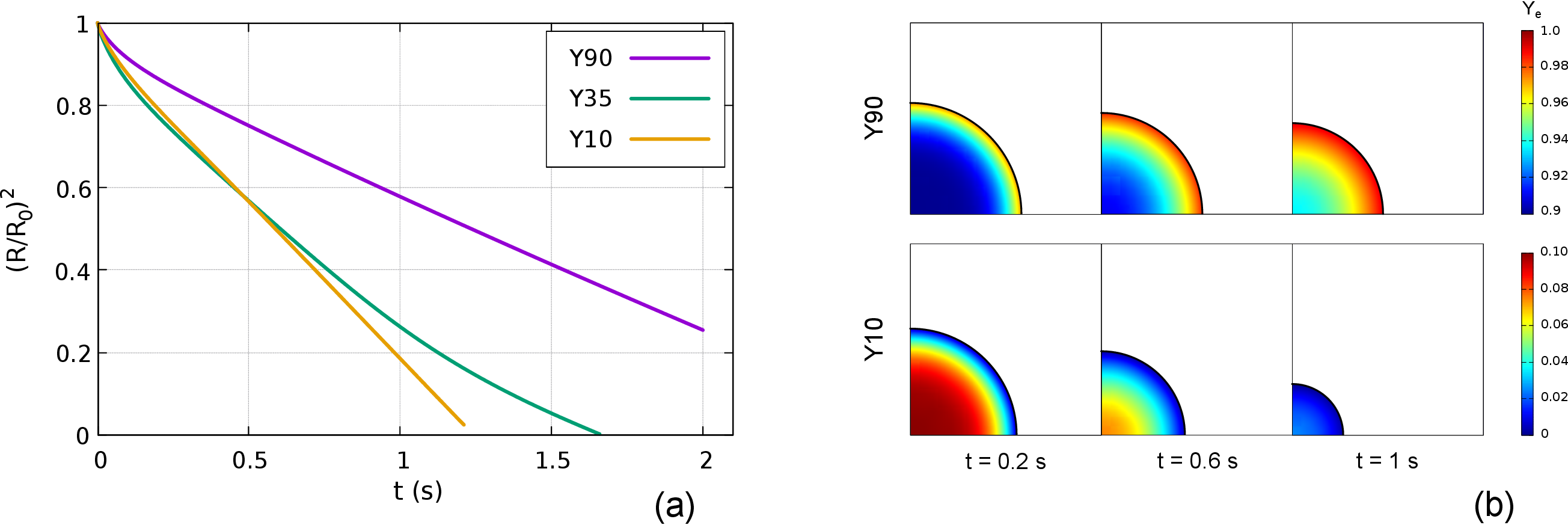}
\caption{Evaporation of ethanol--isooctane mixture droplets. (a) Temporal evolution of the droplet radius for the three initial compositions. (b) Ethanol mass-fraction field inside the liquid droplet for cases Y90 and Y10 at $t=0.2~\mathrm{s}$, $0.6~\mathrm{s}$, and $1~\mathrm{s}$.}
\label{f5.5.1}
\end{figure}

Figure~\ref{f5.5.1}(a) shows the temporal evolution of the droplet radius for the three initial compositions. The evaporation rate depends strongly on the initial mixture composition, with the Y10 droplet evaporating significantly faster than the Y90 droplet. To understand the origin of this behavior, Fig.~\ref{f5.5.1}(b) displays the ethanol mass-fraction field inside the liquid phase at three representative times, $t=0.2~\mathrm{s}$, $0.6~\mathrm{s}$, and $1~\mathrm{s}$, for the two limiting cases Y90 and Y10. These results reveal two qualitatively different evaporation processes. In the Y90 case, the average ethanol mass fraction inside the droplet increases with time, indicating that isooctane evaporates preferentially. In contrast, in the Y10 case, the average ethanol mass fraction decreases continuously, showing that ethanol is then the more volatile component.

\begin{figure}[!h]
\centering
\includegraphics[scale=0.4]{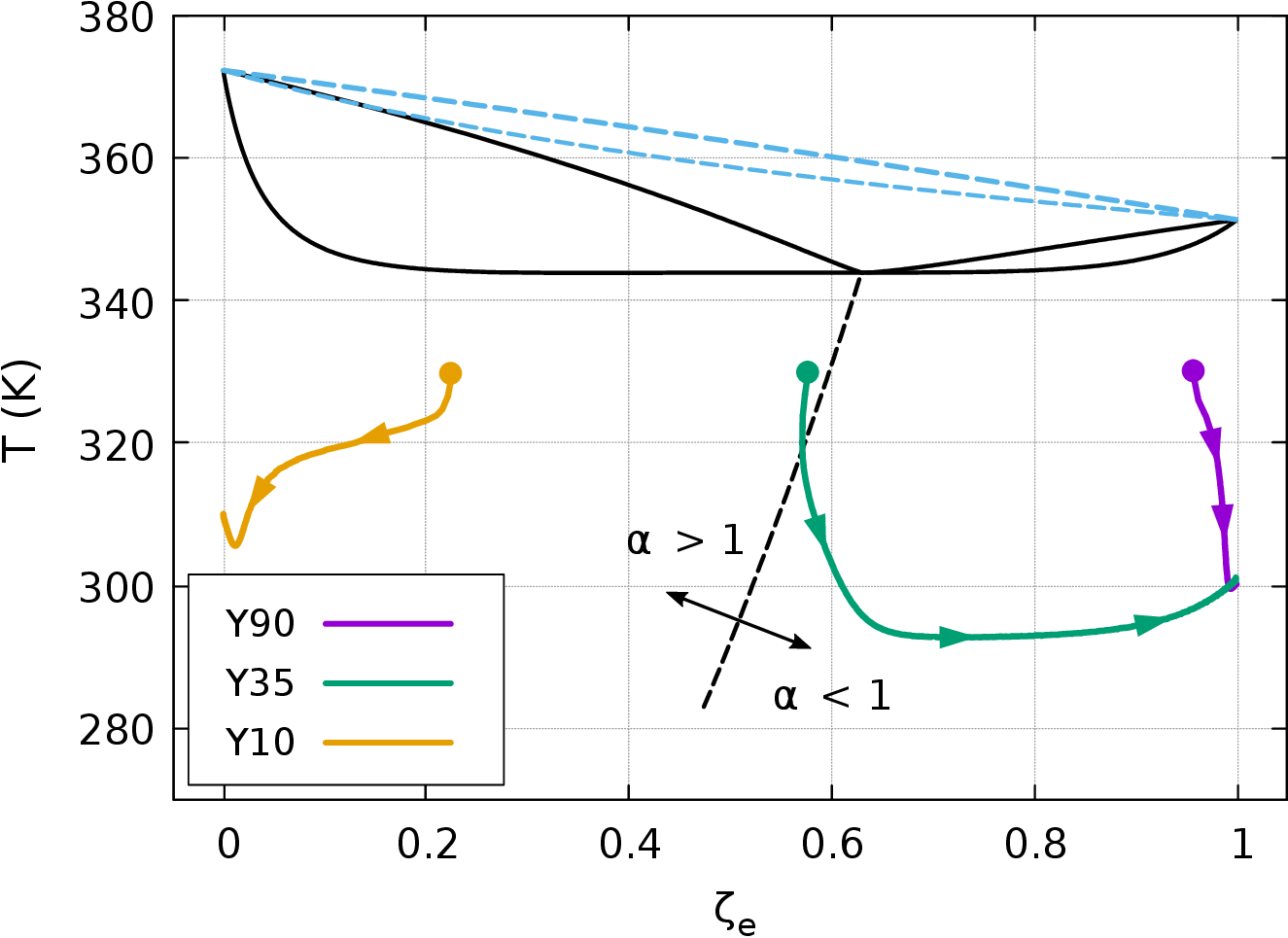}
\caption{Interfacial thermodynamic states of the evaporating ethanol--isooctane droplets in the vapor--liquid equilibrium diagram. The black solid curves denote the equilibrium boundaries predicted by the UNIFAC activity-coefficient model, the blue dashed curves denote the ideal-mixture prediction from Raoult's law, and the black dashed curve marks $\alpha_{e,iso}=1$. The markers show the trajectories of the interfacial states for the three initial droplet compositions.}
\label{f5.5.2}
\end{figure}

At first sight, this result may appear counterintuitive, since pure ethanol is more volatile than pure isooctane. The explanation lies in the non-ideal thermodynamic behavior of the ethanol--isooctane mixture. Because ethanol is strongly polar whereas isooctane is non-polar, the mixture exhibits pronounced non-ideality and possesses an azeotropic point. The corresponding phase diagram at $1~\mathrm{bar}$ is shown in Fig.~\ref{f5.5.2}. The black solid curves denote the vapor--liquid equilibrium boundaries obtained using the UNIFAC activity-coefficient model, while the blue dashed curves correspond to the ideal-mixture prediction based on Raoult's law. The non-ideal phase diagram exhibits an azeotropic point at $T^{az}=343.86~\mathrm{K}$ and $\zeta_{l,e}^{az}=0.631$, whereas no such feature exists for the ideal mixture.

When the ethanol mole fraction in the liquid exceeds the azeotropic value, \textit{i.e.} $\zeta_{l,e}>\zeta_{l,e}^{az}$ as in case Y90, the equilibrium vapor at the interface contains a smaller ethanol fraction than the liquid itself. Isooctane therefore evaporates preferentially, and the ethanol concentration inside the droplet increases. Conversely, when $\zeta_{l,e}<\zeta_{l,e}^{az}$ as in case Y10, the equilibrium vapor is enriched in ethanol, so that ethanol evaporates preferentially and its liquid-phase concentration decreases. This interpretation is further clarified by the relative volatility. Defining the equilibrium constant of component $i$ as $K_i=\zeta_{g,i} / \zeta_{l,i} =\gamma_i p_i^{sat} / P$, the separation factor between ethanol and isooctane is
\begin{equation}
\alpha_{e,iso}
=
\frac{K_e}{K_{iso}}
=
\frac{\gamma_e p_e^{sat}}
{\gamma_{iso} p_{iso}^{sat}}.
\label{e3.67}
\end{equation}
Here, $\alpha_{e,iso}>1$ indicates that ethanol is more volatile than isooctane, whereas $\alpha_{e,iso}<1$ indicates the opposite. In Fig.~\ref{f5.5.2}, the black dashed curve marks the locus $\alpha_{e,iso}=1$, separating the two evaporation regimes. The initial interfacial state of Y90 lies in the region $\alpha_{e,iso}<1$, which explains why isooctane evaporates faster. By contrast, Y10 lies in the region $\alpha_{e,iso}>1$, so that ethanol evaporates faster. The intermediate case Y35 is particularly interesting: because $\alpha_{e,iso}$ depends nonlinearly on both temperature and composition, the interfacial state evolves during evaporation and may cross from the region $\alpha_{e,iso}>1$ into the region $\alpha_{e,iso}<1$. As a result, the ethanol mass fraction first decreases and then increases, which is a distinctive signature of the fully coupled non-ideal multicomponent behavior.

\begin{figure}[!h]
\centering
\includegraphics[scale=0.34]{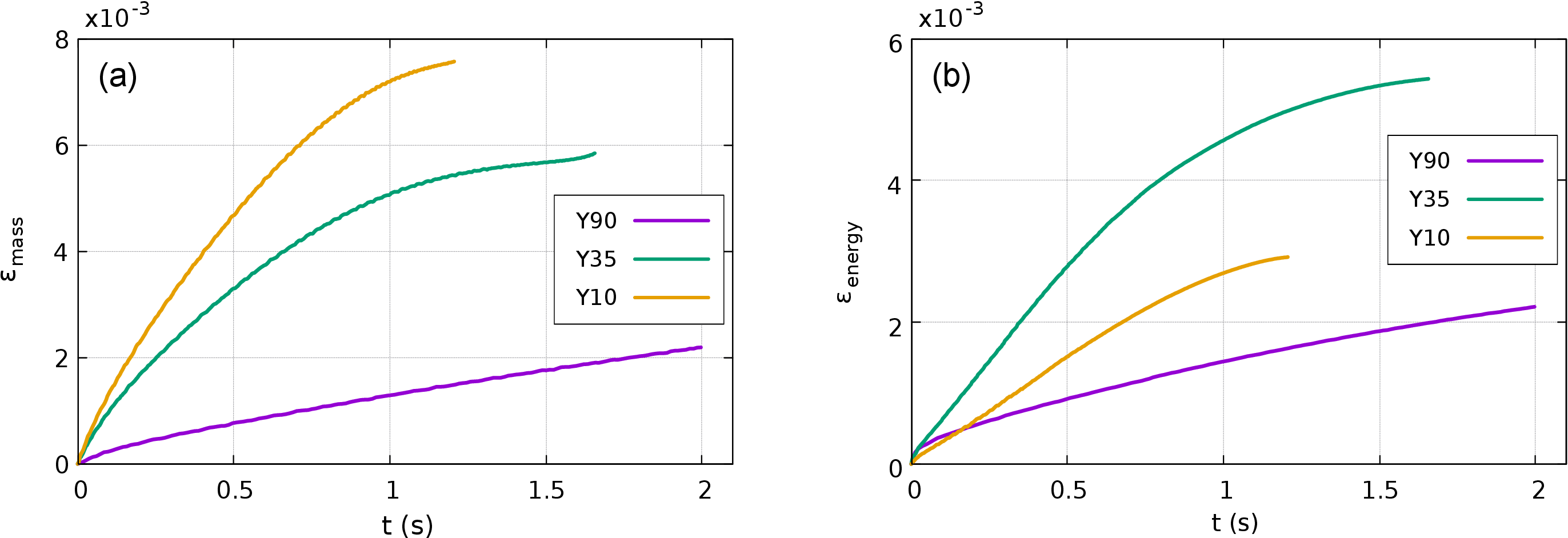}
\caption{Conservation errors for ethanol--isooctane droplet evaporation with different initial compositions: (a) relative mass error and (b) relative energy error.}
\label{f5.5.3}
\end{figure}

Finally, we assess the overall conservation of mass and energy during the evaporation process. Figure~\ref{f5.5.3} shows that both the mass and energy errors remain small for all three initial compositions, with errors below $0.8\%$ throughout the simulations. This confirms that the present numerical method preserves the conservative properties of the coupled multicomponent evaporation problem, even when non-ideal thermodynamic equilibrium strongly modifies the relative evaporation rates of the two components.

\subsection{Evaporation of a sessile water--glycerol droplet}\label{sec5.6}

Finally, we consider the evaporation of a sessile water--glycerol droplet, a configuration that is widely used in experiments and has become a standard benchmark for multicomponent droplet evaporation. This test is intended to validate the present method in a more realistic setting involving a pinned contact line, composition-dependent liquid properties, non-uniform interfacial evaporation, and Marangoni flow driven by composition gradients. Since the volatility of glycerol is negligible compared with that of water, only water is treated as a transferring species. The present case therefore provides a stringent test of the coupled solution of the water-concentration field, the temperature field, and the interfacial mass flux, while retaining the essential physics of a non-ideal multicomponent sessile droplet.

\begin{figure}[!h]
\centering
\includegraphics[scale=0.35]{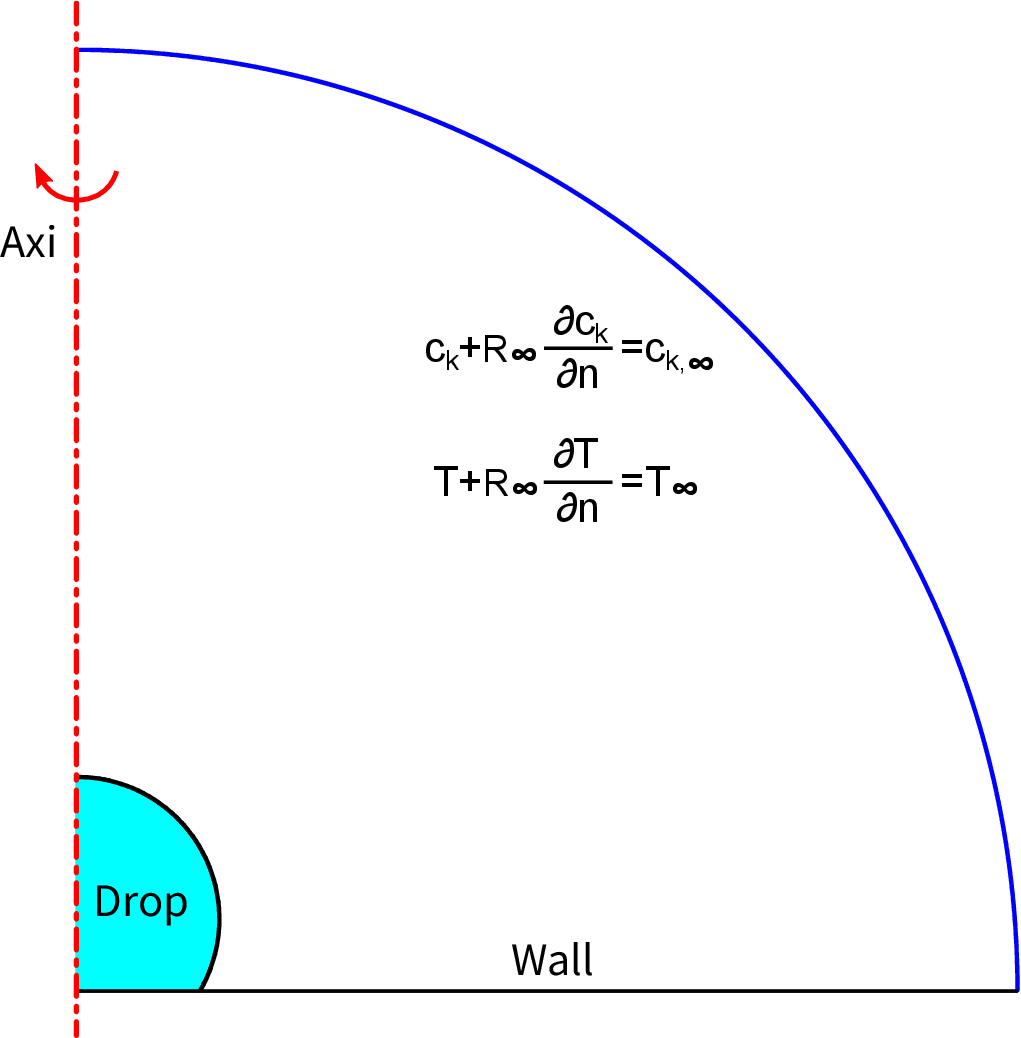}
\caption{Computational configuration for sessile water--glycerol droplet evaporation. The droplet is modeled in an axisymmetric domain with a pinned contact line, a constant-temperature solid substrate, and an outer circular boundary for imposing far-field thermal and water-vapor concentration conditions.}
\label{f5.6.1}
\end{figure}

Figure~\ref{f5.6.1} shows the computational configuration. In a two-dimensional axisymmetric coordinate system, a spherical-cap droplet with an initial volume of $50~\mathrm{nL}$, an initial water mass fraction of $90\%$, and an initial contact angle of $\theta=120^\circ$ is placed on a solid substrate. The computational domain is $2.4~\mathrm{mm}\times2.4~\mathrm{mm}$. To better approximate an unbounded ambient environment, an additional circular outer boundary defined by $x^2+y^2=R_\infty^2$ with $R_\infty=2.4~\mathrm{mm}$ is used to impose the far-field conditions. The wall and ambient temperatures are both fixed at $T_{wall}=T_\infty=25^\circ\mathrm{C}$, while the ambient water-vapor concentration is taken as $c_{v,\infty}=0$, corresponding to zero relative humidity. Following the Dirichlet-to-Neumann mapping approach \citep{givoli2013numerical}, Robin conditions are imposed on the outer boundary in the form
\[
T+R_\infty\frac{\partial T}{\partial n}=T_\infty,
\qquad
c_v+R_\infty\frac{\partial c_v}{\partial n}=c_{v,\infty}.
\]

The spatial resolution corresponds to approximately 206 grid cells across the initial droplet diameter, namely $d/\Delta=206$. A pinned-contact-line model is adopted throughout the simulation, so that the contact-line position remains fixed while the contact angle and the interface shape evolve. The gas-phase properties are prescribed as $\rho_g=1.2~\mathrm{kg/m^3}$, $\mu_g=1.8\times10^{-5}~\mathrm{Pa\cdot s}$, $c_{p,g}=1006~\mathrm{J/(kg\cdot K)}$, $\lambda_g=0.026~\mathrm{W/(m\cdot K)}$, and $D_g=2.6\times10^{-5}~\mathrm{m^2/s}$. By contrast, the liquid-phase properties are allowed to depend on the local water--glycerol composition and are updated during evaporation. The corresponding correlations are taken from the model used by \citet{diddens2017detailed} and from standard property data for aqueous glycerol solutions \citep{takamura2012physical,d2004diffusion}.

\begin{figure}[!h]
\centering
\includegraphics[scale=0.2]{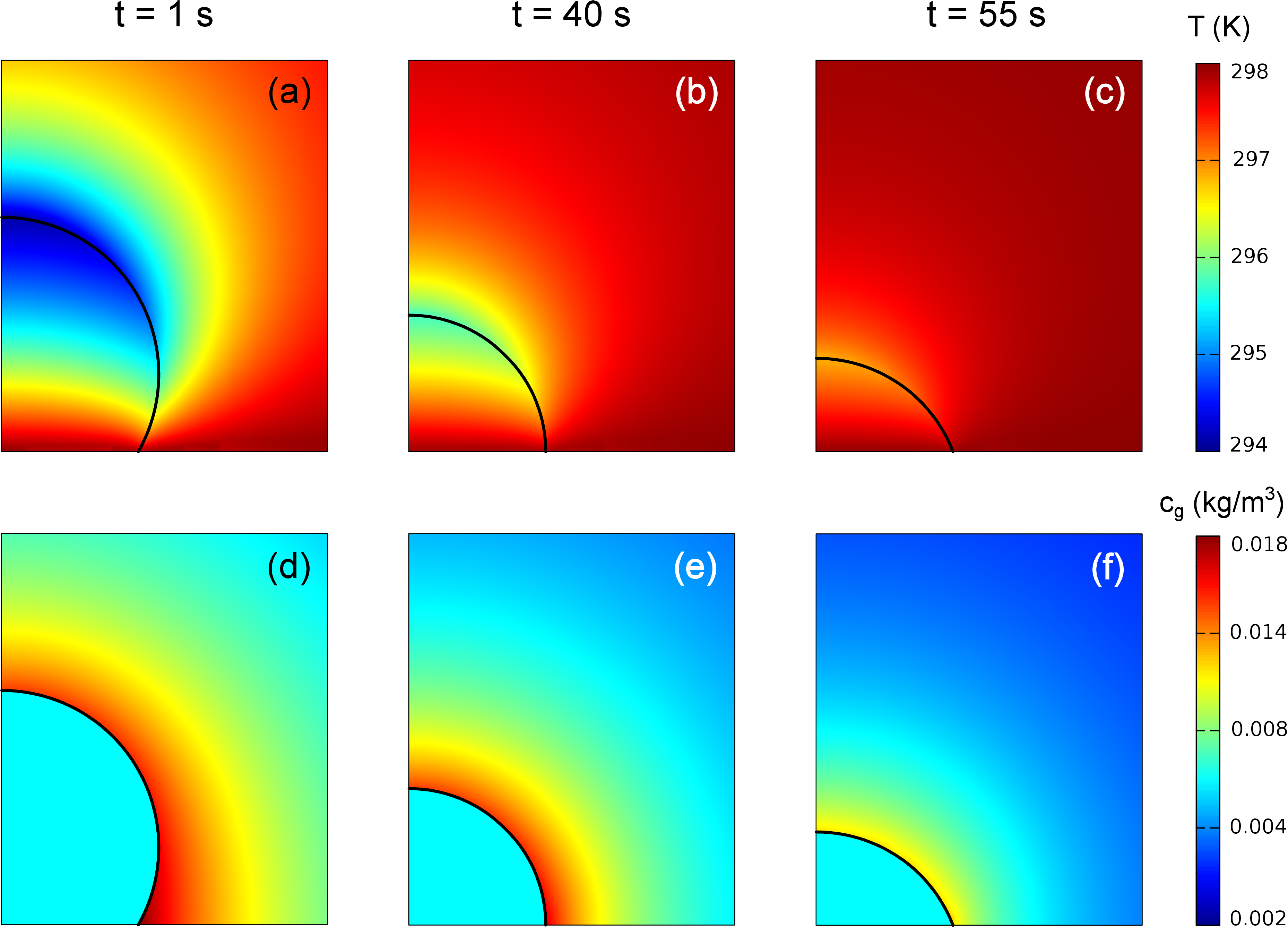}
\caption{Evaporation of a sessile water--glycerol droplet. Droplet shape, temperature field, and water-vapor concentration field are shown at $t=1$, $40$, and $55~\mathrm{s}$, illustrating the evolution of evaporative cooling and vapor transport in the surrounding gas.}
\label{f5.6.2}
\end{figure}

Figure~\ref{f5.6.2} presents the droplet shape together with the temperature field and water-vapor concentration field at three representative times, $t=1$, $40$, and $55~\mathrm{s}$. At the early stage, the contact angle is larger than $90^\circ$, so vapor diffusion near the substrate is geometrically hindered and the local evaporation rate near the contact line is lower than that near the apex of the droplet. As a consequence, evaporative cooling is stronger near the top of the droplet, leading to a lower local temperature and a pronounced vertical temperature gradient, as shown in Fig.~\ref{f5.6.2}(a,d). As evaporation proceeds, the water content inside the droplet decreases continuously. The water-vapor concentration in the surrounding gas and the local evaporation rate both decrease gradually, as shown in Fig.~\ref{f5.6.2}(e,f). Meanwhile, heat supplied from the substrate and the ambient gas progressively compensates for the evaporative cooling, so that the droplet temperature increases again during the later stage, as shown in Fig.~\ref{f5.6.2}(b,c).

\begin{figure}[!h]
\centering
\includegraphics[scale=0.25]{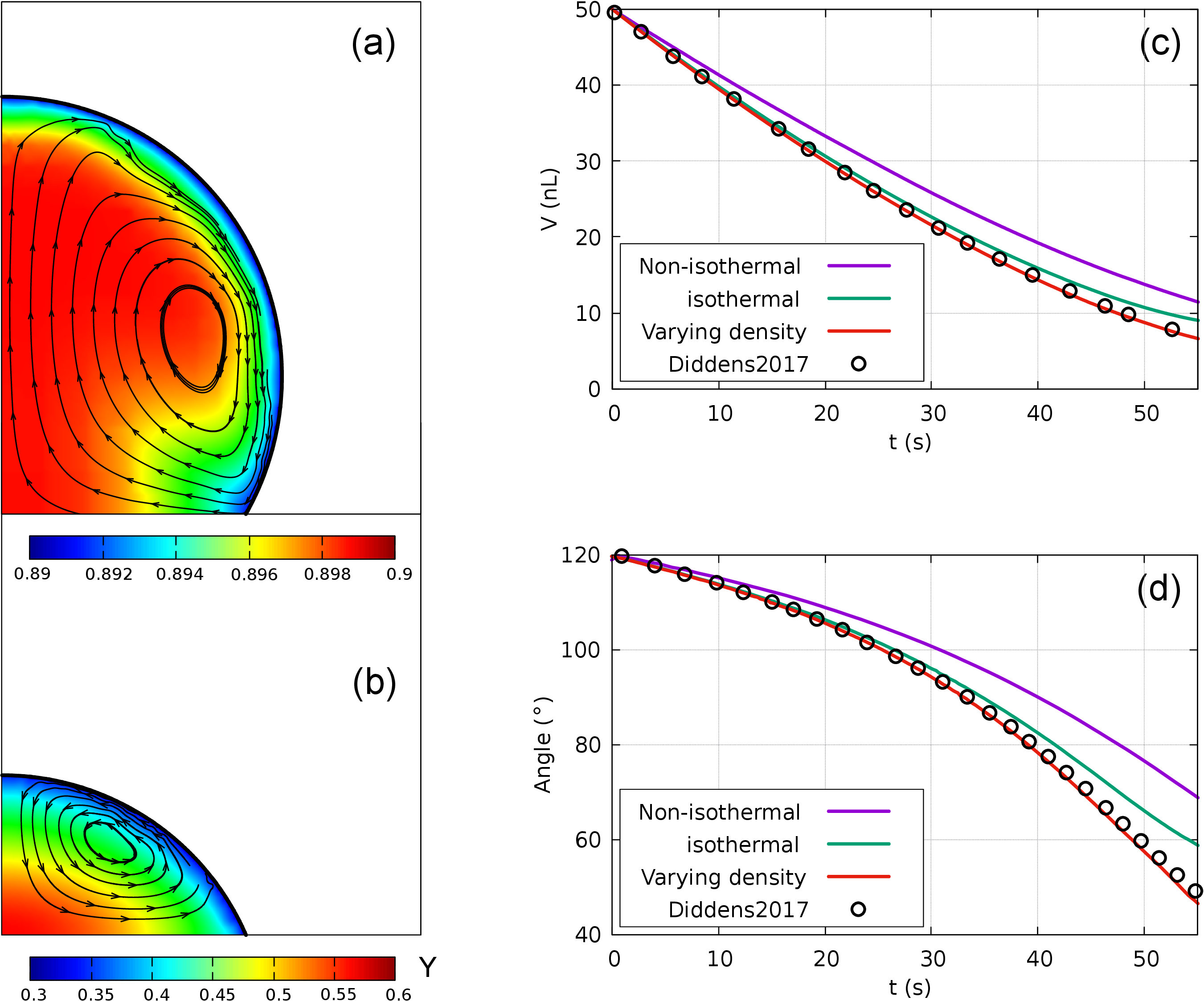}
\caption{Internal transport and global evolution of the sessile water--glycerol droplet. (a,b) Water mass-fraction field and streamlines inside the droplet at $t=1~\mathrm{s}$ and $t=50~\mathrm{s}$. (c,d) Temporal evolution of the droplet volume and contact angle, compared with reference results that account for composition-dependent density variation.}
\label{f5.6.3}
\end{figure}

To further clarify the internal transport mechanism, Fig.~\ref{f5.6.3}(a,b) shows the water mass-fraction field and the streamlines inside the droplet at $t=1~\mathrm{s}$ and $50~\mathrm{s}$. At the beginning of evaporation, the weaker evaporation near the contact line leads to a higher local water concentration there than near the apex, thereby creating a tangential concentration gradient along the interface. Since the surface tension of a water--glycerol mixture increases with the water concentration, the larger surface tension near the contact line drives a Marangoni flow from the apex toward the contact line, as shown in Fig.~\ref{f5.6.3}(a). As the droplet evaporates and the contact angle decreases below $90^\circ$, the evaporation rate becomes larger near the contact line than near the apex, and the Marangoni circulation reverses direction. At later times, the water content becomes small and the viscosity of the mixture rises rapidly, which suppresses both the internal circulation and the overall evaporation rate.

Figure~\ref{f5.6.3}(c,d) further shows the temporal evolution of the droplet volume and contact angle for three model settings: a non-isothermal case, an isothermal case with constant liquid density, and an isothermal case with composition-dependent liquid density. In the isothermal variable-density case, the density treatment follows the same assumptions as those used in the reference finite-element simulations of \citet{diddens2017detailed}, and therefore provides a direct comparison with the reference results. The present isothermal variable-density results agree well with the reference data, confirming that the present Cartesian-grid formulation accurately captures the coupled composition transport and shape evolution of the pinned sessile droplet. The comparison among the three cases further shows that both thermal effects and composition-induced density variations have a noticeable influence on the evaporation dynamics.

\section{Remarks and Conclusions}\label{sec6}

In this work, we have developed a sharp and conservative VOF--finite-volume method for multicomponent liquid--gas mass transfer with deformable interfaces. The method extends our previous single-component phase-change framework to problems in which several species are transferred across the interface and are coupled to the temperature field through latent heat and vapor--liquid equilibrium. The central difficulty is that the interfacial closure is no longer determined by a single mass flux, but by multiple species balances, thermodynamic-equilibrium relations, and a thermal flux-jump condition that must be satisfied simultaneously.

The main contribution of the present work is a fully sharp two-field treatment of both species and temperature transport. The species equations are solved separately in the two phases, and their Robin-type interfacial conditions are imposed directly on the reconstructed VOF interface through an embedded-boundary discretization. More importantly, the temperature equation is also solved in a two-field manner: the liquid-side and gas-side embedded-boundary heat fluxes are coupled through the interfacial flux-jump condition, rather than being regularized into a diffusive one-field source term. This distinguishes the present method from previous VOF-based multicomponent approaches, in which either the interfacial coupling is evaluated explicitly from previous-time-step information or the temperature equation remains treated in a smeared one-field form.

A sequential coupling strategy has also been proposed to close the multicomponent interfacial problem. The partial mass fluxes are first determined from the phase-wise concentration fields; these fluxes are then used to impose the dispersed-phase species conditions, close the heat-flux jump in the temperature equation, and update the thermodynamic-equilibrium state. Together with the geometrical conservative advection of volume, momentum, energy, and species, this formulation preserves the sharpness of interfacial discontinuities while maintaining good mass and energy conservation.

The method has been validated through a set of increasingly complex tests. For static and freely rising dissolving bubbles, it achieves second-order accuracy and accurately predicts the Sherwood number. For multicomponent carbon-dioxide bubble dissolution, it captures the long-time replacement of $\mathrm{CO_2}$ by $\mathrm{N_2}$ and $\mathrm{O_2}$ and agrees well with reduced-model predictions. For n-heptane droplet evaporation, the computed droplet radius, temperature field, and vapor-concentration field agree closely with reference solutions. For ethanol--isooctane droplets, the method reproduces the non-ideal azeotropic behavior and the composition-dependent reversal of relative volatility. Finally, for sessile water--glycerol droplets, it captures the coupled evolution of composition, temperature, evaporation flux, and Marangoni flow in good agreement with reference simulations. In all numerical tests, the conservative character of the method is confirmed by the small mass and energy errors.

Overall, the present results demonstrate that the proposed method provides an accurate, conservative, and robust sharp-interface framework for multicomponent liquid--gas mass transfer. Its main advantage lies in resolving interfacial fluxes and thin concentration or thermal boundary layers without smearing the scalar fields across the interface. The method therefore offers a useful basis for direct numerical simulations of multicomponent bubble dissolution, droplet evaporation, and related interfacial transport problems in environmental, chemical, and energy applications.

\section*{Acknowledgments}

The authors gratefully acknowledge the support of the National Key R\& D Program of China (2022YFE03130000, 2023YFA1011000), and that of the NSFC (W2511004, 12588201 and 12472256). \\

\biboptions{numbers,sort&compress}
\bibliographystyle{model1-num-names}
\bibliography{refs}

\begin{thebibliography}{46}
\expandafter\ifx\csname natexlab\endcsname\relax\def\natexlab#1{#1}\fi
\providecommand{\url}[1]{\texttt{#1}}
\providecommand{\href}[2]{#2}
\providecommand{\path}[1]{#1}
\providecommand{\DOIprefix}{doi:}
\providecommand{\ArXivprefix}{arXiv:}
\providecommand{\URLprefix}{URL: }
\providecommand{\Pubmedprefix}{pmid:}
\providecommand{\doi}[1]{\href{http://dx.doi.org/#1}{\path{#1}}}
\providecommand{\Pubmed}[1]{\href{pmid:#1}{\path{#1}}}
\providecommand{\bibinfo}[2]{#2}
\ifx\xfnm\relax \def\xfnm[#1]{\unskip,\space#1}\fi
\bibitem[{Zhao et~al.(2022)Zhao, Zhang, and Ni}]{zhao2022boiling}
\bibinfo{author}{S.~Zhao}, \bibinfo{author}{J.~Zhang}, \bibinfo{author}{M.-J.
  Ni},
\newblock \bibinfo{title}{Boiling and evaporation model for liquid-gas flows: A
  sharp and conservative method based on the geometrical vof approach},
\newblock \bibinfo{journal}{Journal of Computational Physics}
  \bibinfo{volume}{452} (\bibinfo{year}{2022}) \bibinfo{pages}{110908}.
\bibitem[{Irfan and Muradoglu(2017)}]{irfan2017front}
\bibinfo{author}{M.~Irfan}, \bibinfo{author}{M.~Muradoglu},
\newblock \bibinfo{title}{A front tracking method for direct numerical
  simulation of evaporation process in a multiphase system},
\newblock \bibinfo{journal}{Journal of Computational Physics}
  \bibinfo{volume}{337} (\bibinfo{year}{2017}) \bibinfo{pages}{132--153}.
\bibitem[{Wang and Yang(2019)}]{wang2019vaporization}
\bibinfo{author}{Y.~Wang}, \bibinfo{author}{V.~Yang},
\newblock \bibinfo{title}{Vaporization of liquid droplet with large deformation
  and high mass transfer rate, i: Constant-density, constant-property case},
\newblock \bibinfo{journal}{Journal of Computational Physics}
  \bibinfo{volume}{392} (\bibinfo{year}{2019}) \bibinfo{pages}{56--70}.
\bibitem[{Palmore~Jr and Desjardins(2019)}]{palmore2019volume}
\bibinfo{author}{J.~Palmore~Jr}, \bibinfo{author}{O.~Desjardins},
\newblock \bibinfo{title}{A volume of fluid framework for interface-resolved
  simulations of vaporizing liquid-gas flows},
\newblock \bibinfo{journal}{Journal of Computational Physics}
  \bibinfo{volume}{399} (\bibinfo{year}{2019}) \bibinfo{pages}{108954}.
\bibitem[{Kunkelmann and Stephan(2009)}]{kunkelmann2009cfd}
\bibinfo{author}{C.~Kunkelmann}, \bibinfo{author}{P.~Stephan},
\newblock \bibinfo{title}{Cfd simulation of boiling flows using the
  volume-of-fluid method within openfoam},
\newblock \bibinfo{journal}{Numerical Heat Transfer, Part A: Applications}
  \bibinfo{volume}{56} (\bibinfo{year}{2009}) \bibinfo{pages}{631--646}.
\bibitem[{Fedkiw et~al.(1999)Fedkiw, Aslam, Merriman, Osher
  et~al.}]{fedkiw1999non}
\bibinfo{author}{R.~P. Fedkiw}, \bibinfo{author}{T.~Aslam},
  \bibinfo{author}{B.~Merriman}, \bibinfo{author}{S.~Osher}, et~al.,
\newblock \bibinfo{title}{A non-oscillatory eulerian approach to interfaces in
  multimaterial flows (the ghost fluid method)},
\newblock \bibinfo{journal}{Journal of computational physics}
  \bibinfo{volume}{152} (\bibinfo{year}{1999}) \bibinfo{pages}{457--492}.
\bibitem[{Tanguy et~al.(2014)Tanguy, Sagan, Lalanne, Couderc, and
  Colin}]{tanguy2014benchmarks}
\bibinfo{author}{S.~Tanguy}, \bibinfo{author}{M.~Sagan},
  \bibinfo{author}{B.~Lalanne}, \bibinfo{author}{F.~Couderc},
  \bibinfo{author}{C.~Colin},
\newblock \bibinfo{title}{Benchmarks and numerical methods for the simulation
  of boiling flows},
\newblock \bibinfo{journal}{Journal of Computational Physics}
  \bibinfo{volume}{264} (\bibinfo{year}{2014}) \bibinfo{pages}{1--22}.
\bibitem[{Gibou et~al.(2007)Gibou, Chen, Nguyen, and Banerjee}]{gibou2007level}
\bibinfo{author}{F.~Gibou}, \bibinfo{author}{L.~Chen},
  \bibinfo{author}{D.~Nguyen}, \bibinfo{author}{S.~Banerjee},
\newblock \bibinfo{title}{A level set based sharp interface method for the
  multiphase incompressible navier--stokes equations with phase change},
\newblock \bibinfo{journal}{Journal of Computational Physics}
  \bibinfo{volume}{222} (\bibinfo{year}{2007}) \bibinfo{pages}{536--555}.
\bibitem[{Villegas et~al.(2016)Villegas, Alis, Lepilliez, and
  Tanguy}]{villegas2016ghost}
\bibinfo{author}{L.~R. Villegas}, \bibinfo{author}{R.~Alis},
  \bibinfo{author}{M.~Lepilliez}, \bibinfo{author}{S.~Tanguy},
\newblock \bibinfo{title}{A ghost fluid/level set method for boiling flows and
  liquid evaporation: application to the leidenfrost effect},
\newblock \bibinfo{journal}{Journal of Computational Physics}
  \bibinfo{volume}{316} (\bibinfo{year}{2016}) \bibinfo{pages}{789--813}.
\bibitem[{Haroun et~al.(2010)Haroun, Legendre, and Raynal}]{haroun2010volume}
\bibinfo{author}{Y.~Haroun}, \bibinfo{author}{D.~Legendre},
  \bibinfo{author}{L.~Raynal},
\newblock \bibinfo{title}{Volume of fluid method for interfacial reactive mass
  transfer: Application to stable liquid film},
\newblock \bibinfo{journal}{Chemical Engineering Science} \bibinfo{volume}{65}
  (\bibinfo{year}{2010}) \bibinfo{pages}{2896--2909}.
\bibitem[{Graveleau et~al.(2017)Graveleau, Soulaine, and
  Tchelepi}]{graveleau2017pore}
\bibinfo{author}{M.~Graveleau}, \bibinfo{author}{C.~Soulaine},
  \bibinfo{author}{H.~A. Tchelepi},
\newblock \bibinfo{title}{Pore-scale simulation of interphase multicomponent
  mass transfer for subsurface flow},
\newblock \bibinfo{journal}{Transport in porous media} \bibinfo{volume}{120}
  (\bibinfo{year}{2017}) \bibinfo{pages}{287--308}.
\bibitem[{Maes and Soulaine(2018)}]{maes2018new}
\bibinfo{author}{J.~Maes}, \bibinfo{author}{C.~Soulaine},
\newblock \bibinfo{title}{A new compressive scheme to simulate species transfer
  across fluid interfaces using the volume-of-fluid method},
\newblock \bibinfo{journal}{Chemical Engineering Science} \bibinfo{volume}{190}
  (\bibinfo{year}{2018}) \bibinfo{pages}{405--418}.
\bibitem[{Maes and Soulaine(2020)}]{maes2020unified}
\bibinfo{author}{J.~Maes}, \bibinfo{author}{C.~Soulaine},
\newblock \bibinfo{title}{A unified single-field volume-of-fluid-based
  formulation for multi-component interfacial transfer with local volume
  changes},
\newblock \bibinfo{journal}{Journal of Computational Physics}
  \bibinfo{volume}{402} (\bibinfo{year}{2020}) \bibinfo{pages}{109024}.
\bibitem[{Zanutto et~al.(2022{\natexlab{a}})Zanutto, Paladino, Evrard, van
  Wachem, and Denner}]{zanutto2022modeling1}
\bibinfo{author}{C.~P. Zanutto}, \bibinfo{author}{E.~E. Paladino},
  \bibinfo{author}{F.~Evrard}, \bibinfo{author}{B.~van Wachem},
  \bibinfo{author}{F.~Denner},
\newblock \bibinfo{title}{Modeling of interfacial mass transfer based on a
  single-field formulation and an algebraic vof method considering
  non-isothermal systems and large volume changes},
\newblock \bibinfo{journal}{Chemical Engineering Science} \bibinfo{volume}{247}
  (\bibinfo{year}{2022}{\natexlab{a}}) \bibinfo{pages}{116855}.
\bibitem[{Zanutto et~al.(2022{\natexlab{b}})Zanutto, Evrard, van Wachem,
  Denner, and Paladino}]{zanutto2022modeling}
\bibinfo{author}{C.~P. Zanutto}, \bibinfo{author}{F.~Evrard},
  \bibinfo{author}{B.~van Wachem}, \bibinfo{author}{F.~Denner},
  \bibinfo{author}{E.~E. Paladino},
\newblock \bibinfo{title}{Modeling interfacial mass transfer of highly
  non-ideal mixtures using an algebraic vof method},
\newblock \bibinfo{journal}{Chemical Engineering Science} \bibinfo{volume}{251}
  (\bibinfo{year}{2022}{\natexlab{b}}) \bibinfo{pages}{117458}.
\bibitem[{Bothe and Fleckenstein(2013)}]{bothe2013volume}
\bibinfo{author}{D.~Bothe}, \bibinfo{author}{S.~Fleckenstein},
\newblock \bibinfo{title}{A volume-of-fluid-based method for mass transfer
  processes at fluid particles},
\newblock \bibinfo{journal}{Chemical Engineering Science} \bibinfo{volume}{101}
  (\bibinfo{year}{2013}) \bibinfo{pages}{283--302}.
\bibitem[{Fleckenstein and Bothe(2015)}]{fleckenstein2015volume}
\bibinfo{author}{S.~Fleckenstein}, \bibinfo{author}{D.~Bothe},
\newblock \bibinfo{title}{A volume-of-fluid-based numerical method for
  multi-component mass transfer with local volume changes},
\newblock \bibinfo{journal}{Journal of Computational Physics}
  \bibinfo{volume}{301} (\bibinfo{year}{2015}) \bibinfo{pages}{35--58}.
\bibitem[{Farsoiya et~al.(2023)Farsoiya, Magdelaine, Antkowiak, Popinet, and
  Deike}]{farsoiya2023direct}
\bibinfo{author}{P.~K. Farsoiya}, \bibinfo{author}{Q.~Magdelaine},
  \bibinfo{author}{A.~Antkowiak}, \bibinfo{author}{S.~Popinet},
  \bibinfo{author}{L.~Deike},
\newblock \bibinfo{title}{Direct numerical simulations of bubble-mediated gas
  transfer and dissolution in quiescent and turbulent flows},
\newblock \bibinfo{journal}{Journal of Fluid Mechanics} \bibinfo{volume}{954}
  (\bibinfo{year}{2023}) \bibinfo{pages}{A29}.
\bibitem[{Cipriano et~al.(2024)Cipriano, Frassoldati, Faravelli, Popinet, Cuoci
  et~al.}]{cipriano2024multicomponent}
\bibinfo{author}{E.~Cipriano}, \bibinfo{author}{A.~Frassoldati},
  \bibinfo{author}{T.~Faravelli}, \bibinfo{author}{S.~Popinet},
  \bibinfo{author}{A.~Cuoci}, et~al.,
\newblock \bibinfo{title}{Multicomponent droplet evaporation in a geometric
  volume-of-fluid framework},
\newblock \bibinfo{journal}{Journal of Computational Physics}
  (\bibinfo{year}{2024}) \bibinfo{pages}{112955}.
\bibitem[{Salimi et~al.(2024)Salimi, Scapin, Popescu, Costa, and
  Brandt}]{salimi2024volume}
\bibinfo{author}{S.~Z. Salimi}, \bibinfo{author}{N.~Scapin},
  \bibinfo{author}{E.-R. Popescu}, \bibinfo{author}{P.~Costa},
  \bibinfo{author}{L.~Brandt},
\newblock \bibinfo{title}{A volume-of-fluid method for multicomponent droplet
  evaporation with robin boundary conditions},
\newblock \bibinfo{journal}{Journal of Computational Physics}
  \bibinfo{volume}{514} (\bibinfo{year}{2024}) \bibinfo{pages}{113211}.
\bibitem[{Krishna and Wesselingh(1997)}]{krishna1997maxwell}
\bibinfo{author}{R.~Krishna}, \bibinfo{author}{J.~A. Wesselingh},
\newblock \bibinfo{title}{The maxwell-stefan approach to mass transfer},
\newblock \bibinfo{journal}{Chemical engineering science} \bibinfo{volume}{52}
  (\bibinfo{year}{1997}) \bibinfo{pages}{861--911}.
\bibitem[{Poling et~al.(2001)Poling, Prausnitz, and
  O’connell}]{poling2001properties}
\bibinfo{author}{B.~E. Poling}, \bibinfo{author}{J.~M. Prausnitz},
  \bibinfo{author}{J.~P. O’connell}, \bibinfo{title}{Properties of gases and
  liquids}, \bibinfo{publisher}{McGraw-Hill Education}, \bibinfo{year}{2001}.
\bibitem[{Smith et~al.(2018)Smith, Van~Ness, Abbott, and
  Swihart}]{smith2018introduction}
\bibinfo{author}{J.~M. Smith}, \bibinfo{author}{H.~C. Van~Ness},
  \bibinfo{author}{M.~M. Abbott}, \bibinfo{author}{M.~T. Swihart},
  \bibinfo{title}{Introduction to Chemical Engineering Thermodynamics},
  \bibinfo{edition}{8} ed., \bibinfo{publisher}{McGraw-Hill Education},
  \bibinfo{year}{2018}.
\bibitem[{Deising et~al.(2016)Deising, Marschall, and
  Bothe}]{deising2016unified}
\bibinfo{author}{D.~Deising}, \bibinfo{author}{H.~Marschall},
  \bibinfo{author}{D.~Bothe},
\newblock \bibinfo{title}{A unified single-field model framework for
  volume-of-fluid simulations of interfacial species transfer applied to bubbly
  flows},
\newblock \bibinfo{journal}{Chemical Engineering Science} \bibinfo{volume}{139}
  (\bibinfo{year}{2016}) \bibinfo{pages}{173--195}.
\bibitem[{Farsoiya et~al.(2021)Farsoiya, Popinet, and
  Deike}]{farsoiya2021bubble}
\bibinfo{author}{P.~K. Farsoiya}, \bibinfo{author}{S.~Popinet},
  \bibinfo{author}{L.~Deike},
\newblock \bibinfo{title}{Bubble-mediated transfer of dilute gas in
  turbulence},
\newblock \bibinfo{journal}{Journal of Fluid Mechanics} \bibinfo{volume}{920}
  (\bibinfo{year}{2021}) \bibinfo{pages}{A34}.
\bibitem[{Popinet(2009)}]{popinet2009accurate}
\bibinfo{author}{S.~Popinet},
\newblock \bibinfo{title}{An accurate adaptive solver for
  surface-tension-driven interfacial flows},
\newblock \bibinfo{journal}{Journal of Computational Physics}
  \bibinfo{volume}{228} (\bibinfo{year}{2009}) \bibinfo{pages}{5838--5866}.
\bibitem[{Popinet(2014)}]{popinet2014basilisk}
\bibinfo{author}{S.~Popinet},
\newblock \bibinfo{title}{Basilisk},
\newblock \bibinfo{journal}{URl:http://basilisk.fr.(accessed:10.21.2019)}
  (\bibinfo{year}{2014}).
\bibitem[{Popinet(2015)}]{popinet2015quadtree}
\bibinfo{author}{S.~Popinet},
\newblock \bibinfo{title}{A quadtree-adaptive multigrid solver for the
  serre--green--naghdi equations},
\newblock \bibinfo{journal}{Journal of Computational Physics}
  \bibinfo{volume}{302} (\bibinfo{year}{2015}) \bibinfo{pages}{336--358}.
\bibitem[{Seric et~al.(2018)Seric, Afkhami, and Kondic}]{seric2018direct}
\bibinfo{author}{I.~Seric}, \bibinfo{author}{S.~Afkhami},
  \bibinfo{author}{L.~Kondic},
\newblock \bibinfo{title}{Direct numerical simulation of variable surface
  tension flows using a volume-of-fluid method},
\newblock \bibinfo{journal}{Journal of Computational Physics}
  \bibinfo{volume}{352} (\bibinfo{year}{2018}) \bibinfo{pages}{615--636}.
\bibitem[{Tripathi and Sahu(2018)}]{tripathi2018motion}
\bibinfo{author}{M.~K. Tripathi}, \bibinfo{author}{K.~C. Sahu},
\newblock \bibinfo{title}{Motion of an air bubble under the action of
  thermocapillary and buoyancy forces},
\newblock \bibinfo{journal}{Computers \& Fluids} \bibinfo{volume}{177}
  (\bibinfo{year}{2018}) \bibinfo{pages}{58--68}.
\bibitem[{Francois et~al.(2006)Francois, Cummins, Dendy, Kothe, Sicilian, and
  Williams}]{francois2006balanced}
\bibinfo{author}{M.~M. Francois}, \bibinfo{author}{S.~J. Cummins},
  \bibinfo{author}{E.~D. Dendy}, \bibinfo{author}{D.~B. Kothe},
  \bibinfo{author}{J.~M. Sicilian}, \bibinfo{author}{M.~W. Williams},
\newblock \bibinfo{title}{A balanced-force algorithm for continuous and sharp
  interfacial surface tension models within a volume tracking framework},
\newblock \bibinfo{journal}{Journal of Computational Physics}
  \bibinfo{volume}{213} (\bibinfo{year}{2006}) \bibinfo{pages}{141--173}.
\bibitem[{Scapin et~al.(2020)Scapin, Costa, and Brandt}]{scapin2020volume}
\bibinfo{author}{N.~Scapin}, \bibinfo{author}{P.~Costa},
  \bibinfo{author}{L.~Brandt},
\newblock \bibinfo{title}{A volume-of-fluid method for interface-resolved
  simulations of phase-changing two-fluid flows},
\newblock \bibinfo{journal}{Journal of Computational Physics}
  \bibinfo{volume}{407} (\bibinfo{year}{2020}) \bibinfo{pages}{109251}.
\bibitem[{Malan et~al.(2021)Malan, Malan, Zaleski, and
  Rousseau}]{malan2021geometric}
\bibinfo{author}{L.~Malan}, \bibinfo{author}{A.~Malan},
  \bibinfo{author}{S.~Zaleski}, \bibinfo{author}{P.~Rousseau},
\newblock \bibinfo{title}{A geometric vof method for interface resolved phase
  change and conservative thermal energy advection},
\newblock \bibinfo{journal}{Journal of Computational Physics}
  \bibinfo{volume}{426} (\bibinfo{year}{2021}) \bibinfo{pages}{109920}.
\bibitem[{Weymouth and Yue(2010)}]{weymouth2010conservative}
\bibinfo{author}{G.~D. Weymouth}, \bibinfo{author}{D.~K.-P. Yue},
\newblock \bibinfo{title}{Conservative volume-of-fluid method for free-surface
  simulations on cartesian-grids},
\newblock \bibinfo{journal}{Journal of Computational Physics}
  \bibinfo{volume}{229} (\bibinfo{year}{2010}) \bibinfo{pages}{2853--2865}.
\bibitem[{Marschall et~al.(2012)Marschall, Hinterberger, Sch{\"u}ler, Habla,
  and Hinrichsen}]{marschall2012numerical}
\bibinfo{author}{H.~Marschall}, \bibinfo{author}{K.~Hinterberger},
  \bibinfo{author}{C.~Sch{\"u}ler}, \bibinfo{author}{F.~Habla},
  \bibinfo{author}{O.~Hinrichsen},
\newblock \bibinfo{title}{Numerical simulation of species transfer across fluid
  interfaces in free-surface flows using openfoam},
\newblock \bibinfo{journal}{Chemical engineering science} \bibinfo{volume}{78}
  (\bibinfo{year}{2012}) \bibinfo{pages}{111--127}.
\bibitem[{Sato and Ni{\v{c}}eno(2013)}]{sato2013sharp}
\bibinfo{author}{Y.~Sato}, \bibinfo{author}{B.~Ni{\v{c}}eno},
\newblock \bibinfo{title}{A sharp-interface phase change model for a
  mass-conservative interface tracking method},
\newblock \bibinfo{journal}{Journal of Computational Physics}
  \bibinfo{volume}{249} (\bibinfo{year}{2013}) \bibinfo{pages}{127--161}.
\bibitem[{Bell et~al.(1989)Bell, Colella, and Glaz}]{bell1989second}
\bibinfo{author}{J.~B. Bell}, \bibinfo{author}{P.~Colella},
  \bibinfo{author}{H.~M. Glaz},
\newblock \bibinfo{title}{A second-order projection method for the
  incompressible navier-stokes equations},
\newblock \bibinfo{journal}{Journal of Computational Physics}
  \bibinfo{volume}{85} (\bibinfo{year}{1989}) \bibinfo{pages}{257--283}.
\bibitem[{Boussinesq(1905)}]{boussinesq1905calcul}
\bibinfo{author}{J.~Boussinesq},
\newblock \bibinfo{title}{Calcul du pouvoir refroidissant des courants
  fluides},
\newblock \bibinfo{journal}{Journal de mathematiques pures et appliquees}
  \bibinfo{volume}{1} (\bibinfo{year}{1905}) \bibinfo{pages}{285--332}.
\bibitem[{Levi{\v{c}}(1962)}]{levivc1962physicochemical}
\bibinfo{author}{V.~G. Levi{\v{c}}}, \bibinfo{title}{Physicochemical
  hydrodynamics}, \bibinfo{publisher}{Prentice-Hall}, \bibinfo{year}{1962}.
\bibitem[{Hosoda et~al.(2014)Hosoda, Abe, Hosokawa, and
  Tomiyama}]{Hosoda2014Mass}
\bibinfo{author}{S.~Hosoda}, \bibinfo{author}{S.~Abe},
  \bibinfo{author}{S.~Hosokawa}, \bibinfo{author}{A.~Tomiyama},
\newblock \bibinfo{title}{Mass transfer from a bubble in a vertical pipe},
\newblock \bibinfo{journal}{INTERNATIONAL JOURNAL OF HEAT AND MASS TRANSFER}
  \bibinfo{volume}{69} (\bibinfo{year}{2014}) \bibinfo{pages}{215--222}.
\bibitem[{Hosoda et~al.(2015)Hosoda, Tryggvason, Hosokawa, and
  Tomiyama}]{hosoda2015dissolution}
\bibinfo{author}{S.~Hosoda}, \bibinfo{author}{G.~Tryggvason},
  \bibinfo{author}{S.~Hosokawa}, \bibinfo{author}{A.~Tomiyama},
\newblock \bibinfo{title}{Dissolution of single carbon dioxide bubbles in a
  vertical pipe},
\newblock \bibinfo{journal}{Journal of Chemical Engineering of Japan}
  \bibinfo{volume}{48} (\bibinfo{year}{2015}) \bibinfo{pages}{418--426}.
\bibitem[{Pathak and Raessi(2018)}]{pathak2018steady}
\bibinfo{author}{A.~Pathak}, \bibinfo{author}{M.~Raessi},
\newblock \bibinfo{title}{Steady-state and transient solutions to drop
  evaporation in a finite domain: Alternative benchmarks to the d2 law},
\newblock \bibinfo{journal}{International Journal of Heat and Mass Transfer}
  \bibinfo{volume}{127} (\bibinfo{year}{2018}) \bibinfo{pages}{1147--1158}.
\bibitem[{Givoli(2013)}]{givoli2013numerical}
\bibinfo{author}{D.~Givoli}, \bibinfo{title}{Numerical methods for problems in
  infinite domains}, \bibinfo{publisher}{Elsevier}, \bibinfo{year}{2013}.
\bibitem[{Diddens(2017)}]{diddens2017detailed}
\bibinfo{author}{C.~Diddens},
\newblock \bibinfo{title}{Detailed finite element method modeling of
  evaporating multi-component droplets},
\newblock \bibinfo{journal}{Journal of Computational Physics}
  \bibinfo{volume}{340} (\bibinfo{year}{2017}) \bibinfo{pages}{670--687}.
\bibitem[{Takamura et~al.(2012)Takamura, Fischer, and
  Morrow}]{takamura2012physical}
\bibinfo{author}{K.~Takamura}, \bibinfo{author}{H.~Fischer},
  \bibinfo{author}{N.~R. Morrow},
\newblock \bibinfo{title}{Physical properties of aqueous glycerol solutions},
\newblock \bibinfo{journal}{Journal of Petroleum Science and Engineering}
  \bibinfo{volume}{98} (\bibinfo{year}{2012}) \bibinfo{pages}{50--60}.
\bibitem[{D'Errico et~al.(2004)D'Errico, Ortona, Capuano, and
  Vitagliano}]{d2004diffusion}
\bibinfo{author}{G.~D'Errico}, \bibinfo{author}{O.~Ortona},
  \bibinfo{author}{F.~Capuano}, \bibinfo{author}{V.~Vitagliano},
\newblock \bibinfo{title}{Diffusion coefficients for the binary system
  glycerol+ water at 25 c. a velocity correlation study},
\newblock \bibinfo{journal}{Journal of Chemical \& Engineering Data}
  \bibinfo{volume}{49} (\bibinfo{year}{2004}) \bibinfo{pages}{1665--1670}.

\end{thebibliography}

\appendix
\section{Tangential surface-tension gradient}\label{app1}

A direct way to compute the tangential surface-tension gradient is to use
\begin{equation}
\nabla_t\sigma
=
\nabla\sigma-\boldsymbol{n}(\boldsymbol{n}\cdot\nabla\sigma),
\label{Ae1}
\end{equation}
where $\boldsymbol{n}$ is the unit normal vector of the interface. In interfacial flows, however, this formulation may be inaccurate. When $\sigma$ depends on temperature, the large contrast of thermal properties across the liquid--gas interface may produce a discontinuous temperature gradient, which contaminates the evaluation of $\nabla\sigma$. When $\sigma$ depends on composition, the concentration field is also generally discontinuous across the interface, leading to a similar difficulty. To avoid differentiating discontinuous bulk fields, we compute the tangential gradient of $\sigma$ directly along the reconstructed interface, following the height-function-based approaches of \citet{seric2018direct,tripathi2018motion}.

In two dimensions, the interface has a single tangential direction. The Marangoni force can therefore be written as
\begin{equation}
\boldsymbol{F}_{st}
=
\frac{\partial\sigma}{\partial s}\boldsymbol{t}\delta_s,
\label{Ae2}
\end{equation}
where $\boldsymbol{t}$ is the unit tangent vector, $s$ is the arclength along the interface, and $\delta_s$ is the surface delta function. Within the VOF framework, the interface normal and curvature are evaluated by the height-function method. The same geometrical information is used here to approximate $\partial\sigma/\partial s$.

\begin{figure}[!h]
\centering
\includegraphics[scale=0.6]{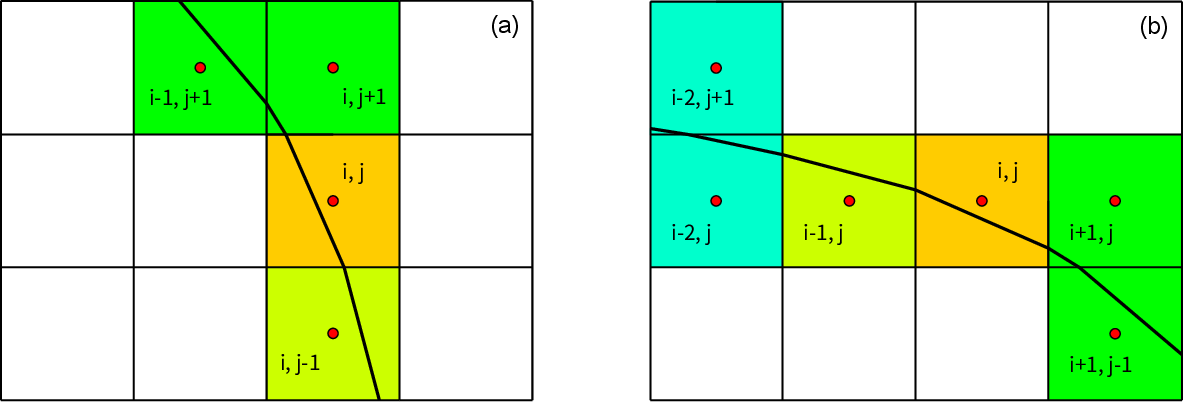}
\caption{Schematic of the height-function-based evaluation of the cell-averaged surface-tension coefficient $\sigma_c$ in interfacial cells, using (a) $x$-oriented and (b) $y$-oriented height functions.}
\label{fa1}
\end{figure}

As illustrated in Fig.~\ref{fa1}, we first define a volume-fraction-weighted average surface-tension coefficient along a row or a column of interfacial cells,
\begin{equation}
\sigma_{c,x}(j)
=
\frac{\sum_i \sigma_{i,j} f_{i,j}}
{\sum_i f_{i,j}},
\qquad
\sigma_{c,y}(i)
=
\frac{\sum_j \sigma_{i,j} f_{i,j}}
{\sum_j f_{i,j}},
\label{Ae3}
\end{equation}
where $\sigma_{i,j}$ and $f_{i,j}$ are the local surface-tension coefficient and volume fraction, respectively. Thus, all interfacial cells in the same row or column share the same averaged value. If only one interfacial cell is present in a given row or column, the averaged value reduces to the local surface-tension coefficient in that cell.

The choice between $\sigma_{c,x}$ and $\sigma_{c,y}$ follows the same orientation criterion as the height-function curvature calculation. If $|n_x|>|n_y|$, the interface is represented locally by an $x$-oriented height function $x=h_x(y)$, and $\sigma_{c,x}$ is used. The tangential derivative is then approximated by
\begin{equation}
\left(
\frac{\partial\sigma}{\partial s}
\right)_{i,j}
=
\frac{(\sigma_{c,x})_{j+1}-(\sigma_{c,x})_{j-1}}{ds},
\qquad
ds
=
2\Delta
\sqrt{
1+
\left(
\frac{\partial h_x}{\partial y}
\right)^2
}.
\label{Ae4}
\end{equation}
Conversely, if $|n_y|>|n_x|$, the interface is represented locally by a $y$-oriented height function $y=h_y(x)$, and $\sigma_{c,y}$ is used:
\begin{equation}
\left(
\frac{\partial\sigma}{\partial s}
\right)_{i,j}
=
\frac{(\sigma_{c,y})_{i+1}-(\sigma_{c,y})_{i-1}}{ds},
\qquad
ds
=
2\Delta
\sqrt{
1+
\left(
\frac{\partial h_y}{\partial x}
\right)^2
}.
\label{Ae5}
\end{equation}
Here, $\Delta$ is the grid size, and $ds$ is the corresponding interfacial arclength between the two neighboring height-function positions.

Finally, the scalar tangential derivative is projected onto the Cartesian components of the unit tangent vector. For the $x$-oriented height function, the tangent vector is chosen along the positive $y$ direction as $\boldsymbol{t} = (\partial h_x/\partial y,\;1)/
\sqrt{1+(\partial h_x/\partial y)^2}$, whereas for the $y$-oriented height function it is chosen along the positive $x$ direction as $\boldsymbol{t} = \left(1,\;\partial h_y/\partial x
\right)/\sqrt{1+(\partial h_y/\partial x)^2}$. The Marangoni force is then decomposed into its Cartesian components and added to the momentum equation as
\begin{equation}
F_{st,x}
=
\frac{\partial\sigma}{\partial s}t_x\delta_s,
\label{Ae6}
\end{equation}
and
\begin{equation}
F_{st,y}
=
\frac{\partial\sigma}{\partial s}t_y\delta_s,
\label{Ae7}
\end{equation}
where $t_x$ and $t_y$ are the two components of $\boldsymbol{t}$. This procedure evaluates the surface-tension gradient along the interface itself and avoids differentiating discontinuous temperature or concentration fields across the liquid--gas interface.

\section{Saturation vapor-pressure correlations}\label{app2}

The saturation vapor pressures used in the droplet-evaporation simulations are evaluated from the Antoine equations,
\begin{subequations}
\begin{align}
\mathrm{n\mbox{-}heptane:}\qquad
\log_{10} P^{sat}
&=
4.02832
-
\frac{1268.636}{T-56.199},
\label{Be1}
\\[0.8em]
\mathrm{ethanol:}\qquad
\log_{10} P^{sat}
&=
5.33675
-
\frac{1648.22}{T+230.918-273.15},
\label{Be2}
\\[0.8em]
\mathrm{isooctane:}\qquad
\log_{10} P^{sat}
&=
3.93646
-
\frac{1257.85}{T+220.767-273.15}.
\label{Be3}
\end{align}
\end{subequations}
Here, $P^{sat}$ is expressed in bar and $T$ in Kelvin.

\end{document}